\documentclass[twocolumn,floatfix,aps,prd,nofootinbib,showpacs]{revtex4}

\usepackage{graphicx}

\begin{document}

\def\agt{\mathrel{\raise.3ex\hbox{$>$}\mkern-14mu\lower0.6ex\hbox{$\sim$}}}
\def\alt{\mathrel{\raise.3ex\hbox{$<$}\mkern-14mu\lower0.6ex\hbox{$\sim$}}}

\newcommand{\beq}{\begin{equation}}
\newcommand{\eeq}{\end{equation}}
\newcommand{\beqn}{\begin{eqnarray}}
\newcommand{\eeqn}{\end{eqnarray}}
\newcommand{\pa}{\partial}
\newcommand{\vp}{\varphi}
\newcommand{\varep}{\varepsilon}
\newcommand{\ep}{\epsilon}
\newcommand{\comp}{(M/R)_\infty} 
\def\bI{\hbox{$\,I\!\!\!$--}}

\title{Merger of binary neutron stars with realistic
  equations of state in full general relativity}

\author{Masaru Shibata$^1$, Keisuke Taniguchi$^2$, and K\=oji Ury\=u$^3$}

\affiliation{$^1$ Graduate School of Arts and Sciences, 
University of Tokyo, Komaba, Meguro, Tokyo 153-8902, Japan \\
$^2$ Department of Physics, University of Illinois at Urbana-Champaign,
Urbana, IL 61801 \\
$^3$ Astrophysical Sector, SISSA, via Beirut 2/4, Trieste 34013, Italy
}

\begin{abstract}
~\\
We present numerical results of three-dimensional simulations for
the merger of binary neutron stars in full general relativity. Hybrid
equations of state are adopted to mimic realistic nuclear equations of
state. In this approach, we divide the equations of state into two
parts as $P=P_{\rm cold}+P_{\rm th}$. $P_{\rm cold}$ is the cold part
for which we assign a fitting formula for realistic equations of state
of cold nuclear matter slightly modifying the formula developed by
Haensel and Potekhin.  We adopt the SLy and FPS equations of
state for which the maximum allowed ADM mass of cold and spherical
neutron stars is $\approx 2.04M_{\odot}$ and $1.80M_{\odot}$,
respectively. $P_{\rm th}$ denotes the
thermal part which is written as $P_{\rm th}=(\Gamma_{\rm th}-1)\rho
(\varep-\varep_{\rm cold})$, where $\rho$, $\varep$, $\varep_{\rm
cold}$, and $\Gamma_{\rm th}$ are the baryon rest-mass density, total
specific internal energy, specific internal energy of the cold part,
and the adiabatic constant, respectively.  Simulations are performed
for binary neutron stars of the total ADM mass in the range between
$2.4M_{\odot}$ and $2.8M_{\odot}$ with the rest-mass ratio $Q_M$ to be
in the range $0.9 \alt Q_M \leq 1$. It is found that if the total ADM
mass of the system is larger than a threshold $M_{\rm thr}$, a black hole is
promptly formed in the merger irrespective of the mass ratios. In the
other case, the outcome is a hypermassive neutron star of a large
ellipticity, which results from the large adiabatic index
of the realistic equations of state adopted. The value of $M_{\rm thr}$ 
depends on the equation of state: $M_{\rm thr} \sim 2.7M_{\odot}$ and
$\sim 2.5M_{\odot}$ for the SLy and FPS equations of state, respectively. 
Gravitational waves are 
computed in terms of a gauge-invariant wave extraction technique. In
the formation of the hypermassive neutron star, quasiperiodic
gravitational waves of a large amplitude and of frequency between 3
and 4 kHz are emitted. The estimated emission time scale is $\alt 100$
ms, after which the hypermassive neutron star collapses to a black
hole. Because of the long emission time, the effective amplitude may
be large enough to be detected by advanced laser interferometric 
gravitational wave detectors if the distance to the source is smaller
than $\sim 100$ Mpc.  Thermal properties of the outcome formed after 
the merger are also analyzed to approximately estimate the neutrino
emission energy.
\end{abstract}

\maketitle

\section{Introduction}

Binary neutron stars \cite{HT,Stairs} inspiral as a
result of the radiation reaction of gravitational waves, and
eventually merge. The most optimistic scenario based mainly on a
recent discovery of binary system PSRJ0737-3039 \cite{NEW} suggests
that such mergers may occur approximately once per year within a
distance of about 50 Mpc \cite{BNST}. Even the most conservative
scenario predicts an event rate approximately once per year within a
distance of about 100 Mpc \cite{BNST}. This indicates that 
the detection rate of gravitational waves by the advanced 
LIGO will be $\sim 40$--600 yr$^{-1}$. Thus, the merger of
binary neutron stars is one of the most promising sources for 
kilometer-size laser interferometric detectors \cite{KIP,Ando}. 

Hydrodynamic simulations employing full general relativity provide the
best approach for studying the merger of binary neutron stars. Over
the last few years, numerical methods for solving coupled equations of
the Einstein and hydrodynamic equations have been developed
\cite{gr3d,bina,bina2,other,Font,STU,marks,illinois,Baiotti}
and now such simulations are feasible with an accuracy high enough
for yielding scientific results \cite{STU}. With the current
implementation, radiation reaction of gravitational waves in the merger
of binary neutron stars can be taken into account within $\sim 1\%$ error 
in an appropriate computational setting \cite{STU}.  This fact
illustrates that it is now a robust tool for the detailed theoretical
study of astrophysical phenomena and gravitational waves emitted.

So far, all the simulations for the merger of binary neutron stars in
full general relativity have been performed adopting an ideal equation
of state \cite{bina,bina2,STU,marks,illinois}. (But, see \cite{OUPT}
for a simulation in an approximately general relativistic gravity.) 
For making better models of the merger which can be used for 
quantitative comparison with observational data, it is necessary to
adopt more realistic equations of state as the next step.  Since the
lifetime (from the birth to the merger) of observed binary neutron
stars is longer than $\sim 100$ million yrs \cite{Stairs}, the thermal
energy per nucleon in each neutron star will be much lower than the
Fermi energy of neutrons \cite{ST,Tsuruta} at the onset of the merger.
This implies that for modeling the binary neutron stars just before
the merger, it is appropriate to use cold nuclear equations of
state. During the merger, shocks will be formed and the kinetic energy
will be converted to the thermal energy. However, previous studies
have indicated that the shocks are not very strong in the merger of
binary neutron stars, in contrast to those in iron core collapse of
massive stars. The reason is that the approaching velocity at the
first contact of two neutron stars is much smaller than the orbital
velocity and the sound speed of nuclear matter (e.g., \cite{STU}).
This implies that the pressure and the internal energy associated with
the finite thermal energy (temperature) are not still as large as
those of the cold part.  From this reason, we adopt a hybrid equation
of state in which the finite-temperature part generated by shocks is
added as a correction using a simple prescription (see Sec. II B).  On
the other hand, realistic equations of state are assigned to the cold
part \cite{PR,DH,HP}.

The motivation which stimulates us to perform new simulations is that
the stiffness and adiabatic index of the realistic equations of state
are quite different from those in the $\Gamma$-law equation of state
with $\Gamma=2$ (hereafter referred to as the $\Gamma=2$ equation of
state) which has been widely adopted so far (e.g., \cite{STU})
\footnote{In this paper, we distinguish the stiffness and
the magnitude of the adiabatic index clearly. 
We mention ``the equation of state is softer (stiffer)''
when the pressure at a given density is smaller (larger) than another. 
Thus, even if the adiabatic index is larger for 
the supranuclear density, the equation of state
may be softer in the case that the pressure is smaller. }. 
It can be expected that these differences will modify the 
properties of the merger quantitatively 
as described in the following. 

Since the realistic equations of state are softer than the $\Gamma=2$ one, 
each neutron star becomes more compact (cf. Fig. \ref{FIG2}). 
This implies that the merger will set in at a more compact state which
is reached after more energy and angular momentum are already dissipated 
by gravitational radiation. Namely, compactness of 
the system at the onset of the merger is larger. 
This will modify the dynamics of the merger, and accordingly, 
the threshold mass for prompt black hole formation
(hereafter $M_{\rm thr}$) will be changed.

The adiabatic index of the equations of state is also different from
that for the $\Gamma=2$ equation of state. This 
will modify the shape of the hypermassive neutron stars 
\footnote{The hypermassive neutron star is defined as
a differentially rotating neutron star for which the total baryon rest-mass
is larger than the
maximum allowed value of rigidly rotating neutron stars for a
given equation of state: See \cite{BSS} for definition.}, 
which are formed after the merger in the case that 
the total mass is smaller than $M_{\rm thr}$. Previous Newtonian and post 
Newtonian studies \cite{RS,C,FR} have indicated that for smaller adiabatic
index of the equations of state, the degree of 
the nonaxial symmetry of the formed neutron star becomes smaller. However,
if its value is sufficiently large, the formed neutron star can be
ellipsoidal. As a result of 
this change, the amplitude of gravitational waves emitted from the formed 
neutron star is significantly changed. 
Since the adiabatic index of the realistic equations of state is much 
larger than that of the $\Gamma=2$ equation of state for supranuclear
density \cite{PR,DH,HP}, the significant modification in the shape of the 
hypermassive neutron stars and in the amplitude of gravitational
waves emitted from them is expected. 

The paper is organized as follows. In Sec. II A--C, 
basic equations, gauge conditions, methods for extracting 
gravitational waves, and quantities used in the analysis for
numerical results are reviewed. 
Then, the hybrid equations of state adopted in this paper are
described in Sec. II D. 
In Sec. III, after briefly describing the computational setting and
the method for computation of initial condition, 
the numerical results are presented. We 
pay particular attention to the merger process, the outcome, and
gravitational waveforms. Section IV is devoted to a summary. 
Throughout this paper, we adopt the geometrical 
units in which $G=c=1$ where $G$ and $c$ 
are the gravitational constant and the speed of light.  
Latin and Greek indices denote spatial components ($x, y, z$) 
and space-time components ($t, x, y, z$), respectively:
$r \equiv \sqrt{x^2+y^2+z^2}$. 
$\delta_{ij}(=\delta^{ij})$ denotes the Kronecker delta. 

\section{Formulation}

\subsection{Summary of formulation}

Our formulation and numerical scheme for fully general relativistic
simulations in three spatial dimensions 
are the same as in~\cite{STU}, to which the reader may refer for
details of basic equations and successful numerical results.

The fundamental variables for the hydrodynamics are 
$\rho$: rest-mass density, 
$\varep$ : specific internal energy, 
$P$ : pressure, $u^{\mu}$ : four velocity, and
\beqn
v^i ={dx^i \over dt}={u^i \over u^t},
\eeqn
where subscripts $i, j, k, \cdots$ denote $x, y$ and $z$, and 
$\mu$ the spacetime components.
The fundamental variables for geometry are 
$\alpha$: lapse function, $\beta^k$: shift vector, 
$\gamma_{ij}$: metric in three-dimensional spatial hypersurface,
$\gamma\equiv e^{12\phi}={\rm det}(\gamma_{ij})$, 
$\tilde \gamma_{ij}=e^{-4\phi}\gamma_{ij}$: conformal three-metric, and 
$K_{ij}$: extrinsic curvature. 

For a numerical implementation of the hydrodynamic equations, 
we define a weighted density, 
a weighted four-velocity, and a specific energy defined, respectively, by
\beqn
&&\rho_* \equiv \rho \alpha u^t e^{6\phi}, \\
&&\hat u_i \equiv h u_i, \\
&& \hat e \equiv h\alpha u^t - {P \over \rho \alpha u^t},
\eeqn
where $h=1+\varepsilon+P/\rho$ denotes the specific enthalpy. 
General relativistic hydrodynamic equations are written
into the conservative form for variables $\rho_*$, $\rho_* \hat u_i$,
and $\rho_* \hat e$, and solved using a high-resolution shock-capturing
scheme \cite{Font}. 
In our approach, the transport terms such as $\pa_i (\cdots)$ are computed by 
an approximate Riemann solver with third-order (piecewise parabolic)
spatial interpolation with a Roe-type averaging \cite{shiba2d}.
At each time step, $\alpha u^t$ is determined by solving
an algebraic equation derived from the normalization $u^{\mu}u_{\mu}=-1$, and
then, the primitive variables such as $\rho$, $\epsilon$, and $v^i$ are
updated. An atmosphere of small density 
$\rho \sim 10^9~{\rm g/cm^3}$ is added uniformly 
outside neutron stars at $t=0$, 
since the vacuum is not allowed in the shock-capturing scheme. 
The integrated mass of the atmosphere is at most $1\%$ of the total mass
in the present simulation. Furthermore, we add a friction term for a
matter of low density $\sim 10^9~{\rm g/cm^3}$ to 
avoid infall of such atmosphere toward the central region. 
Hence, the effect of the atmosphere for the evolution 
of binary neutron stars is very small. 

The Einstein evolution equations are solved using a version of the
BSSN formalism following previous papers \cite{SN,gr3d,bina2,STU}: 
We evolve $\tilde \gamma_{ij}$, $\phi$, 
$\tilde A_{ij} \equiv e^{-4\phi}(K_{ij}-\gamma_{ij} K_k^{~k})$,
and the trace of the extrinsic curvature $K_k^{~k}$ 
together with three auxiliary functions
$F_i\equiv \delta^{jk}\pa_{j} \tilde \gamma_{ik}$ using an
unconstrained free evolution code. The latest version of our
formulation and numerical method is described in \cite{STU}.
The point worthy to note is that the equation for $\phi$ is
written to a conservative form similar to the continuity equation, 
and solving this improves the accuracy of the conservation of the ADM mass and
angular momentum significantly. 

As the time slicing condition, 
an approximate maximal slice (AMS) condition $K_k^{~k} \approx 0$ 
is adopted following previous papers \cite{bina2}.
As the spatial gauge condition, we adopt a
hyperbolic gauge condition as in \cite{S03,STU}. 
Successful numerical results for the merger of binary neutron stars
in these gauge conditions are presented in \cite{STU}. 
In the presence of a black hole, the location is
determined using an apparent horizon finder for which the method
is described in \cite{AH}. 

Following previous works, we adopt binary neutron stars in
quasiequilibrium circular orbits as the initial condition.  In
computing the quasiequilibrium state, we use the so-called conformally flat 
formalism for the Einstein equation \cite{WM}. A solution in
this formalism satisfies the constraint equations in general relativity, 
and hence, it can be used for the initial condition. The irrotational 
velocity field is assumed since it is considered to be a good 
approximation for coalescing binary neutron stars in nature 
\cite{CBS}. The coupled equations of the field and hydrostatic 
equations \cite{irre} are solved by a pseudospectral method developed by
Bonazzola, Gourgoulhon, and Marck \cite{GBM}. Detailed numerical
calculations have been done by Taniguchi and part of the numerical
results are presented in \cite{TG}.

\subsection{Extracting gravitational waves}

Gravitational waves are computed in terms of the gauge-invariant 
Moncrief variables in a flat spacetime \cite{moncrief} as
we have been carried out in our series of paper (e.g., \cite{gw3p2,STU,SS3}).
The detailed equations are describe in \cite{STU,SS3} to which the reader 
may refer. In this method, we split the metric in the wave zone into the
flat background and linear perturbation. Then, 
the linear part is decomposed using the tensor spherical harmonics and 
gauge-invariant variables are constructed for each mode of
eigen values $(l,m)$.
The gauge-invariant variables of $l \geq 2$ can be 
regarded as gravitational waves in the wave zone, and hence, 
we focus on such mode. In the merger of binary neutron stars
of nearly equal mass, the even-parity mode of $(l, |m|)=(2, 2)$ is much
larger than other modes. Thus, in the following, we pay attention only 
to this mode. 

Using the gauge-invariant variables, the luminosity and the angular
momentum flux of gravitational waves can be defined by
\beqn
&&{dE \over dt}={r^2 \over 32\pi}\sum_{l,m}\Bigl[
|\pa_t R_{lm}^{\rm E}|^2+|\pa_t R_{lm}^{\rm O}|^2 \Bigr],
\label{dedt} \\
&&{dJ \over dt}={r^2 \over 32\pi}\sum_{l,m}\Bigl[
 |m(\pa_t R_{lm}^{\rm E}) R_{lm}^{\rm E} |
+|m(\pa_t R_{lm}^{\rm O}) R_{lm}^{\rm O} | \Bigr], 
\label{dJdt} 
\eeqn
where $R_{lm}^{\rm E}$ and $R_{lm}^{\rm O}$ are the
gauge-invariant variables of even and odd parities. 
The total radiated energy and angular momentum are obtained by
the time integration of $dE/dt$ and $dJ/dt$.

To search for the characteristic frequencies of gravitational waves,
the Fourier spectra are computed by 
\beq
\bar R_{lm}(f)=\int e^{2\pi i f t} R_{lm}(t)dt,
\eeq
where $f$ denotes a frequency of gravitational waves. 
Using the Fourier spectrum, the energy power spectrum is defined as 
\beq
{dE \over df}={\pi \over 4}r^2 \sum_{l\geq 2, m\geq 0}
|\bar R_{lm}(f) f|^2 ~~~(f > 0), \label{power}
\eeq
where for $m\not=0$, we define 
\beq
\bar R_{lm}(f)
\equiv \sqrt{|\bar R_{l m}(f)|^2 + |\bar R_{l -m}(f)|^2}~~(m>0), 
\eeq
and use $|\bar R_{lm}(-f)|=|\bar R_{lm}(f)|$ for
deriving Eq. (\ref{power}). 

We also use a quadrupole formula which is described in \cite{SS1,SS2,SS3}.
As shown in \cite{SS1}, a kind of quadrupole formula can provide 
approximate gravitational waveforms from oscillating compact stars.
In this paper, the applicability is tested for the
merger of binary neutron stars. 

In quadrupole formulas, gravitational waves are computed from 
\beqn
h_{ij}=\biggl[P_{i}^{~k} P_{j}^{~l}-{1 \over 2}P_{ij}P^{kl}\biggr]
\biggl({2 \over r}{d^2\bI_{kl} \over dt^2}\biggr),\label{quadf}
\eeqn
where $\bI_{ij}$ and $P_{ij}=\delta_{ij}-n_i n_j$ ($n_i=x^i/r$)
denote a tracefree quadrupole moment and a projection tensor. 

In fully general relativistic and dynamical spacetimes, 
there is no unique definition for the quadrupole moment $I_{ij}$.
Following \cite{SS1,SS2,SS3}, we choose the formula as 
\beq
I_{ij} = \int \rho_* x^i x^j d^3x. 
\eeq
Then, using the continuity equation, 
the first time derivative is computed as 
\beq
\dot I_{ij} = \int \rho_* (v^i x^j +x^i v^j)d^3x.
\eeq
To compute $\ddot I_{ij}$, we 
carry out the finite differencing of the numerical result 
for $\dot I_{ij}$.

In this paper, we focus only on $l=2$ mass quadrupole modes. 
Then, the gravitational waveforms are written as
\beqn
&&h_+=
{1 \over r}\biggl[ \sqrt{{5 \over 64\pi}}
\{ R_{22+}(1+\cos^2\theta)\cos(2\varphi) \nonumber \\
&&~~~~~~~~~~~~~~~~~~+R_{22-}(1+\cos^2\theta)\sin(2\varphi) \} \nonumber \\
&&~~~~~~~~~~
+ \sqrt{{15 \over 64\pi}}R_{20} \sin^2\theta \biggr],\label{eq34} \\
&&h_{\times}={2 \over r} \sqrt{{5 \over 64\pi}}
\Bigl[ -R_{22+} \cos\theta \sin(2\varphi) \nonumber \\
&&~~~~~~~~~~~~~~~~~~~+R_{22-} \cos\theta \cos(2\varphi)\Bigr], \label{eq35}
\eeqn
in the gauge-invariant wave extraction technique, and
\beqn
&&h_+=
{1 \over r} \biggl[ {\ddot I_{xx}-\ddot I_{yy} \over 2}
(1+\cos^2\theta)\cos(2\varphi) \nonumber \\
&&~~~~~~~~~~~+\ddot I_{xy}(1+\cos^2 \theta) \sin(2\varphi) \nonumber \\
&&~~~~~~~~~~~+\biggl( \ddot I_{zz}-{\ddot I_{xx}+\ddot I_{yy} \over 2} \biggr)
\sin^2\theta \biggr],\\
&&h_{\times}=
{2 \over r} \biggl[ -{\ddot I_{xx}-\ddot I_{yy} \over 2}
\cos\theta \sin(2\varphi)\nonumber \\
&&~~~~~~~~~~~~~
+\ddot I_{xy} \cos\theta \cos(2\varphi) \biggr], \label{quadform}
\eeqn
in the quadrupole formula. In Eqs. (\ref{eq34}) and (\ref{eq35}),
we use the variables defined by 
\beqn
&&R_{22\pm} \equiv {R_{22}^{\rm E} \pm R_{2~-2}^{\rm E} \over \sqrt{2}}r,\\
&&R_{20} \equiv R_{20}^{\rm E} r.
\eeqn
For the derivation of $h_+$ and $h_{\times}$, 
we assume that the wave part of the
spatial metric in the wave zone is written as 
\beqn
dl^2&&=dr^2+r^2[(1+h_+)d\theta^2+\sin^2\theta(1-h_+)d\varphi^2 \nonumber \\
&&~~~~~~~~~~~~~
+2 \sin\theta h_{\times} d\theta d\varphi], 
\eeqn
and set $R_{2~\pm 1}^{\rm E}=0$ and 
$I_{xz}=I_{yz}=0$ since we assume the reflection symmetry
with respect to the equatorial plane. 

In the following, we present
\beqn
&& R_+=\sqrt{{5 \over 16\pi}}R_{22+},\\
&& R_{\times}=\sqrt{{5 \over 16\pi}}R_{22-},
\eeqn
in the gauge-invariant wave extraction method,
and as the corresponding variables, 
\beqn
&& A_+=\ddot I_{xx}-\ddot I_{yy},\\
&& A_{\times}=2 \ddot I_{xy},
\eeqn
in the quadrupole formula.
These have the unit of length and provide the amplitude of a given mode
measured by an observer located in the most optimistic direction. 

\subsection{Definitions of quantities and methods for calibration}

In numerical simulations, we refer to the total baryon rest-mass, 
the ADM mass, and the angular momentum of the system, which are given by 
\beqn
M_* &&\equiv \int \rho_* d^3x, \\
M &&\equiv -{1 \over 2\pi} 
\oint_{r\rightarrow\infty} \pa_i \psi dS_i \nonumber \\
&&=\int \biggl[ \rho_{\rm H} e^{5\phi} +{e^{5\phi} \over 16\pi}
\biggl(\tilde A_{ij} \tilde A^{ij}-{2 \over 3}(K_k^{~k})^2 \nonumber \\
&&~~~~~~~~~~~~~~~~~~~~~~~~~~~~~~-\tilde R_k^{~k} 
e^{-4\phi}\biggr)\biggr]d^3x, \label{eqm00}\\
J &&\equiv {1 \over 8\pi}\oint_{r\rightarrow\infty} 
\varphi^ i \tilde A_i^{~j} e^{6\phi} dS_j \nonumber \\
&&=\int e^{6\phi}\biggl[J_i \varphi^i  
+{1 \over 8\pi}\biggl( \tilde A_i^{~j} \pa_j \varphi^i
-{1 \over 2}\tilde A_{ij}\varphi^k\pa_k \tilde \gamma^{ij}
\nonumber \\
&&~~~~~~~~~~~~~~~~~~~~~~~~~~
+{2 \over 3}\varphi^j \pa_j K_k^{~k} \biggr) \biggr]d^3x,
\label{eqj00}
\eeqn
where $dS_j=r^2 \pa_j r d(\cos\theta)d\varphi$, 
$\varphi^j=-y(\pa_x)^j + x(\pa_y)^j$, $\psi=e^{\phi}$, 
$\rho_{\rm H}=\rho \alpha u^t \hat e$, $J_i=\rho \hat u_i$, and
$\tilde R_k^{~k}$ denotes the Ricci scalar with respect to
$\tilde \gamma_{ij}$. 
To derive the expressions for $M$ and $J$ in the form of
volume integral, the Gauss law is used. 
Here, $M_*$ is a conserved quantity. 
We also use the notations $M_{*1}$ and $M_{*2}$ which 
denote the baryon rest-mass of the primary and secondary neutron stars, 
respectively. In terms of them, the baryon rest-mass ratio is defined 
by $Q_M=M_{*2}/M_{*1} (\leq 1)$. 

In numerical simulation, 
$M$ and $J$ are computed using the volume integral
shown in Eqs. (\ref{eqm00}) and (\ref{eqj00}).
Since the computational domain is finite, 
they are not constant and decrease 
after gravitational waves propagate to the outside of the
computational domain during time evolution.
Therefore, in the following, they are referred to as 
the ADM mass and the angular momentum computed in the finite domain 
(or simply as $M$ and $J$, which decrease with time).

The decrease rates of $M$ and $J$ should be equal to the emission rates 
of the energy and the angular momentum by gravitational radiation
according to the conservation law.
Denoting the radiated energy and angular momentum
from the beginning of the simulation to the time $t$ 
as $\Delta E(t)$ and $\Delta J(t)$, the conservation relations are
written as
\beqn
&&M(t) + \Delta E(t) =M_0,\label{eqm01}\\
&&J(t) + \Delta J(t) =J_0,\label{eqj01}
\eeqn
where $M_0$ and $J_0$ are the initial values of $M$ and $J$.
We check if these conservation laws hold during the simulation.

Significant violation of the conservation laws 
indicates that the radiation reaction of gravitational waves
is not taken into account accurately. During the merger of
binary neutron stars, the angular momentum is dissipated by several $10\%$, 
and thus, the dissipation effect plays an important role in the evolution
of the system. Therefore, it is required to confirm that 
the radiation reaction is computed accurately. 

The violation of the Hamiltonian constraint 
is locally measured by the equation as 
\beqn
\displaystyle
f_{\psi} &&\equiv 
\Bigl|\tilde \Delta \psi - {\psi \over 8}\tilde R_k^{~k} 
+ 2\pi \rho_{\rm H} \psi^5  \nonumber \\
&&~~~~~~~~~+{\psi^5 \over 8} \Bigl(\tilde A_{ij} \tilde A^{ij}
-{2 \over 3}(K_k^{~k})^2\Bigr)\Bigr|  \nonumber \\
&&~~~\biggl[|\tilde \Delta \psi | + |{\psi \over 8}\tilde R_k^{~k}| 
+ |2\pi \rho_{\rm H} \psi^5|  \nonumber \\
&&~~~~~~~~~+{\psi^5 \over 8} \Bigl(|\tilde A_{ij} \tilde A^{ij}|+
{2 \over 3}(K_k^{~k})^2\Bigr)\biggr]^{-1}. 
\eeqn
Following \cite{shiba2d}, we define and monitor a global quantity as 
\beq
H \equiv {1 \over M_*} \int \rho_* f_{\psi} d^3x. \label{vioham}
\eeq
Hereafter, this quantity will be referred to as the averaged violation
of the Hamiltonian constraint.

\subsection{Equations of state}

Since the lifetime of binary neutron stars from the birth to the merger
is longer than $\sim 100$ million yrs for the observed systems \cite{Stairs},
the temperature of each neutron star will be very low ($\alt 10^5$ K)
\cite{ST,Tsuruta} at the onset of merger; i.e., the thermal energy
per nucleon is much smaller than the Fermi energy of neutrons. 
This implies that for modeling the binary neutron stars
just before the merger, it is appropriate to use cold nuclear 
equations of state. On the other hand, during the merger, 
shocks will be formed and the kinetic energy will be converted to the 
thermal energy to increase the temperature.
However, previous studies have indicated 
that the shocks in the merger are not strong enough to increase
the thermal energy to the level as large as the Fermi energy
of neutrons, since the approaching velocity at the first contact
of two neutron stars is much smaller than the orbital velocity
and the sound speed of nuclear matter.
This implies that the pressure and the internal energy associated with
the finite temperature are not still as large as those of the cold
part. From this reason, we adopt a hybrid equation of state. 

\begin{table*}[t]
\vspace{-4mm}
\begin{center}
\begin{tabular}{lllllll} \hline
& $i$ & $p_i$ (SLy) & $p_i$ (FPS) &
$i$ & $p_i$ (SLy) & $p_i$ (FPS) \\ \hline
&1 & 0.1037 & 0.15806 & 9 & $9\times 10^5$ & $9 \times 10^5$ \\ \hline
&2 & 0.1956 & 0.220   & 10 & 4 & 5 \\ \hline
&3 & 39264  & 5956.4  & 11 & 0.75 & 0.75 \\ \hline
&4 & 1.9503 & 1.633   & 12 & 0.057 & 0.0627 \\ \hline
&5 & 254.83 & 170.68  & 13 & 0.138 & 0.1387 \\ \hline
&6 & 1.3823 & 1.1056  & 14 & 0.84 & 0.56 \\ \hline
&7 & $-1.234$ & $-0.703$  & 15 & 0.338 & 0.308 \\ \hline
&8 & $1.2 \times 10^5$  & $2 \times 10^4$     & & & \\ \hline
\end{tabular}
\caption{The values of $p_{i}$ we choose in units of $c=G=M_{\odot}=1$. 
}
\end{center}
\vspace{-5mm}
\end{table*}

In this equation of state,
we write the pressure and the specific internal energy in the form
\beqn
P=P_{\rm cold} + P_{\rm th},\\
\varep=\varep_{\rm cold} + \varep_{\rm th},
\eeqn
where $P_{\rm cold}$ and $\varep_{\rm cold}$ are the cold (zero-temperature)
parts, and are written as functions of $\rho$. 
$P_{\rm th}$ and $\varep_{\rm th}$ are the thermal
(finite-temperature) parts. During the simulation, 
$\rho$ and $\varep$ are computed from hydrodynamic variables
$\rho_*$ and $\hat e$. Thus, $\varep_{\rm th}$ is determined by
$\varep-\varep_{\rm cold}$. 

For the cold parts, we assign realistic equations of state
for zero-temperature nuclear matter. In this paper, 
we adopt the SLy \cite{DH} and
FPS equations of state \cite{PR}. These 
are tabulated as functions of the baryon rest-mass 
density for a wide density range 
from $\sim 10~{\rm g/cm^3}$ to $\sim 10^{16}~{\rm g/cm^3}$. 
To simplify numerical implementation for simulation,
we make fitting formulae from the tables of equations of state, 
slightly modifying the original approach proposed in  \cite{HP}. 

In our approach, we first make a fitting formula for $\varep_{\rm cold}$ as
\beqn
&&\varep_{\rm cold}(\rho)
=[(1+p_1\rho^{p_2}+p_3\rho^{p_4})(1+p_5 \rho^{p_6})^{p_7}
-1]\nonumber \\
&&~~~~~~~~~~~~~ \times f(-p_8\rho+p_{10})\nonumber \\
&&~~~~~~~~~~~~~ +p_{12} \rho^{p_{13}}f(p_8\rho-p_{10})f(-p_9\rho+p_{11})
\nonumber \\
&&~~~~~~~~~~~~~  +p_{14}\rho^{p_{15}}f(p_9\rho-p_{11}),
\eeqn
where 
\beqn
f(x)={1 \over e^x +1}. 
\eeqn
The coefficients $p_i~(i=$1--15)
denote constants, and are listed in Table I. In
making the formula, we focus only on the density
for $\rho \geq 10^{10}~{\rm g/cm^3}$ in this work,
since the matter of lower density does not play an
important role in the merger. 
Then, the pressure is computed from the thermodynamic relation
in the zero-temperature limit 
\beqn
P_{\rm cold} =\rho^2 {d \varep_{\rm cold} \over d\rho}. 
\eeqn
With this approach, the accuracy of the fitting for the pressure
is not as good as that in \cite{HP}. However,
the first law of the thermodynamics is completely satisfied in
contrast to that in \cite{HP}.

\begin{figure*}[thb]
\vspace{-4mm}
\begin{center}
  (a)\includegraphics[width=3.in]{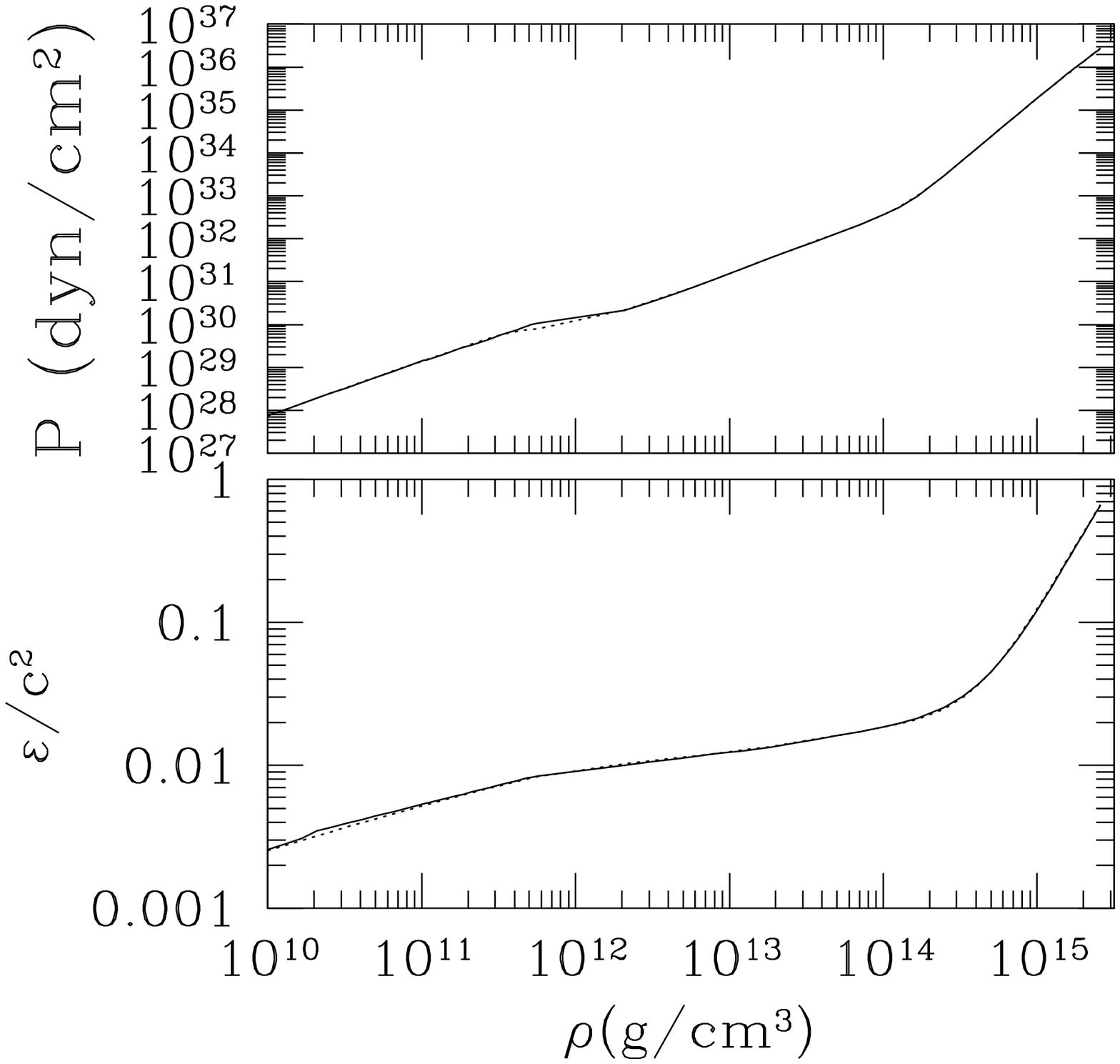}
  ~~(b)\includegraphics[width=3.in]{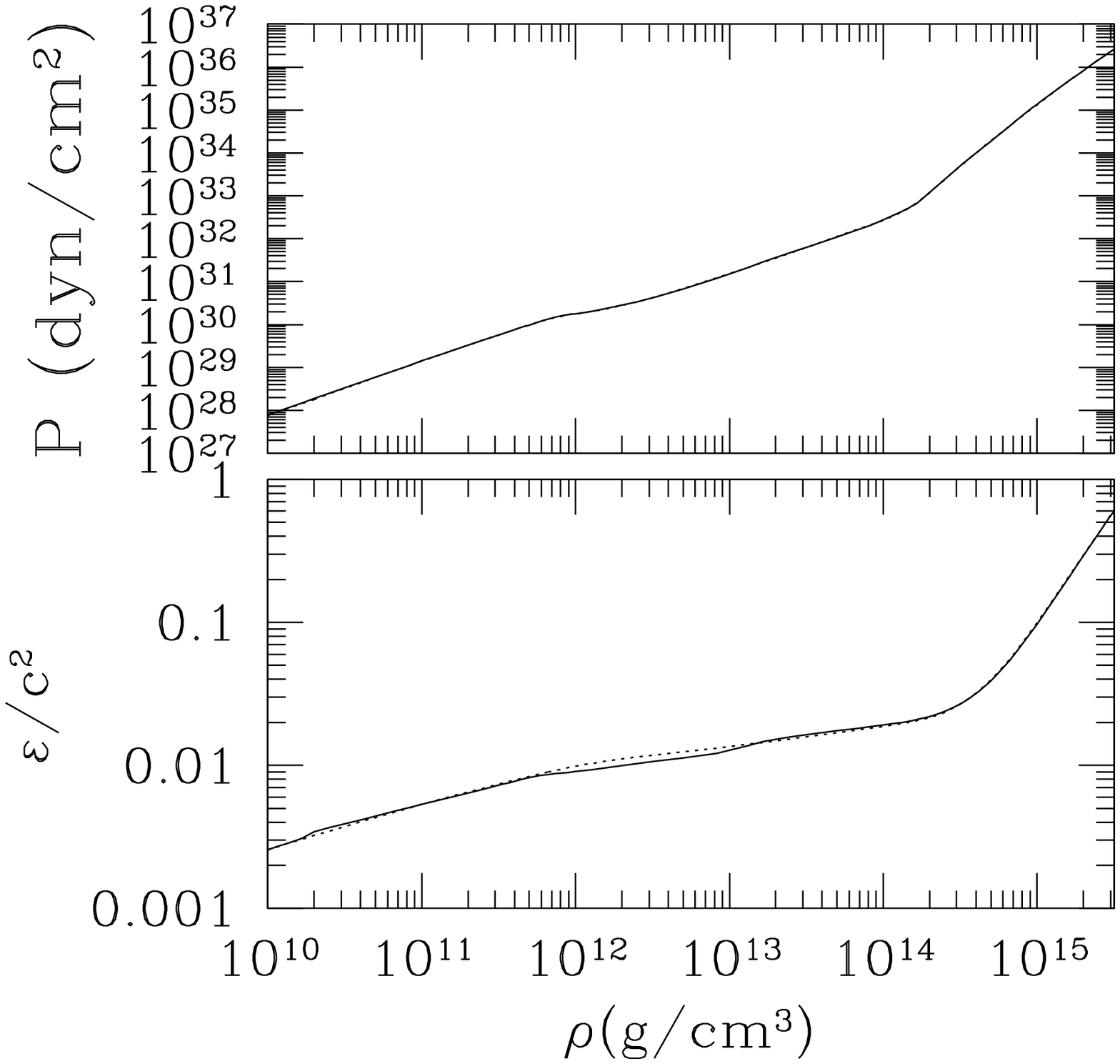}
\end{center}
\vspace{-2mm}
\caption{Pressure and specific internal energy
as functions of baryon rest-mass density $\rho$ 
(a) for the SLy and (b) for the FPS equations of state.
The solid and dotted curves denote the results by
fitting formulae and numerical data tabulated, respectively. 
\label{FIG1} }
\end{figure*}

In Fig. 1, we compare $P_{\rm cold}$ and $\varep_{\rm cold}$
calculated by the fitting formulae (solid curves) with the numerical data 
tabulated (dotted curves) \footnote{
The tables for the SLy and FPS equations of state, which were involved
in the LORENE library in Meudon group (http://www.lorene.obspm.fr),
were implemented by Haensel and Zdunik. 
}.
It is found that two results agree approximately.
The relative error between two is within $\sim 10\%$ for
$\rho > 10^{10}~{\rm g/cm^3}$ 
and $\alt 2\%$ for supranuclear density with
$\rho \agt 2 \times 10^{14}~{\rm g/cm^3}$.

\begin{figure*}[thb]
\vspace{-4mm}
\begin{center}
  (a)\includegraphics[width=3.in]{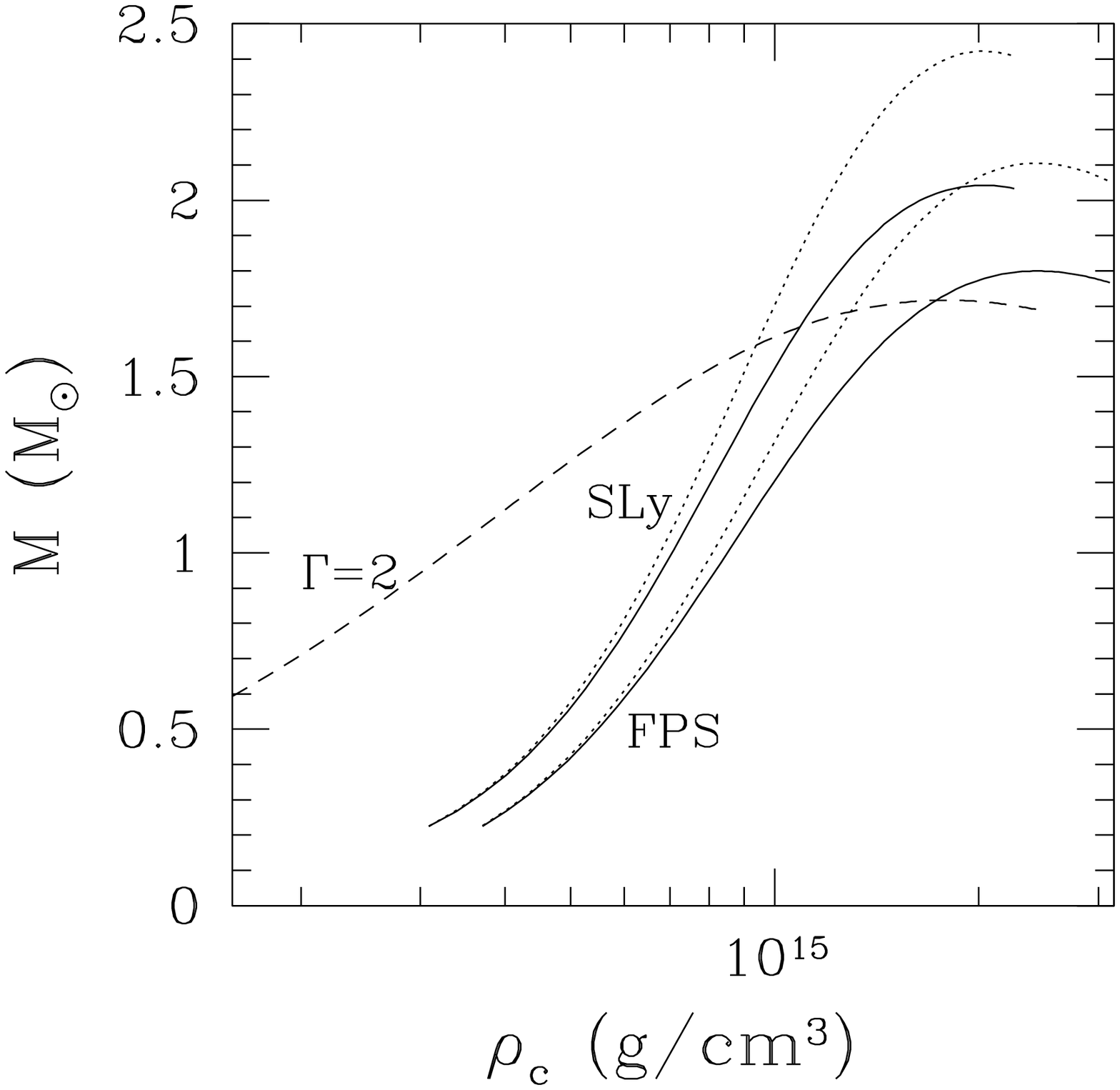}
  ~~(b)\includegraphics[width=3.in]{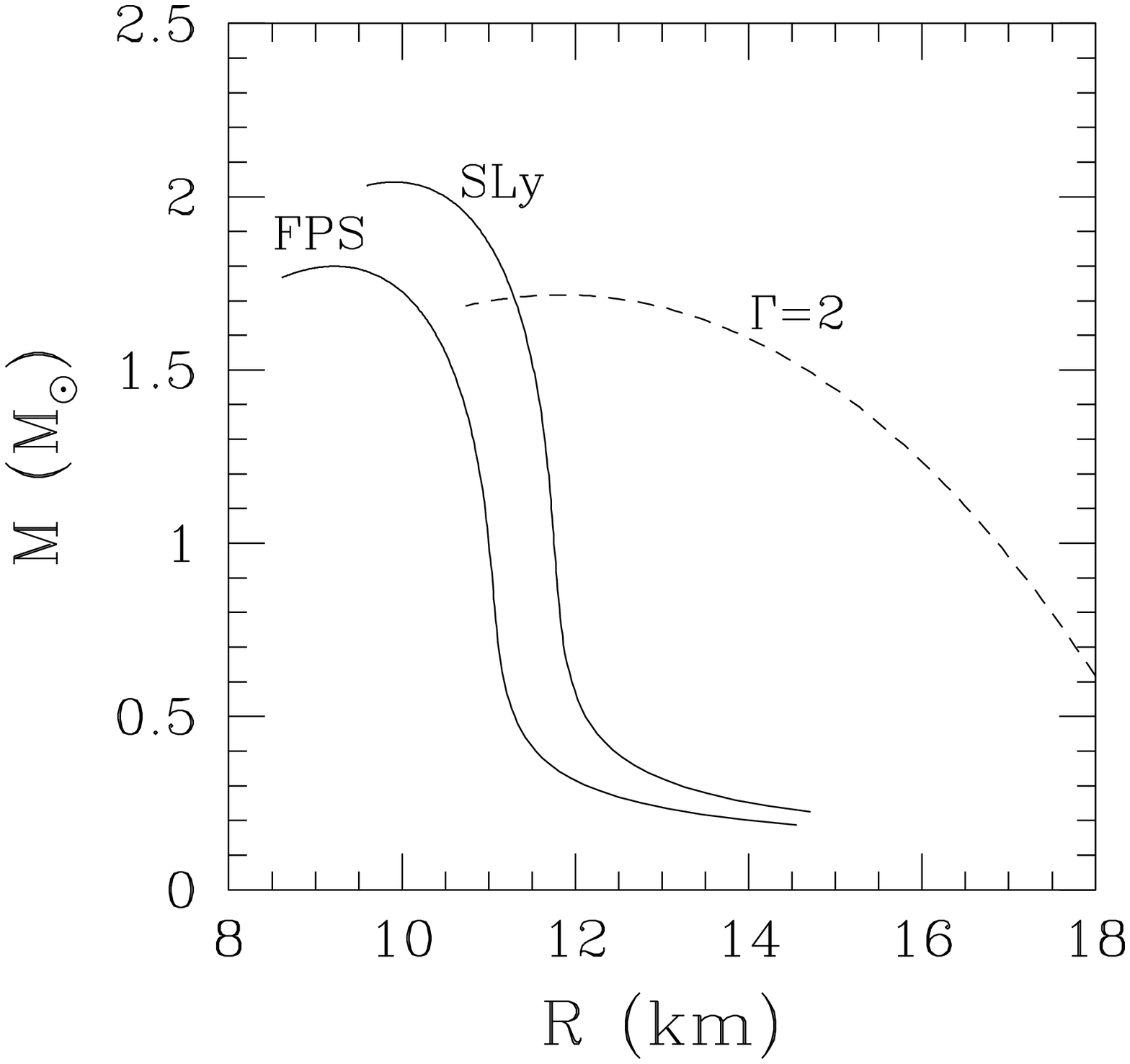}
\end{center}
\vspace{-2mm}
\caption{
(a) ADM mass (solid curves) and total baryon rest-mass
(dotted curves) as functions of central baryon
rest-mass density $\rho_c$ and 
(b) relation between the circumferential radius and the ADM mass 
for cold and spherical neutron stars
in equilibrium. 'FPS' and 'SLy' denote the sequences
for the FPS and SLy equations of state, respectively. 
The relations for the $\Gamma=2$ polytropic equation of state
with $K_{\rm p}=1.6 \times 10^5$ in the cgs unit are also drawn by the dashed
curves. 
\label{FIG2} }
\end{figure*}

In Fig. \ref{FIG2}, we show the relations among the ADM mass $M$,
the total baryon rest-mass $M_*$, the central density $\rho_c$, and the
circumferential radius $R$ for cold and spherical neutron stars
in the SLy and FPS equations
of state. For comparison, we present the results for
the $\Gamma=2$ polytropic equation of state $P=K_{\rm p} \rho^2$
which was adopted in \cite{STU}. In the polytropic equations of state,
there exists a degree of freedom for the choice of the polytropic
constant $K_{\rm p}$. Here, for getting approximately
the same value of the maximum ADM mass for cold and spherical
neutron stars as that of the realistic equations of state, we set 
\beqn
K_{\rm p}=1.6\times 10^5 ~({\rm cgs~unit}). 
\eeqn
In this case, the maximum ADM mass is about $1.72 M_{\odot}$.
We note that for the $\Gamma=2$ equation of state, 
the ADM mass $M$, the circumferential radius $R$, and the density 
can be rescaled by changing the value of
$K_{\rm p}$ using the following rule: 
\beqn
M \propto K_{\rm p}^{1/2},~~~
R \propto K_{\rm p}^{1/2},~~~{\rm and}~~~
\rho \propto K_{\rm p}^{-1}. 
\eeqn
Hence, the mass and the radius 
are arbitrarily rescaled although the compactness
$M/R$ is invariant in the rescaling. 

Figure \ref{FIG2} shows that in the realistic equations of state,
the central density and the circumferential radius are in a narrow range
for the ADM mass between $\sim 0.8M_{\odot}$ and 
$\sim 1.5M_{\odot}$. Also, it is found that neutron stars in the 
realistic equations of state are more compact than those in the 
$\Gamma=2$ polytropic equation of state for a given mass. 
Namely, the realistic equations of state are {\em softer} than the 
$\Gamma=2$ one. On the other hand, the adiabatic index $d\ln P/d\ln \rho$
for the realistic equations of state is much larger than 2
for the supranuclear density \cite{PR,DH,HP}. 
These properties result in quantitatively different results in the merger of
two neutron stars from those found in the previous work \cite{STU}. 

The thermal part of the pressure $P_{\rm th}$ 
is related to the specific thermal energy 
$\varepsilon_{\rm th}\equiv \varepsilon-\varepsilon_{\rm cold}$ as 
\beq
P_{\rm th}=(\Gamma_{\rm th}-1)\rho \varepsilon_{\rm th}, 
\eeq
where $\Gamma_{\rm th}$ is an adiabatic constant. 
As a default, we set $\Gamma_{\rm th}=2$ taking into account
the fact that the equations of state for high-density nuclear matter
are fairly stiff. (We note that for the ideal nonrelativistic Fermi gas,
$\Gamma_{\rm th} \approx 5/3$ \cite{Chandra}. For the nuclear matter,
it is reasonable to consider that it is much larger than this value.)
To investigate the dependence of the numerical results
on the value, we also choose $\Gamma_{\rm th}=1.3$ and 1.65. 
The thermal part of the pressure plays an important role
when shocks are formed during the evolution. For the
smaller value of $\Gamma_{\rm th}-1$, local 
conversion rate of the kinetic energy to
the thermal energy at the shocks should be smaller.

\section{Numerical results}

\subsection{Initial condition and computational setting}

\begin{table*}[tb]
\vspace{-4mm}
\begin{center}
\begin{tabular}{cccccccccccc} \hline
\hspace{-2mm} Model \hspace{-2mm} &
Each ADM mass &
$\rho_{\rm max} $  & $Q_M$ &  $M_*$  & 
\hspace{-3mm} $M_0$ \hspace{-2mm} &  $q_0$  & 
\hspace{-3mm} $P_{0}$ \hspace{-2mm} &
\hspace{-3mm} $C_0$  \hspace{-2mm} &  
\hspace{-3mm}$Q_{*}$ \hspace{-2mm} &
\hspace{-3mm}$f_0$ \hspace{-2mm} \\ \hline
SLy1212 &1.20, 1.20 & 8.03, 8.03 & 1.00
& 2.605 & 2.373 & 0.946 & 2.218 & 0.103 & 1.075 & 0.902  \\ 
SLy1313 &1.30, 1.30 & 8.57, 8.57 & 1.00
& 2.847 & 2.568 & 0.922 & 2.110 & 0.112 & 1.175 & 0.948  \\ 
SLy135135 &1.35, 1.35 & 8.86, 8.86 & 1.00
& 2.969 & 2.666 & 0.913 & 2.083 & 0.116 & 1.225 & 0.960  \\ 
SLy1414 &1.40, 1.40 & 9.16, 9.16 & 1.00
& 3.093 & 2.763 & 0.902 & 2.012 & 0.122 & 1.277 & 0.994  \\ 
SLy125135 &1.25, 1.35 & 8.29, 8.86 & 0.9179
& 2.847 & 2.568 & 0.921 & 2.110 & 0.112 & 1.175 & 0.948  \\ 
SLy135145 &1.35, 1.45 & 8.85, 9.48 & 0.9226
& 3.094 & 2.763 & 0.901 & 2.013 & 0.122 & 1.277 & 0.994  \\ 
\hline
FPS1212 &1.20, 1.20 & 9.93, 9.93 & 1.00
& 2.624 & 2.371 & 0.925 & 1.980 & 0.111 & 1.251 & 1.010  \\ 
FPS125125 &1.25, 1.25 & 10.34, 10.34 & 1.00
& 2.746 & 2.469 & 0.914 & 1.935 & 0.116 & 1.309 & 1.034  \\ 
FPS1313 &1.30, 1.30 & 10.79, 10.79 & 1.00
& 2.869 & 2.566 & 0.903 & 1.882 & 0.121 & 1.368 & 1.063  \\ 
FPS1414 &1.40, 1.40 & 11.76, 11.76 & 1.00
& 3.120 & 2.760 & 0.882 & 1.750 & 0.134 & 1.487 & 1.143  \\ \hline
\end{tabular}
\caption{
A list of several quantities for quasiequilibrium initial data.
The ADM mass of each star when they are in isolation, 
the maximum density for each star, 
the baryon rest-mass ratio $Q_M \equiv M_{*2}/M_{*1}$, 
the total baryon rest-mass, 
the total ADM mass $M_{0}$, nondimensional spin parameter 
$q_0=J_0/M_{0}^2$, orbital period $P_{0}$, 
the orbital compactness [$C_0\equiv (M_{0}\Omega)^{2/3}$], 
the ratio of the total baryon rest-mass to 
the maximum allowed mass for a spherical and cold star 
($Q_{*}\equiv M_*/M_{*~\rm max}^{\rm sph}$), and 
the frequency of gravitational waves $f_0=2/P_0$. 
The density, mass, period, and wave frequency 
are shown in units of $10^{14}{\rm g/cm^3}$,
$M_{\odot}$, ms, and kHz, respectively. 
}
\end{center}
\end{table*}

\begin{table*}[tb]
\vspace{-4mm}
\begin{center}
\begin{tabular}{ccccccccc} \hline
\hspace{-2mm} Model \hspace{-2mm} & $\Gamma_{\rm th}$ &
Grid number & $L$ & $\Delta$ & $\lambda_{0}$ & $f_{\rm merger}$
& $\lambda_{\rm merger}$ & Product ~~~\\ \hline
SLy1212b   &2   & (377, 377, 189) & 77.8 & 0.414& 333 &3.1& 97 &NS\\ 
SLy1313a   &2   & (633, 633, 317) & 130.8& 0.414& 316 &3.2& 94 &NS\\ 
SLy1313b   &2   & (377, 377, 189) & 77.8 & 0.414& 316 &3.2& 94 &NS\\ 
SLy1313c   &1.3 & (377, 377, 189) & 77.8 & 0.414& 316 &3.7& 81 &NS
$\rightarrow$ BH\\ 
SLy1313d   &1.65& (377, 377, 189) & 77.8 & 0.414& 316 &3.4& 88 &NS\\ 
SLy135135b &2   & (377, 377, 189) & 77.8 & 0.414& 316 &3.6& 83 &NS
$\rightarrow$ BH\\ 
SLy1414a   &2   & (633, 633, 317) & 130.8& 0.414& 302 &---& ---&BH\\ 
SLy125135a &2   & (633, 633, 317) & 130.8& 0.414& 316 &3.2& 94 &NS\\ 
SLy135145a &2   & (633, 633, 317) & 130.8& 0.414& 302 &---& ---&BH\\ 
\hline
FPS1212b   &2   & (377, 377, 189) & 69.5 & 0.370& 297 &3.5& 86 & 
NS $\rightarrow$ BH\\ 
FPS125125b   &2   & (377, 377, 189) & 69.5 & 0.370& 297 & &  & 
NS $\rightarrow$ BH\\ 
FPS1313b   &2   & (377, 377, 189) & 69.5 & 0.370& 282 &---&--- &BH\\ 
FPS1414b   &2   & (377, 377, 189) & 69.5 & 0.370& 262 &---&--- &BH\\ \hline
\end{tabular}
\caption{
A list of computational setting.
$\Gamma_{\rm th}$, $L$, $\Delta$, $\lambda_0$, and $f_{\rm merger}$ 
denote the adiabatic index for the thermal part, the location of outer
boundaries along each axis, the grid spacing, 
the wave length of gravitational waves at $t=0$,
and the frequency of gravitational waves from the formed hypermassive
neutron stars, respectively. $\lambda_{\rm merger}$ denotes
the wave length of gravitational waves from the formed hypermassive
neutron stars $\lambda_{\rm merger}=c/f_{\rm merger}$. 
The length and the frequency 
are shown in units of km and kHz. In the last column, 
the outcome is shown. NS implies that a hypermassive neutron star
is produced and remains stable at $t \sim 10$ ms. 
NS$\rightarrow$BH implies that
a hypermassive neutron star is formed first, but as a result of 
gravitational radiation reaction, it collapses to a black hole 
in $t \alt 10$ ms. 
BH implies that a black hole is promptly formed.
}
\end{center}
\end{table*}

\begin{figure*}[thb]
\vspace{-4mm}
\begin{center}
  \includegraphics[width=2.2in]{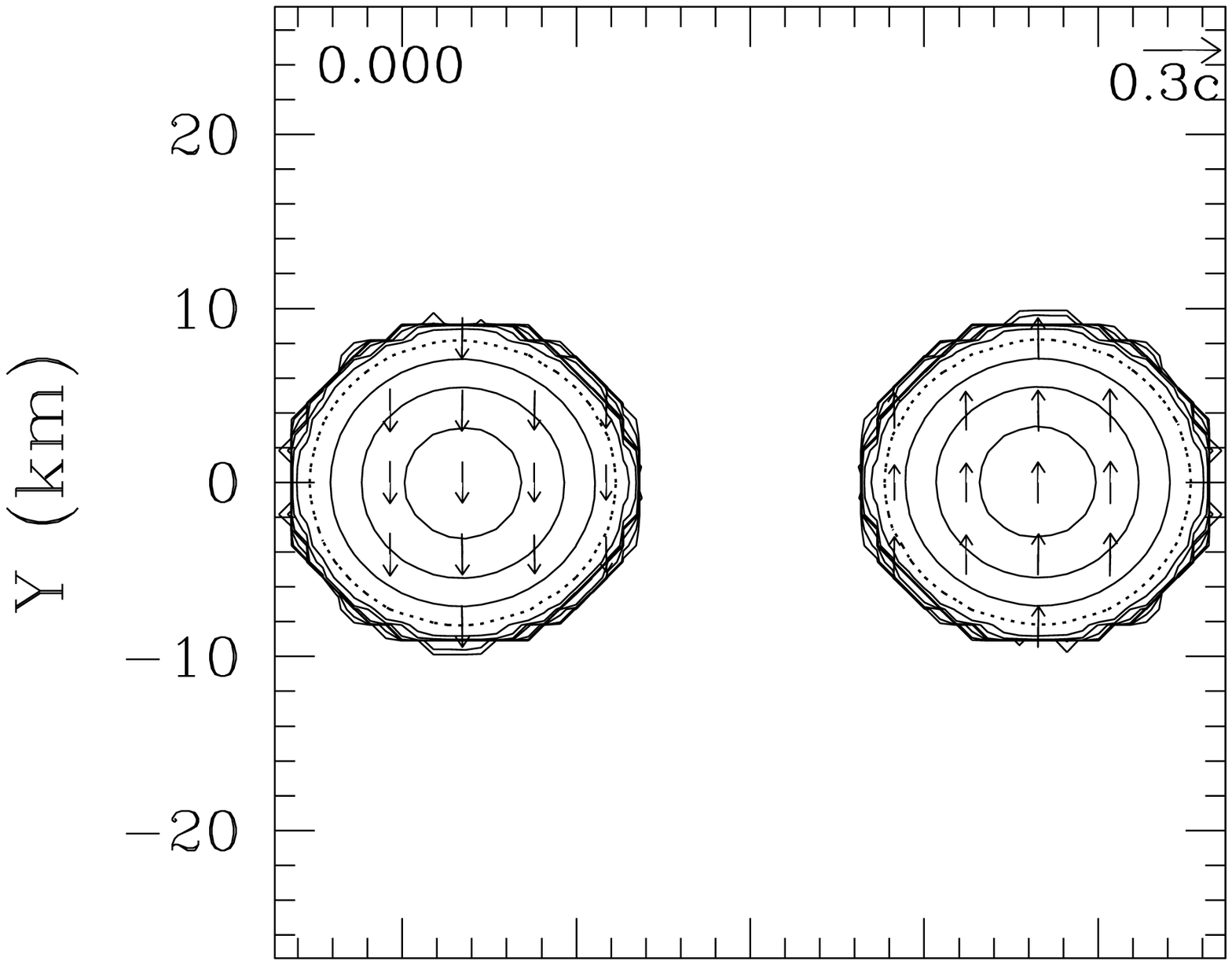}
  \hspace{-1.65cm}\includegraphics[width=2.2in]{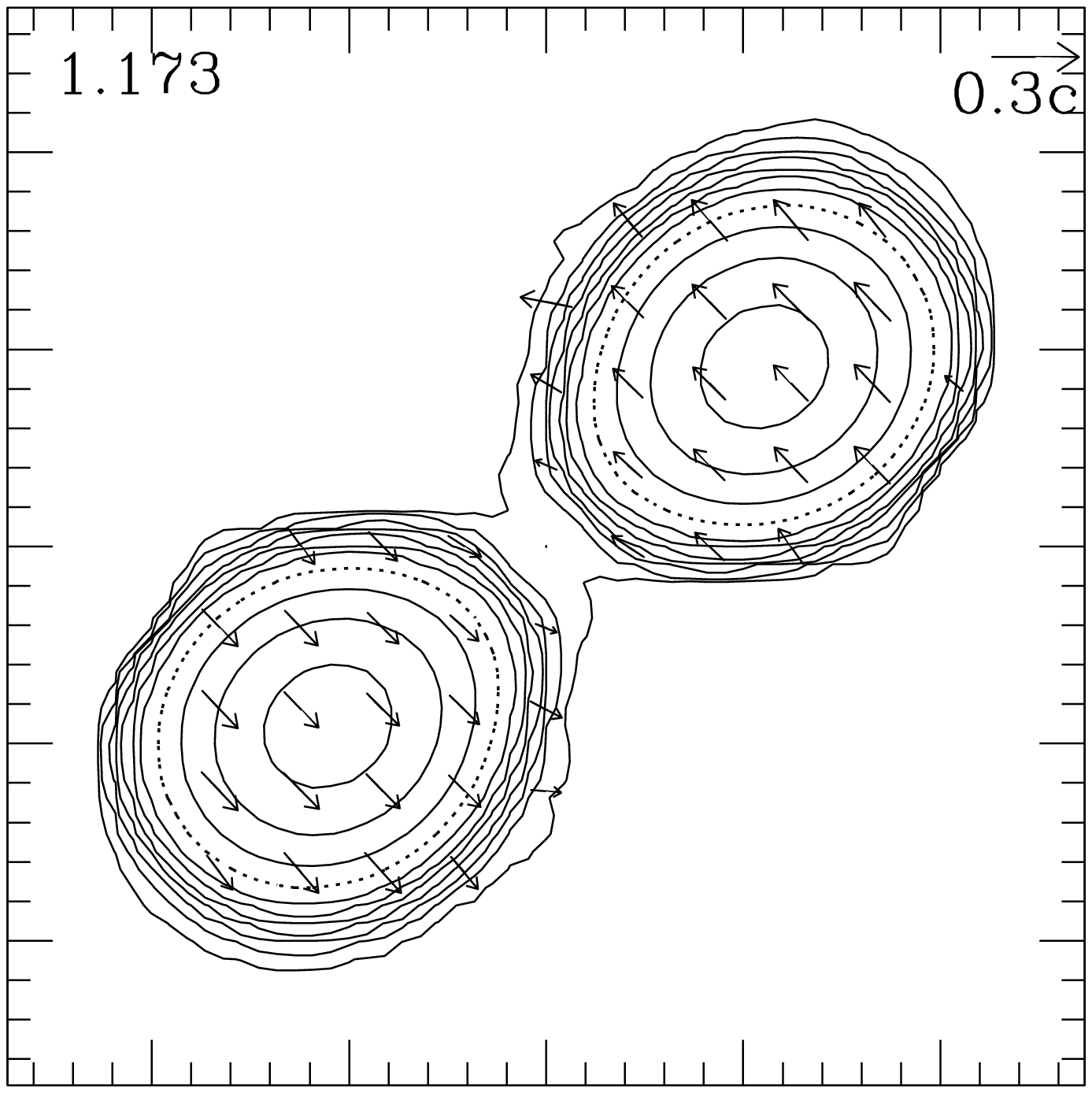} 
  \hspace{-1.65cm}\includegraphics[width=2.2in]{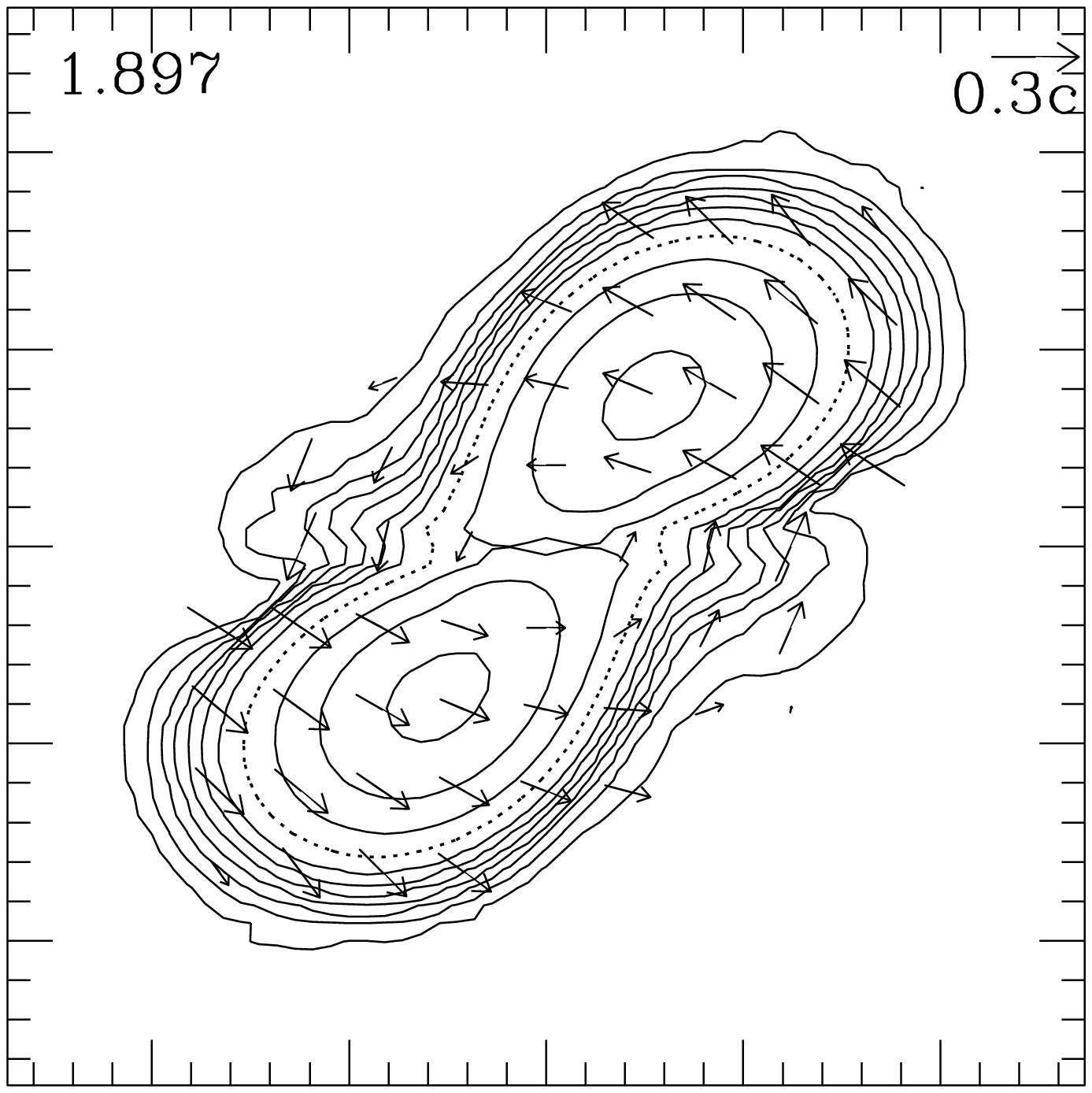} \\
\vspace{-1.65cm}
  \hspace{1mm}\includegraphics[width=2.2in]{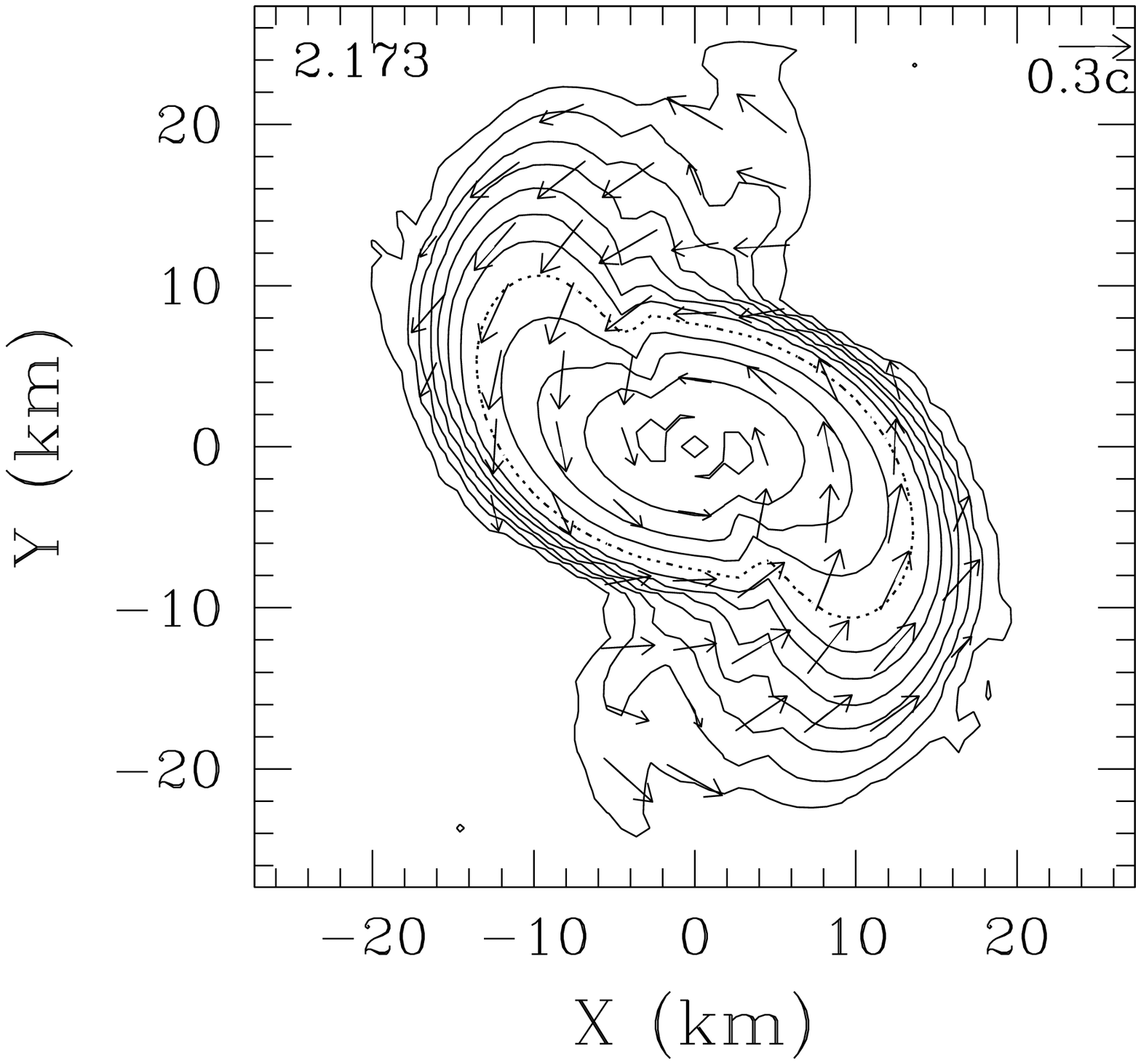} 
  \hspace{-1.65cm}\includegraphics[width=2.2in]{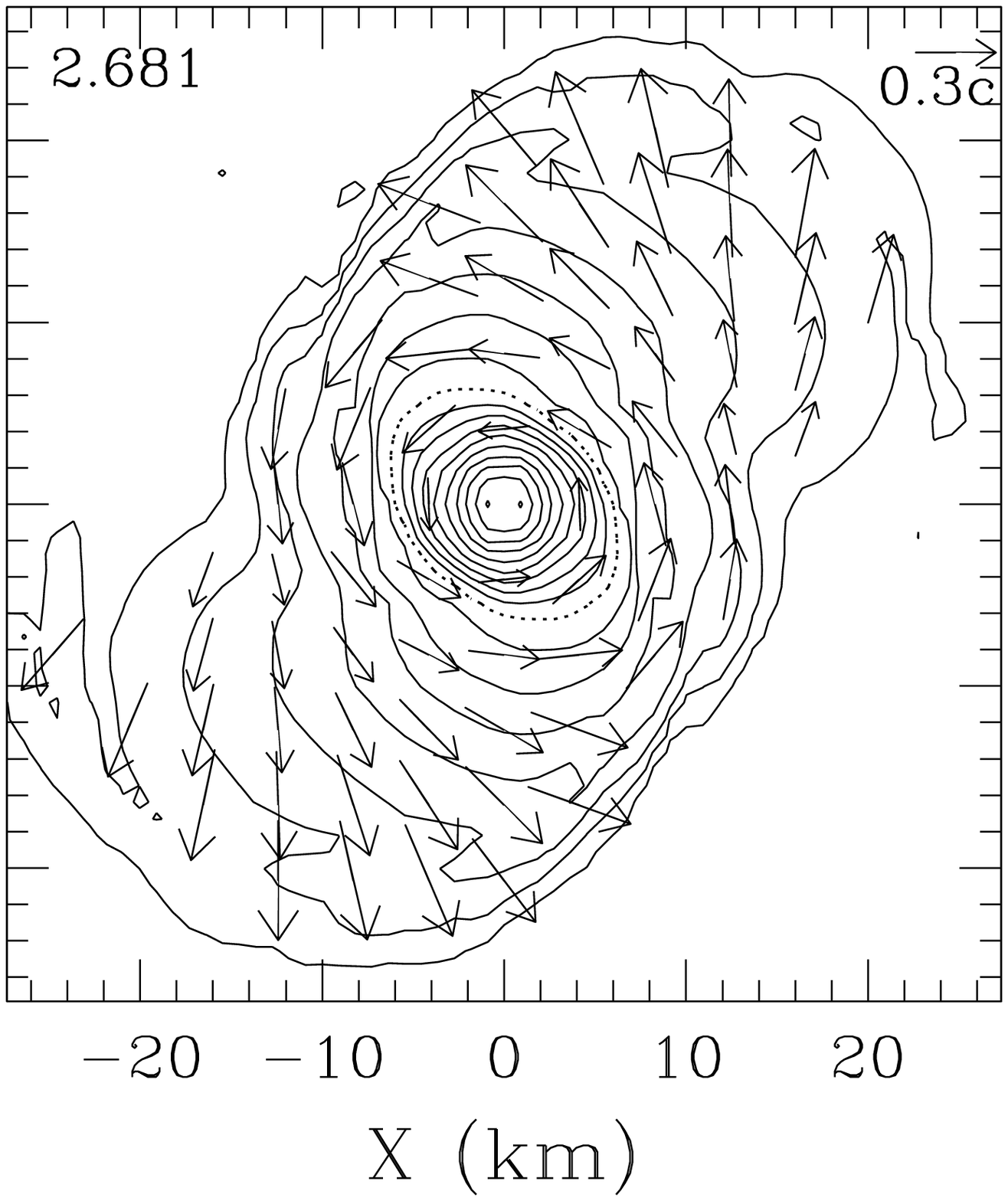}
  \hspace{-1.65cm}\includegraphics[width=2.2in]{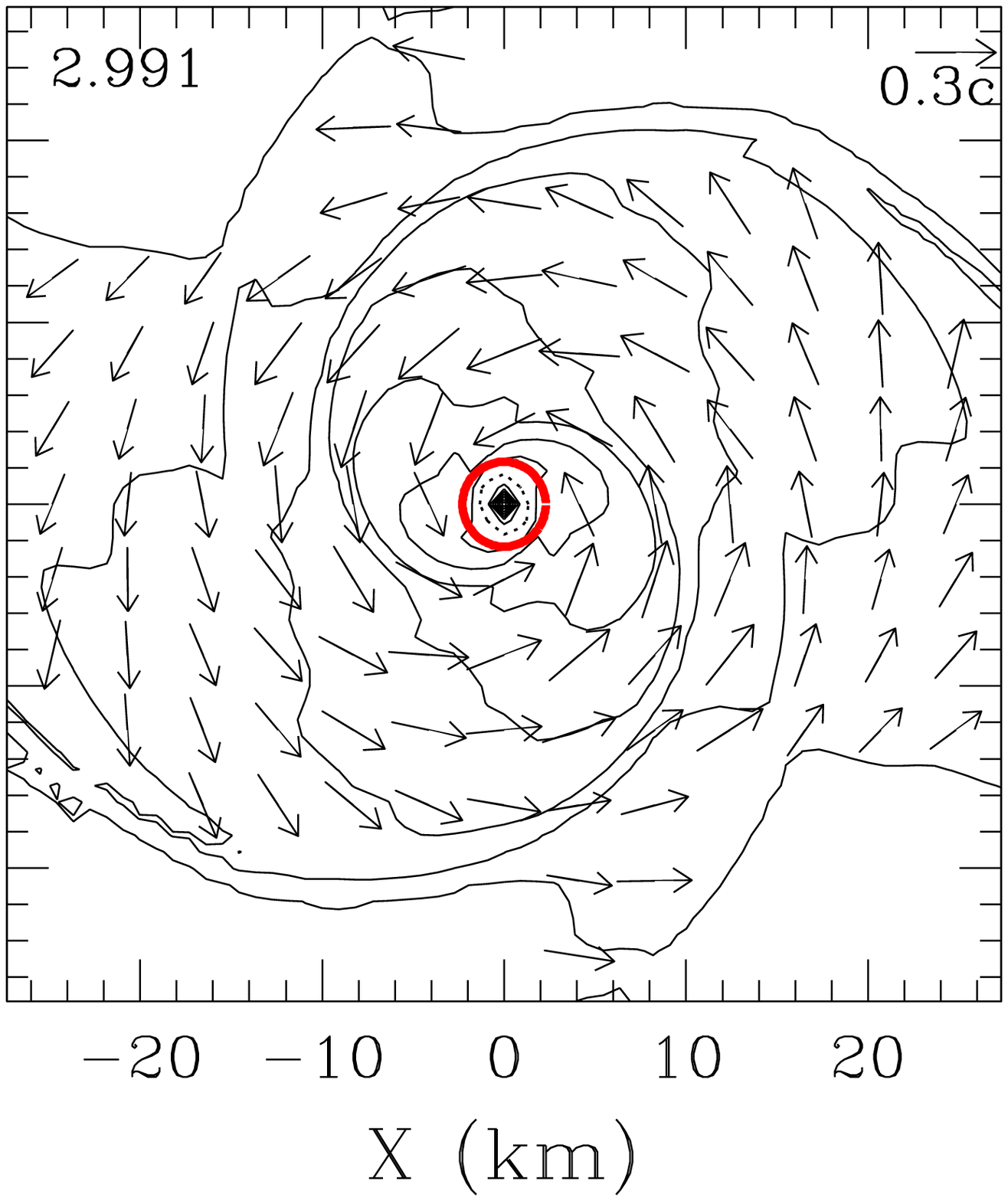}
\vspace{-4mm}
\caption{\small
Snapshots of the density contour curves for $\rho$ 
in the equatorial plane for model SLy1414a.
The solid contour curves are drawn for
$\rho=2\times 10^{14} \times i ~{\rm g/cm^3}~(i=2 \sim 10)$ and for 
$2\times 10^{14} \times 10^{-0.5 i}~{\rm g/cm^3}~(i=1 \sim 7)$. 
The dotted curves denote $2 \times 10^{14}~{\rm g/cm^3}$. 
The number in the upper left-hand side denotes the elapsed time
from the beginning of the simulation in units of ms. 
The initial orbital period in this case is 2.012 ms. 
Vectors indicate the local velocity field $(v^x,v^y)$, and the scale 
is shown in the upper right-hand corner.
The thick circle in the last panel of radius $r \sim 2$ km
denotes the location of the apparent horizon. 
\label{FIG3}}
\end{center}
\end{figure*}

\begin{figure*}[thb]
\vspace{-4mm}
\begin{center}
  \includegraphics[width=2.2in]{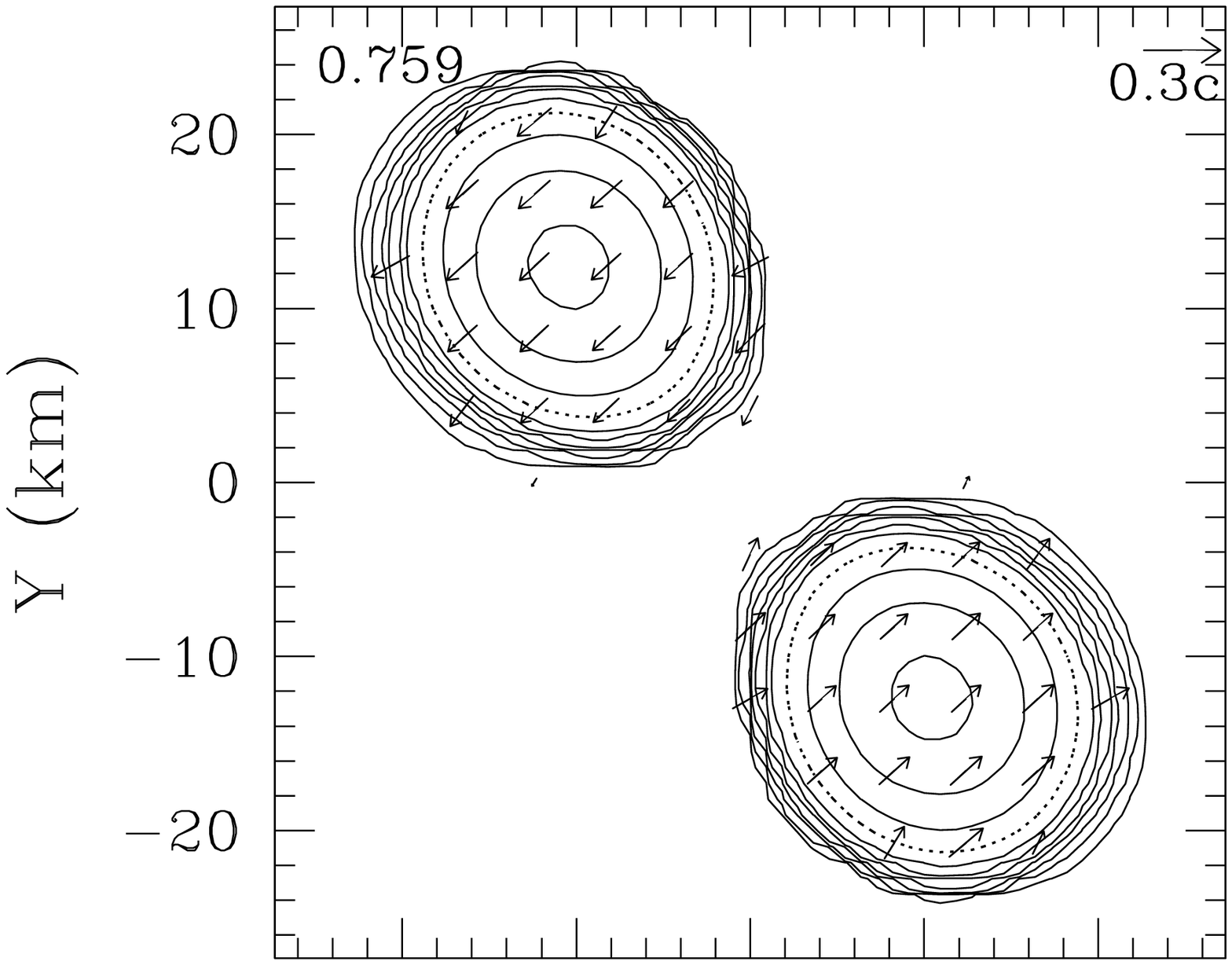}
  \hspace{-1.65cm}\includegraphics[width=2.2in]{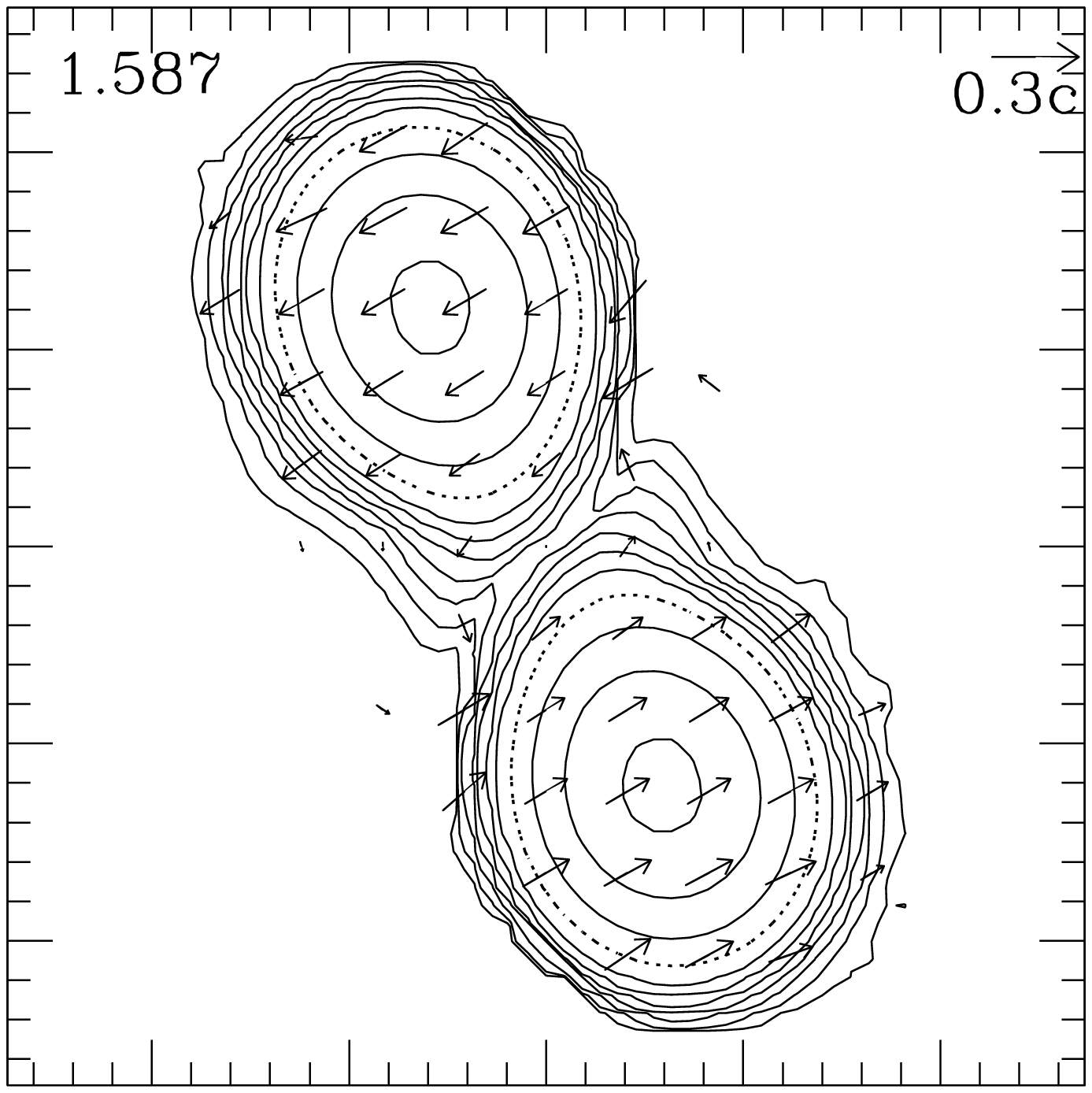} 
  \hspace{-1.65cm}\includegraphics[width=2.2in]{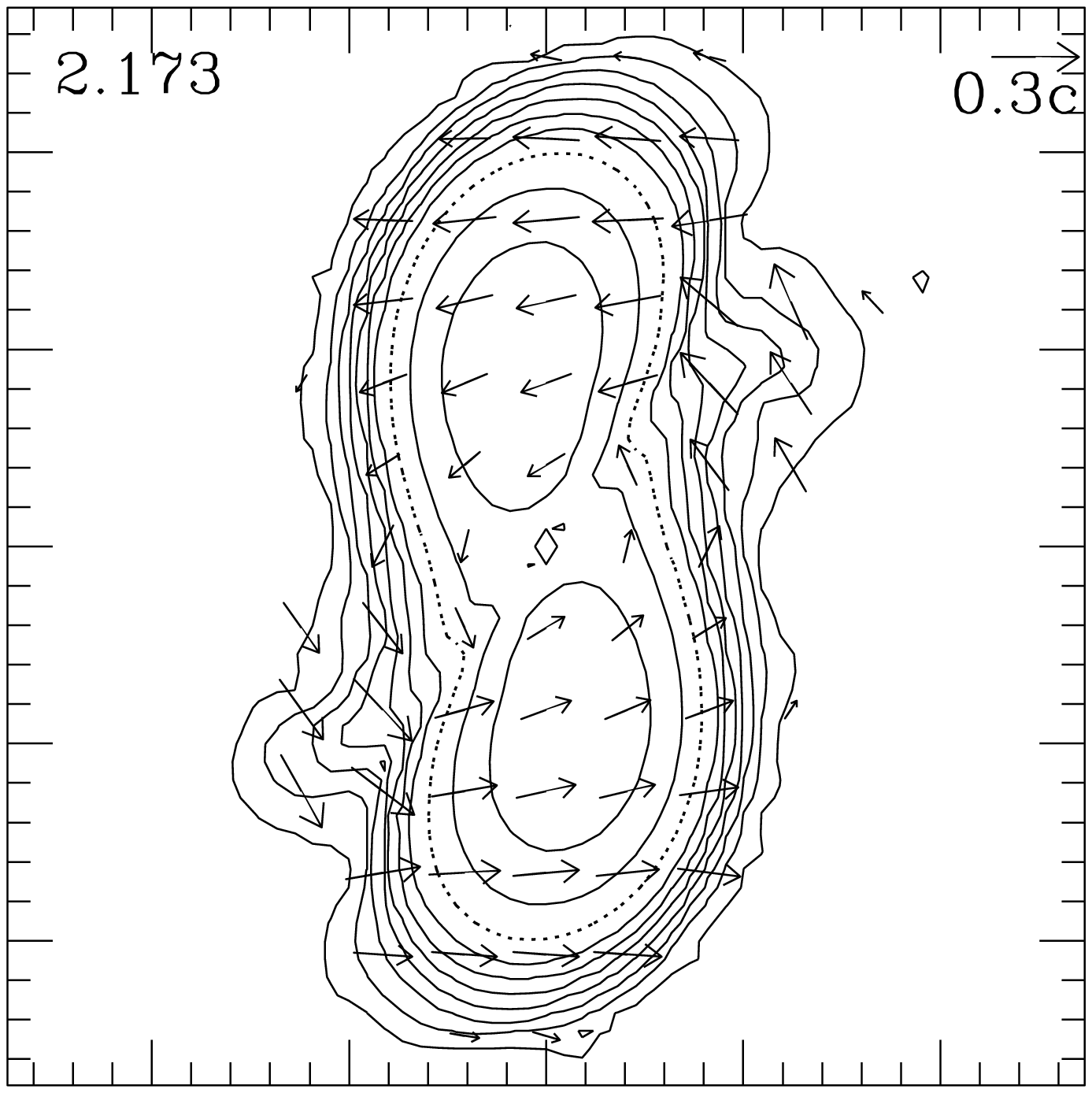} \\
\vspace{-1.65cm}
  \includegraphics[width=2.2in]{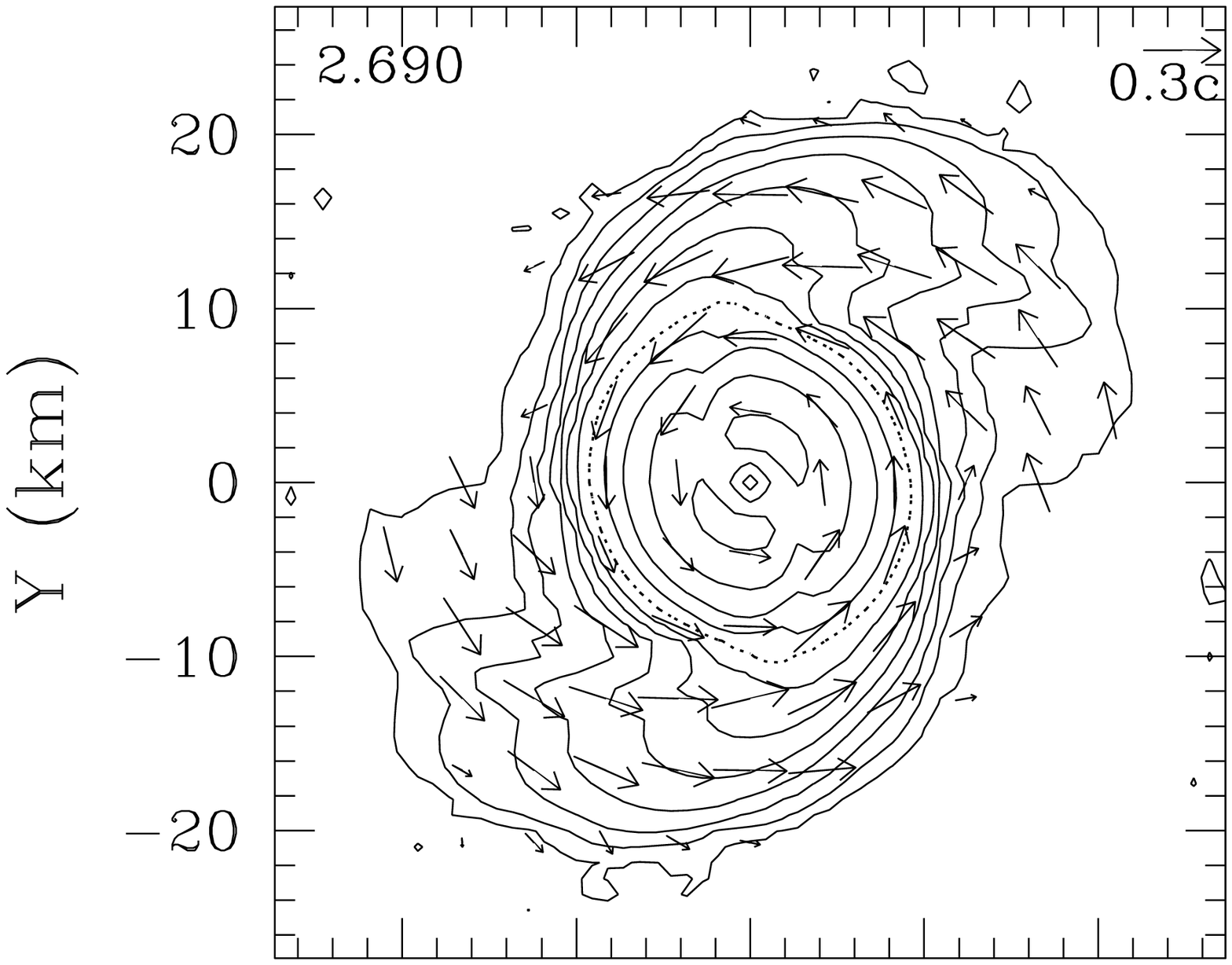}
  \hspace{-1.65cm}\includegraphics[width=2.2in]{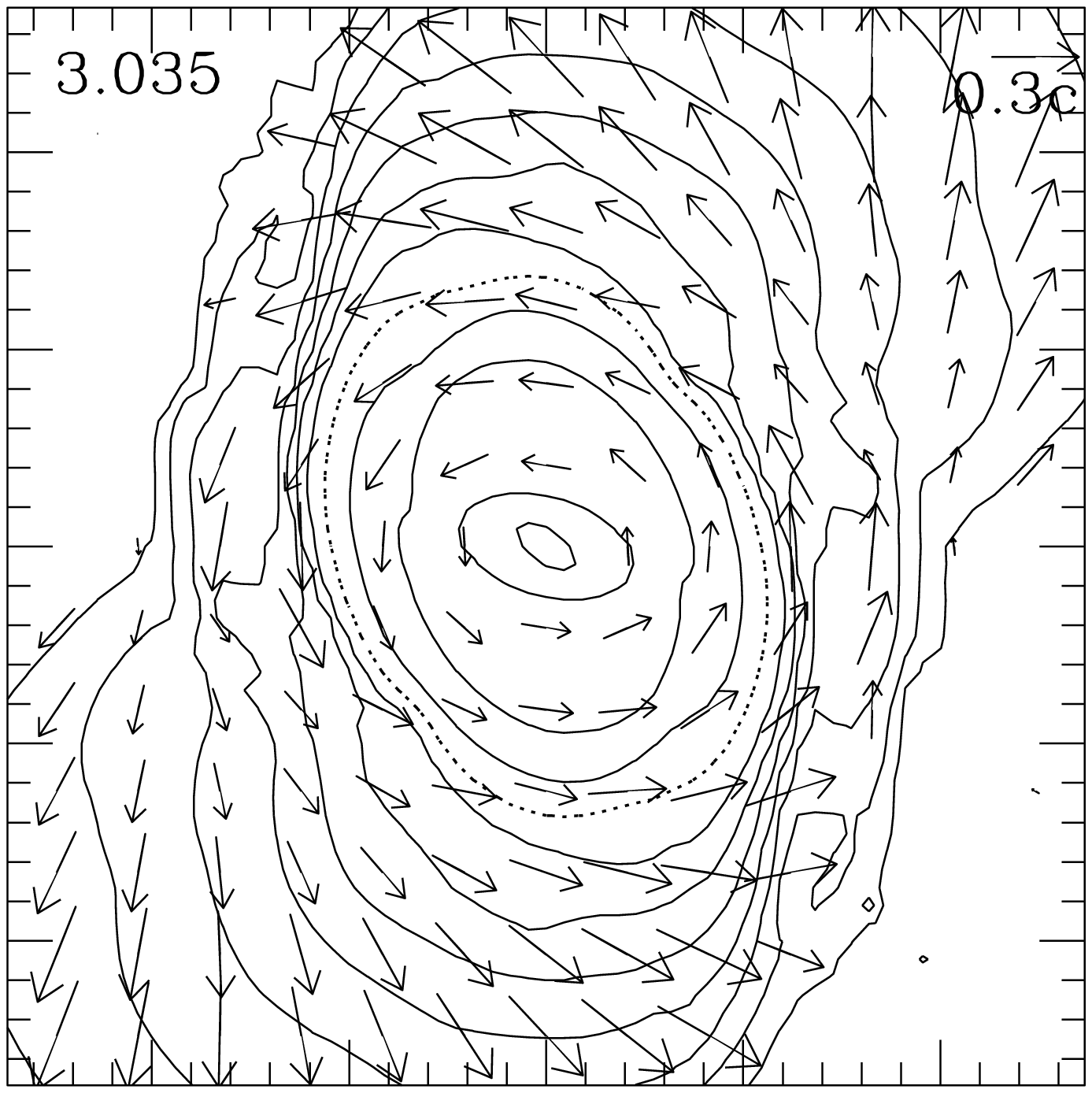} 
  \hspace{-1.65cm}\includegraphics[width=2.2in]{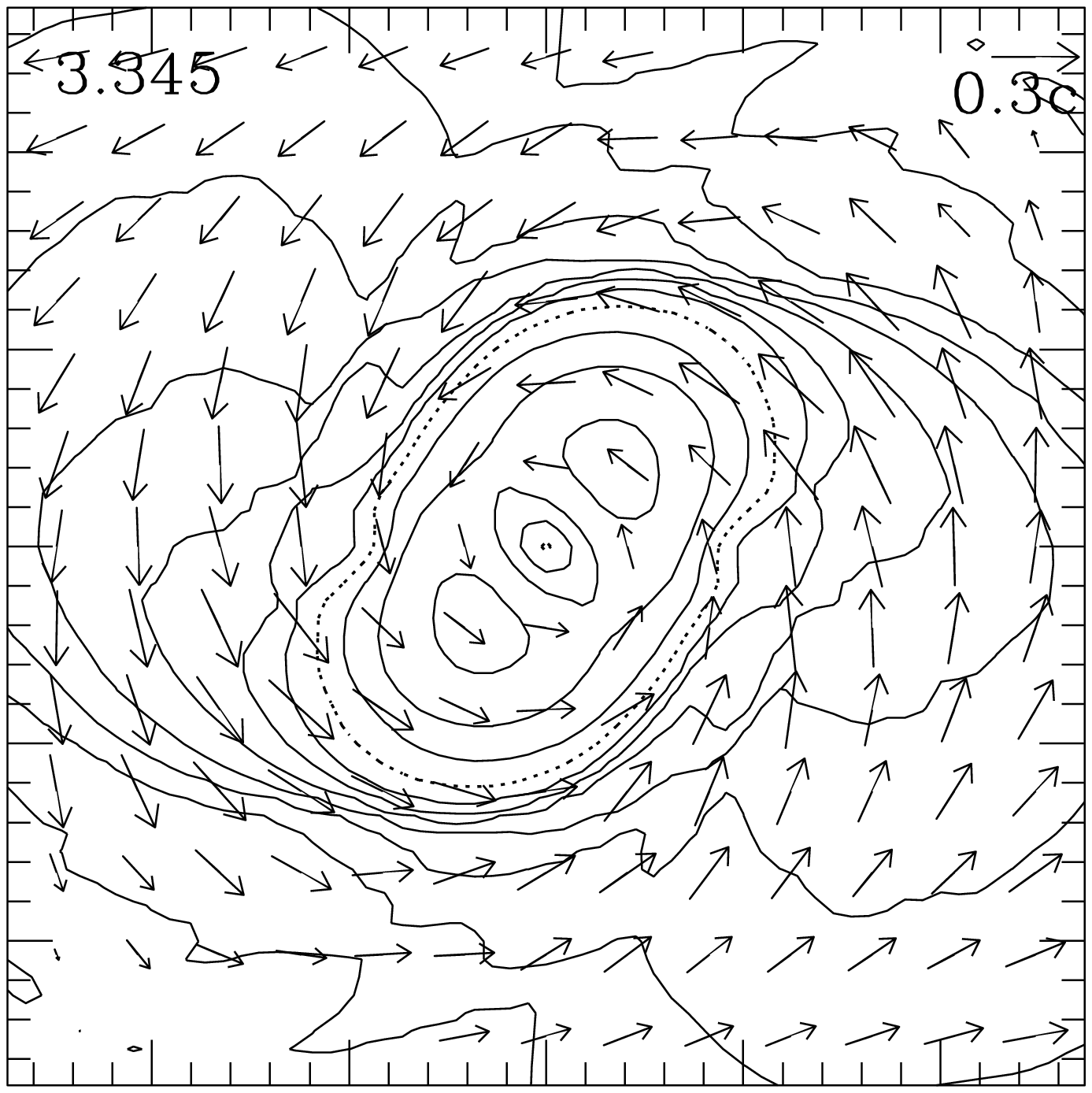} \\
\vspace{-1.63cm}
  \hspace{1mm}\includegraphics[width=2.2in]{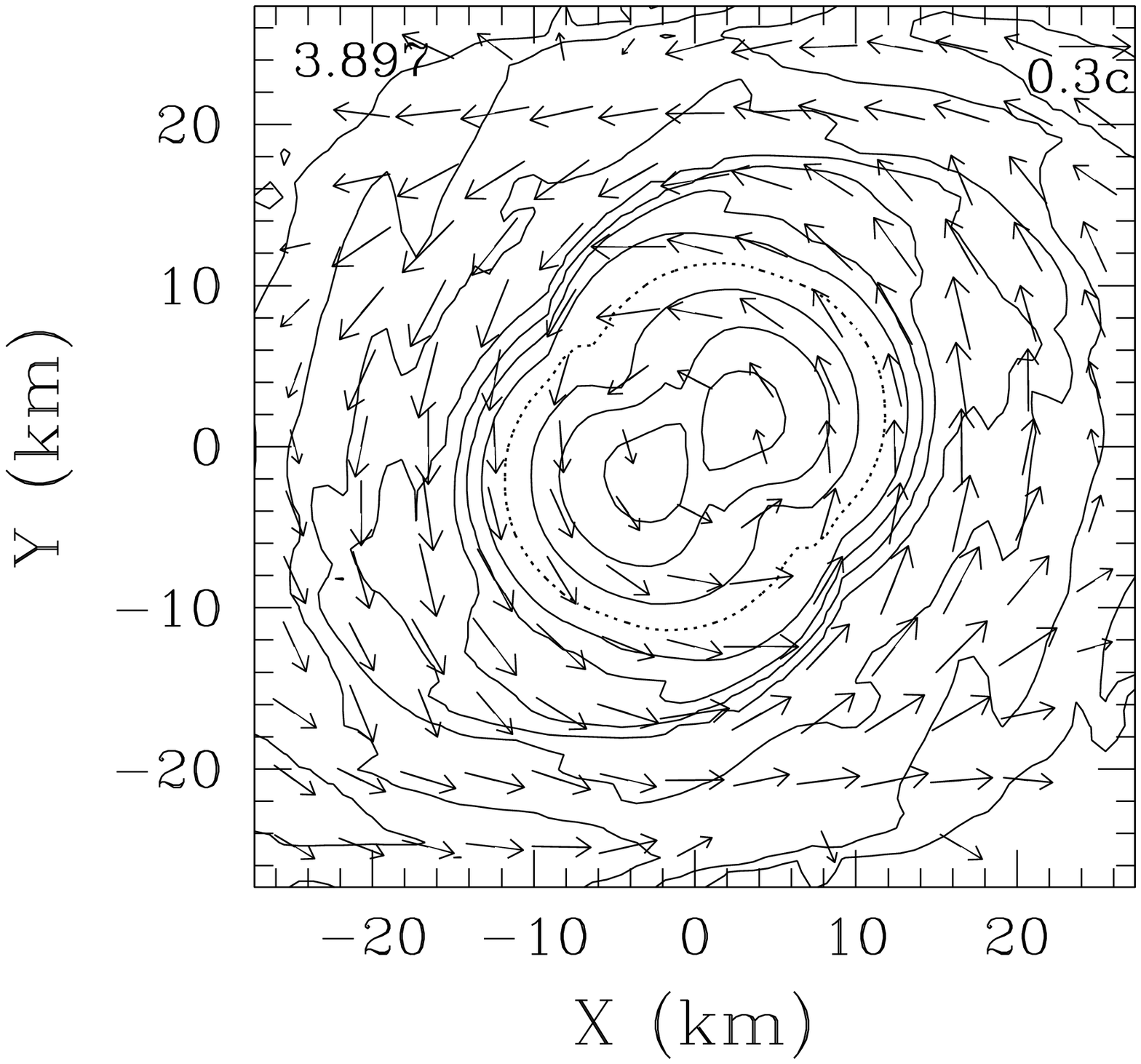}
  \hspace{-1.65cm}\includegraphics[width=2.2in]{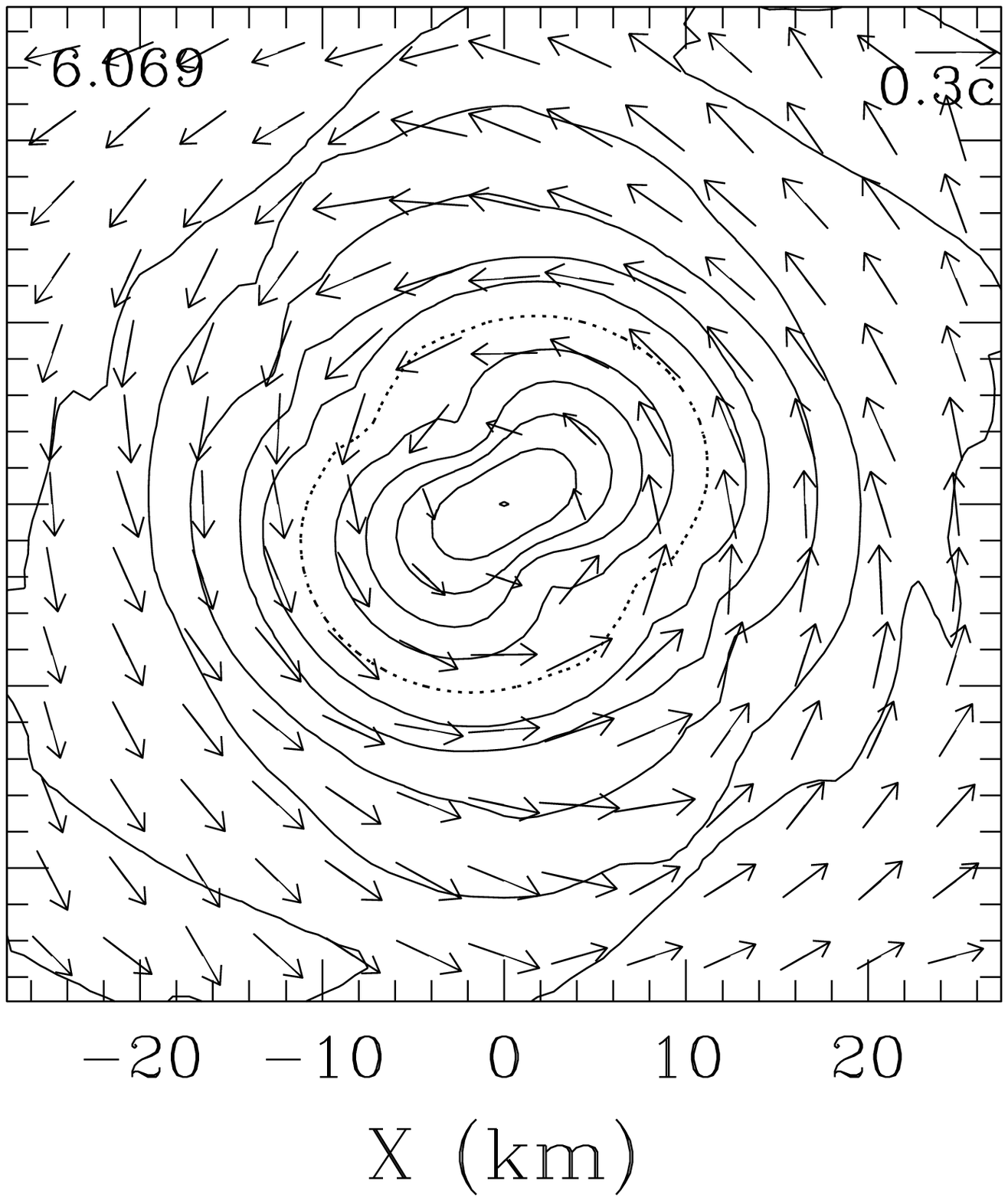} 
  \hspace{-1.65cm}\includegraphics[width=2.2in]{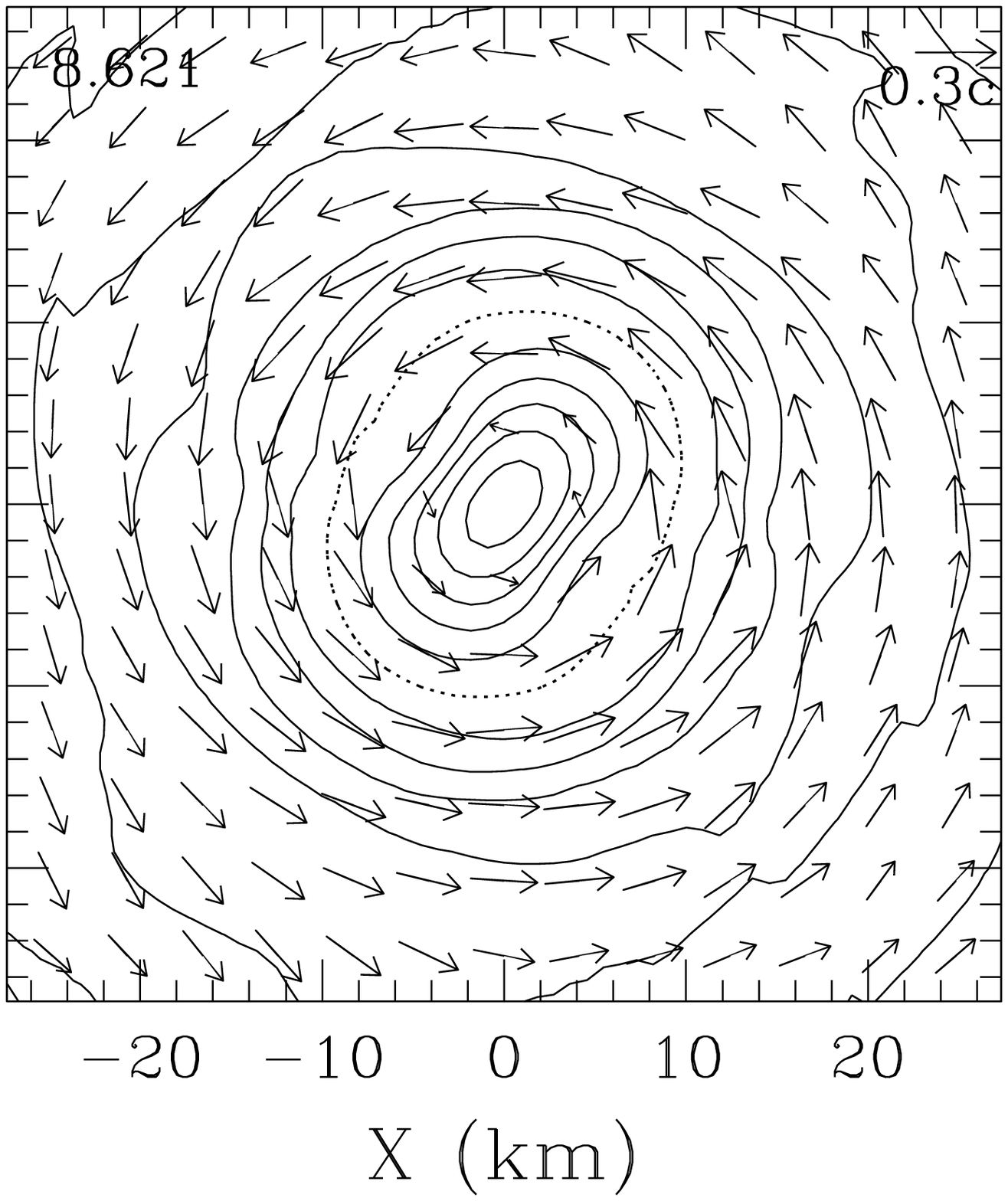}
\vspace{-4mm}
\caption{\small
The same as Fig. \ref{FIG3} but for model SLy1313a. 
The initial orbital period is 2.110 ms in this case. 
\label{FIG4}}
\end{center}
\end{figure*}

\begin{figure*}[p]
\vspace{-4mm}
\begin{center}
  \includegraphics[width=2.2in]{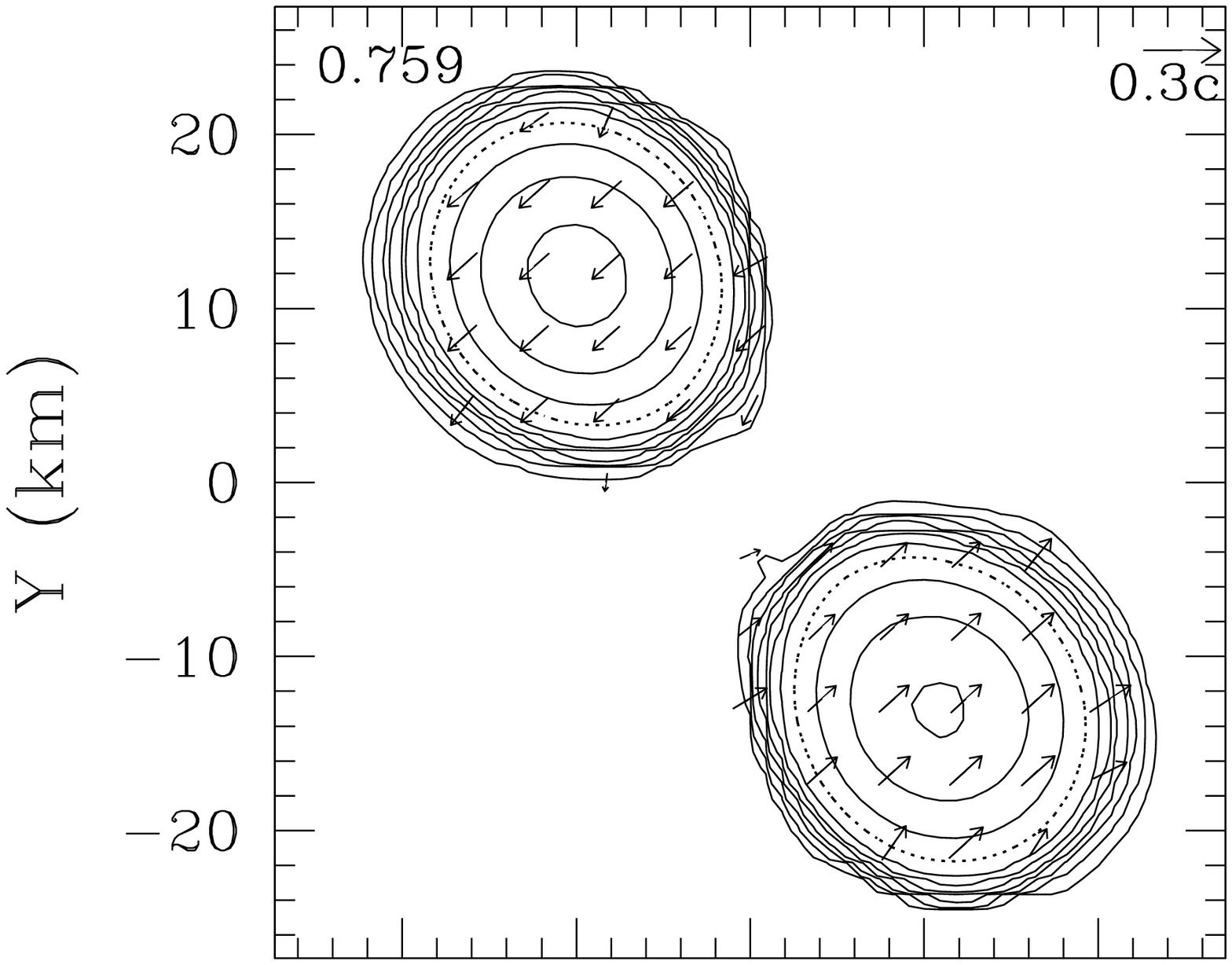}
  \hspace{-1.65cm}\includegraphics[width=2.2in]{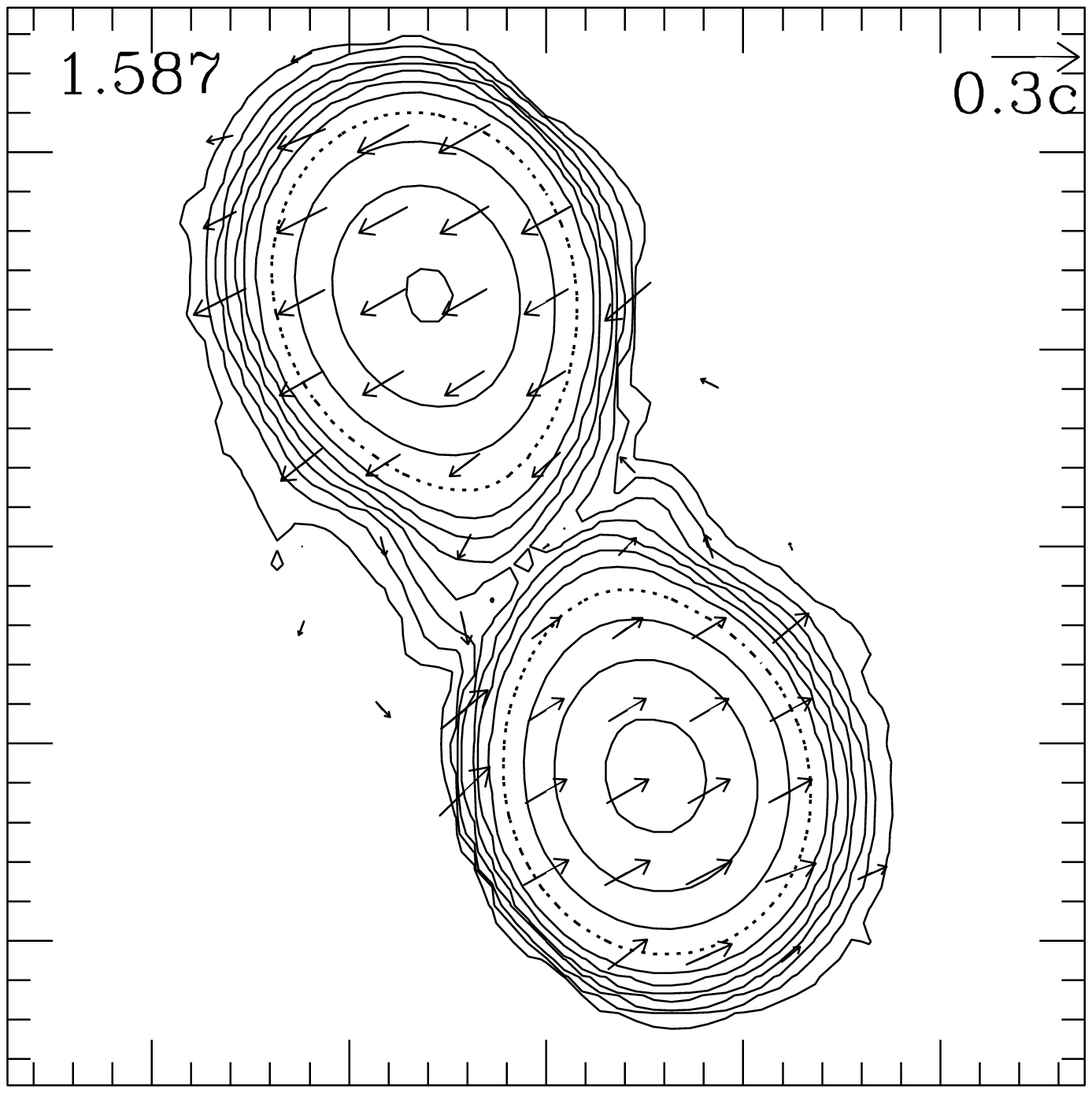} 
  \hspace{-1.65cm}\includegraphics[width=2.2in]{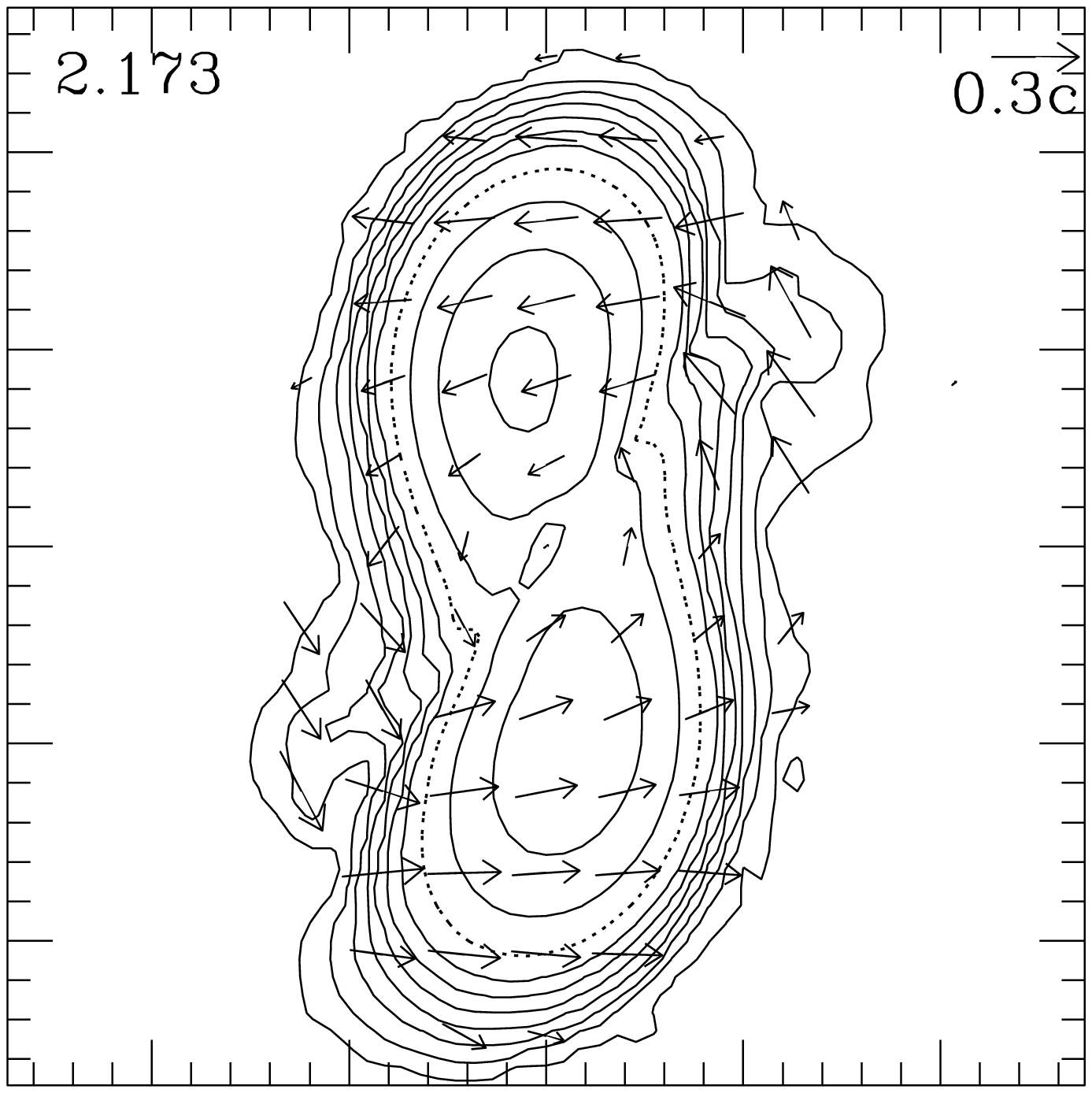} \\
\vspace{-1.65cm}
  \includegraphics[width=2.2in]{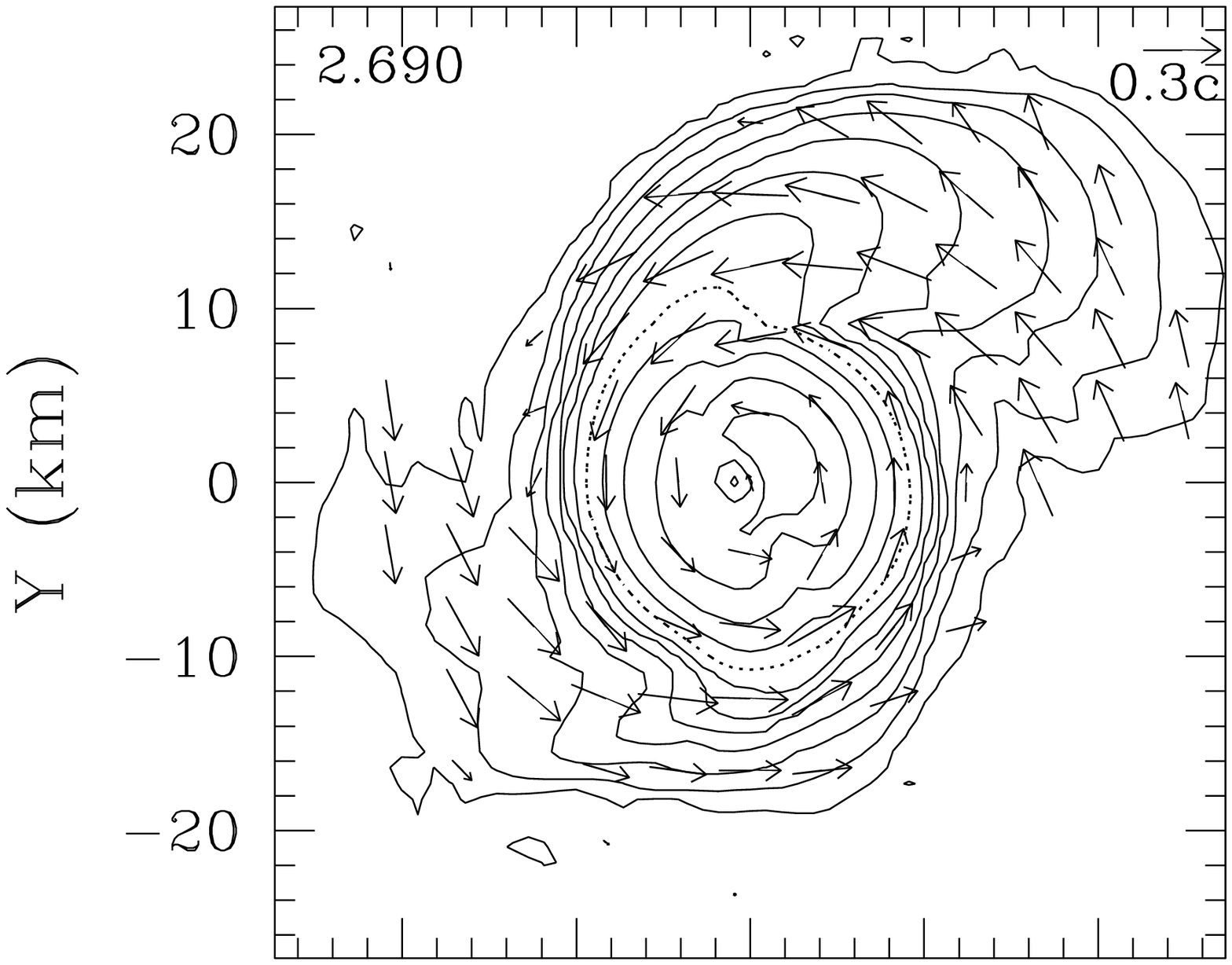}
  \hspace{-1.65cm}\includegraphics[width=2.2in]{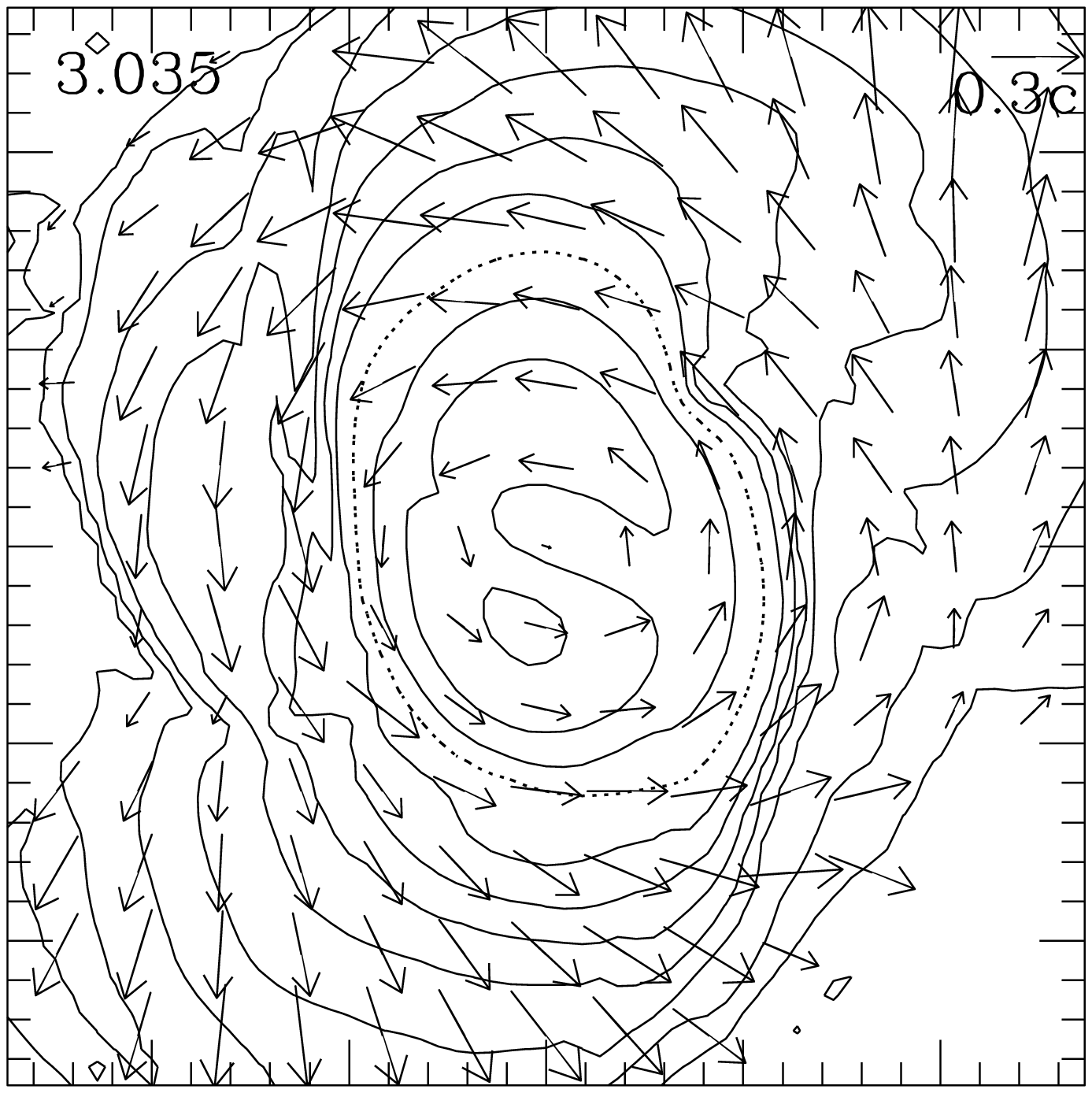} 
  \hspace{-1.65cm}\includegraphics[width=2.2in]{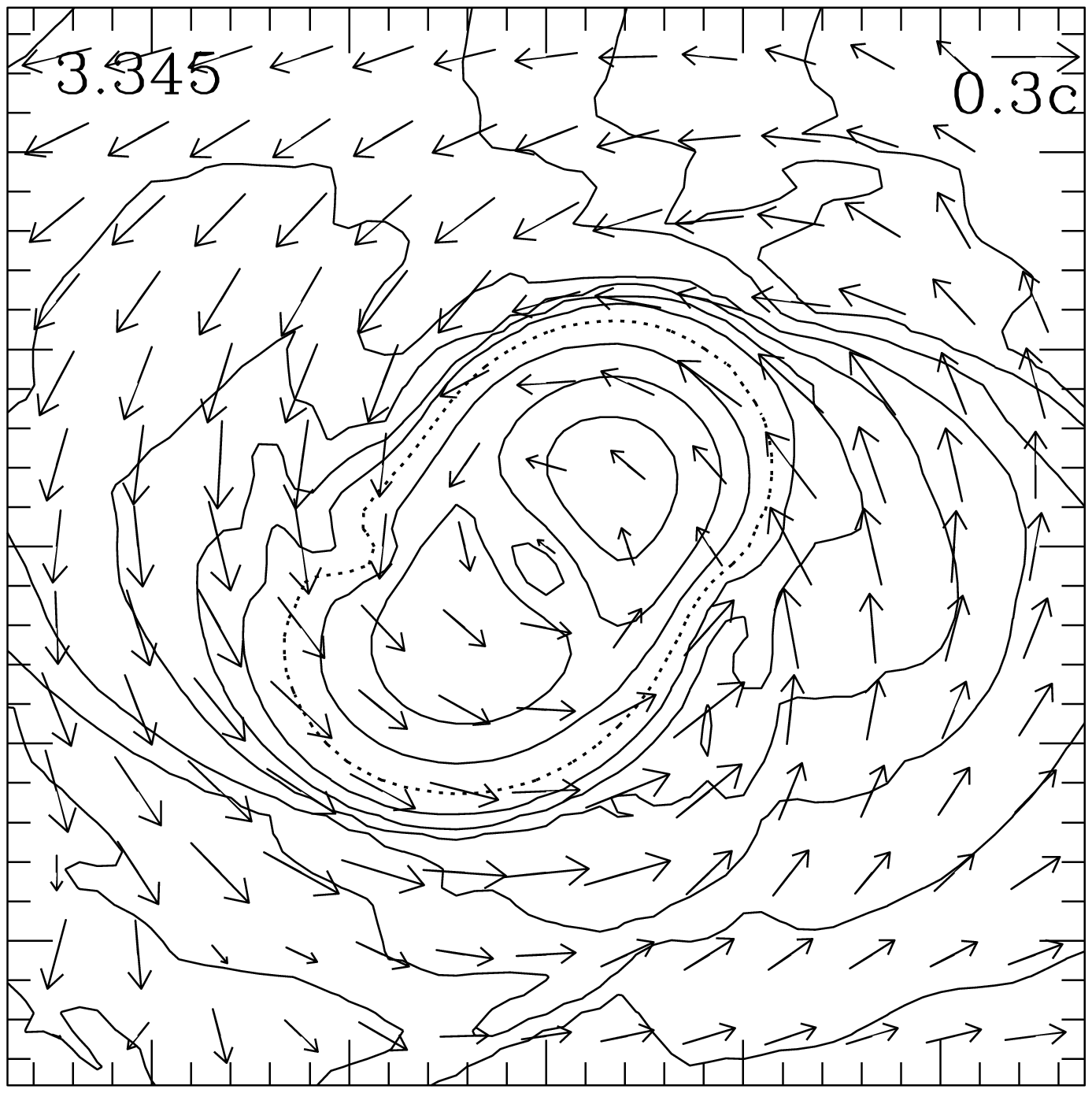} \\
\vspace{-1.63cm}
  \hspace{1mm}\includegraphics[width=2.2in]{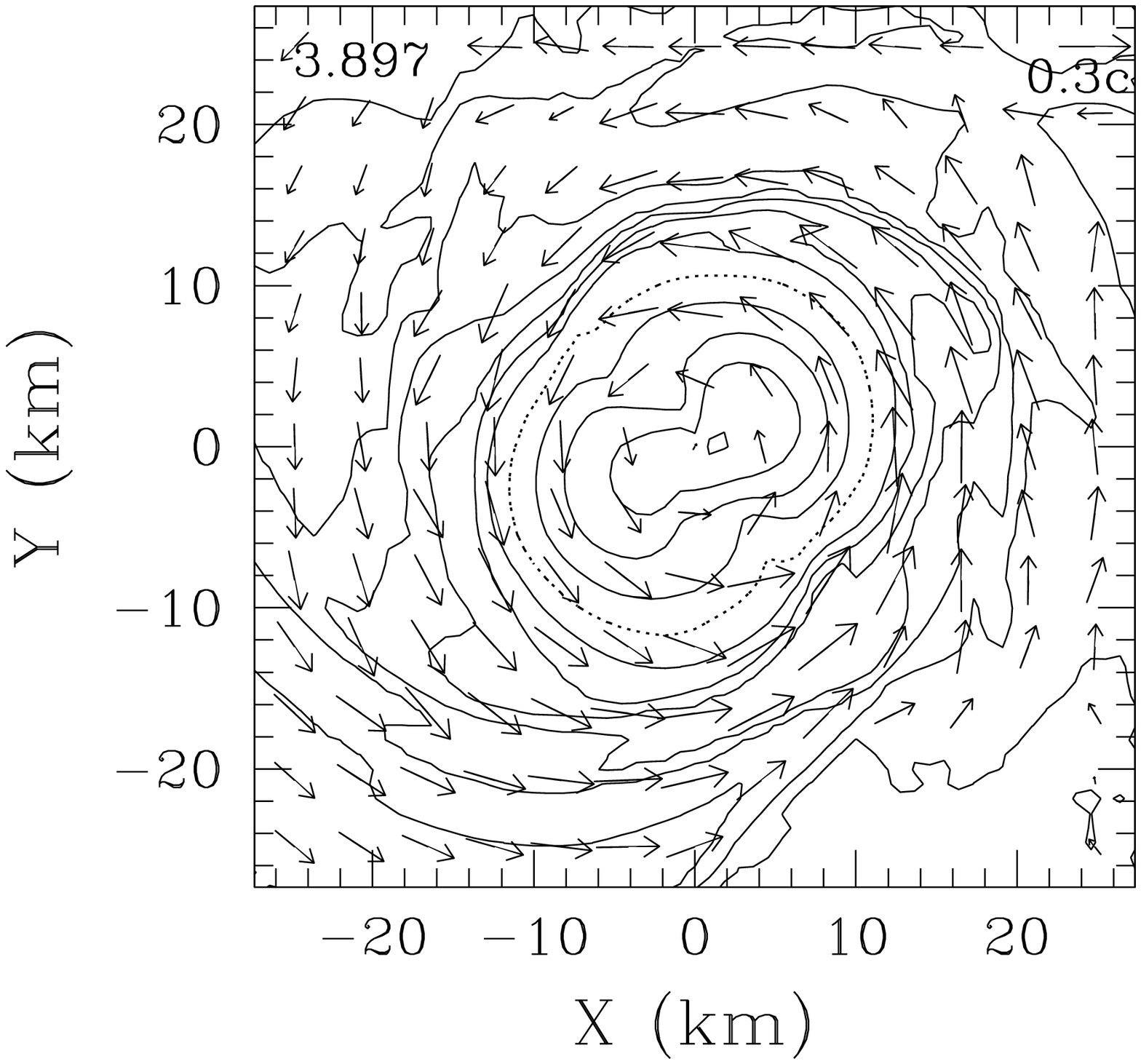}
  \hspace{-1.65cm}\includegraphics[width=2.2in]{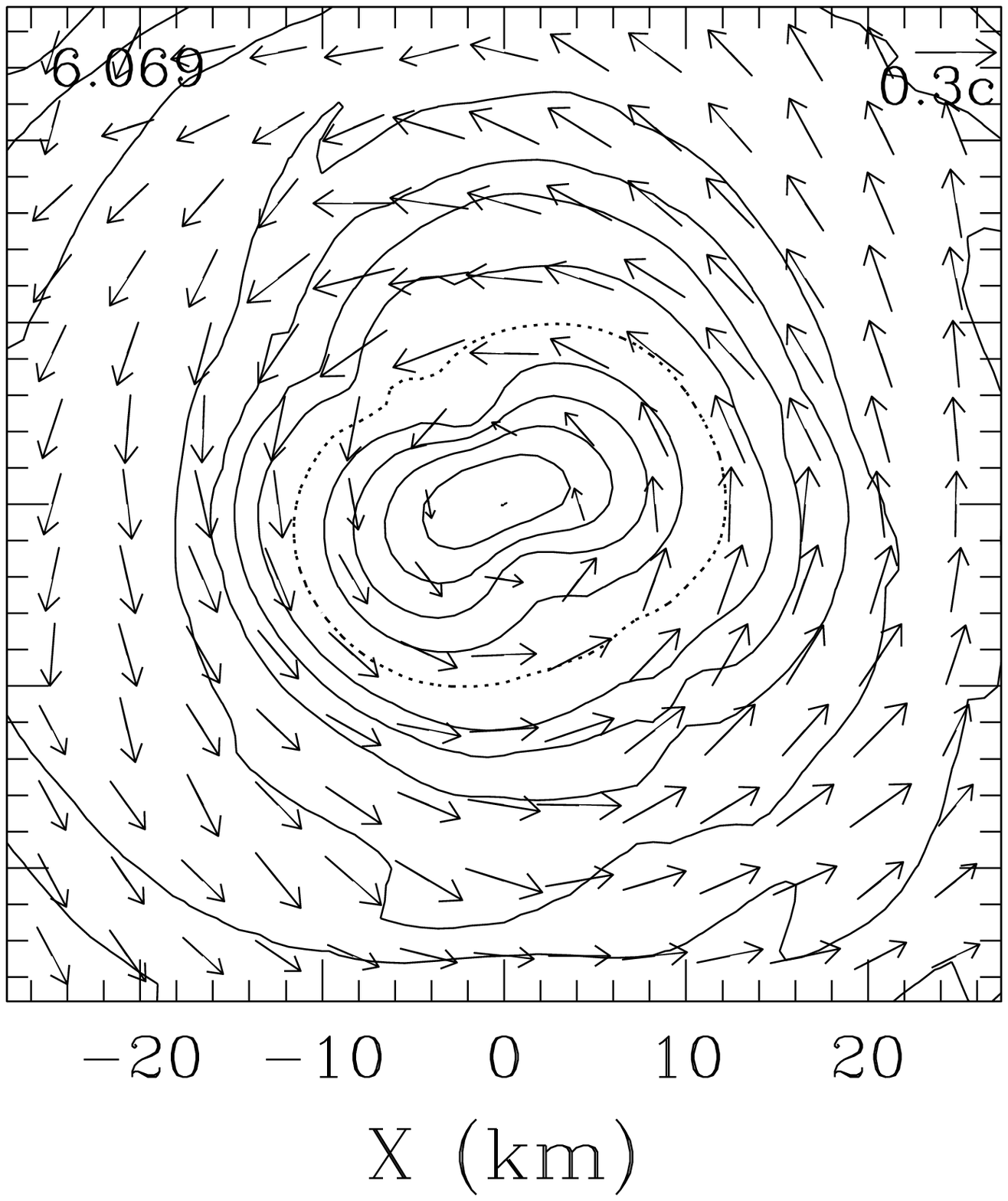} 
  \hspace{-1.65cm}\includegraphics[width=2.2in]{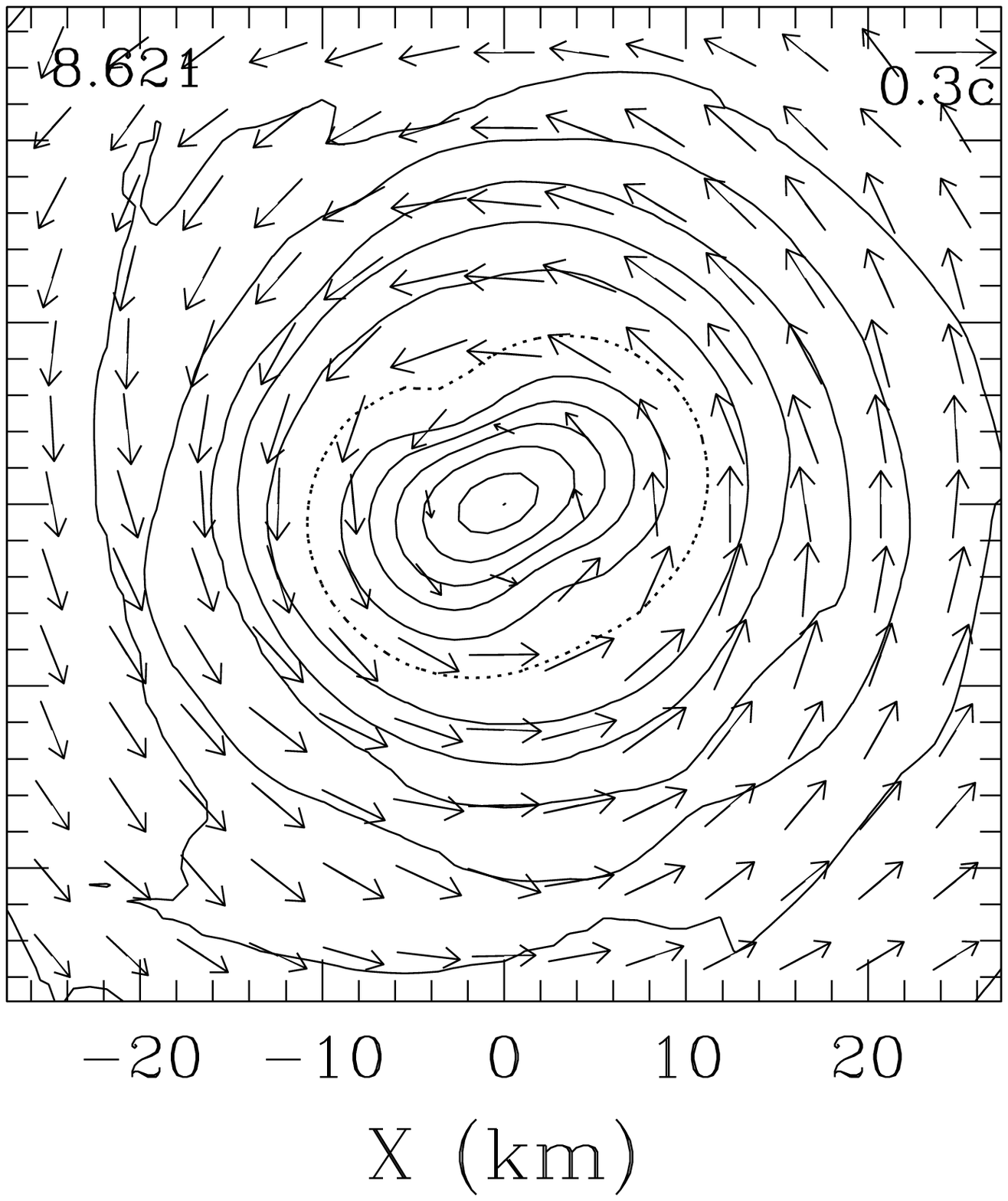}
\vspace{-4mm}
\caption{\small
The same as Fig. \ref{FIG3} but for model SLy125135a. 
The initial orbital period is 2.110 ms in this case. 
\label{FIG5}}
\end{center}
\end{figure*}

\begin{figure*}[p]
\vspace{-10mm}
\begin{center}
  (a)\includegraphics[width=2.7in]{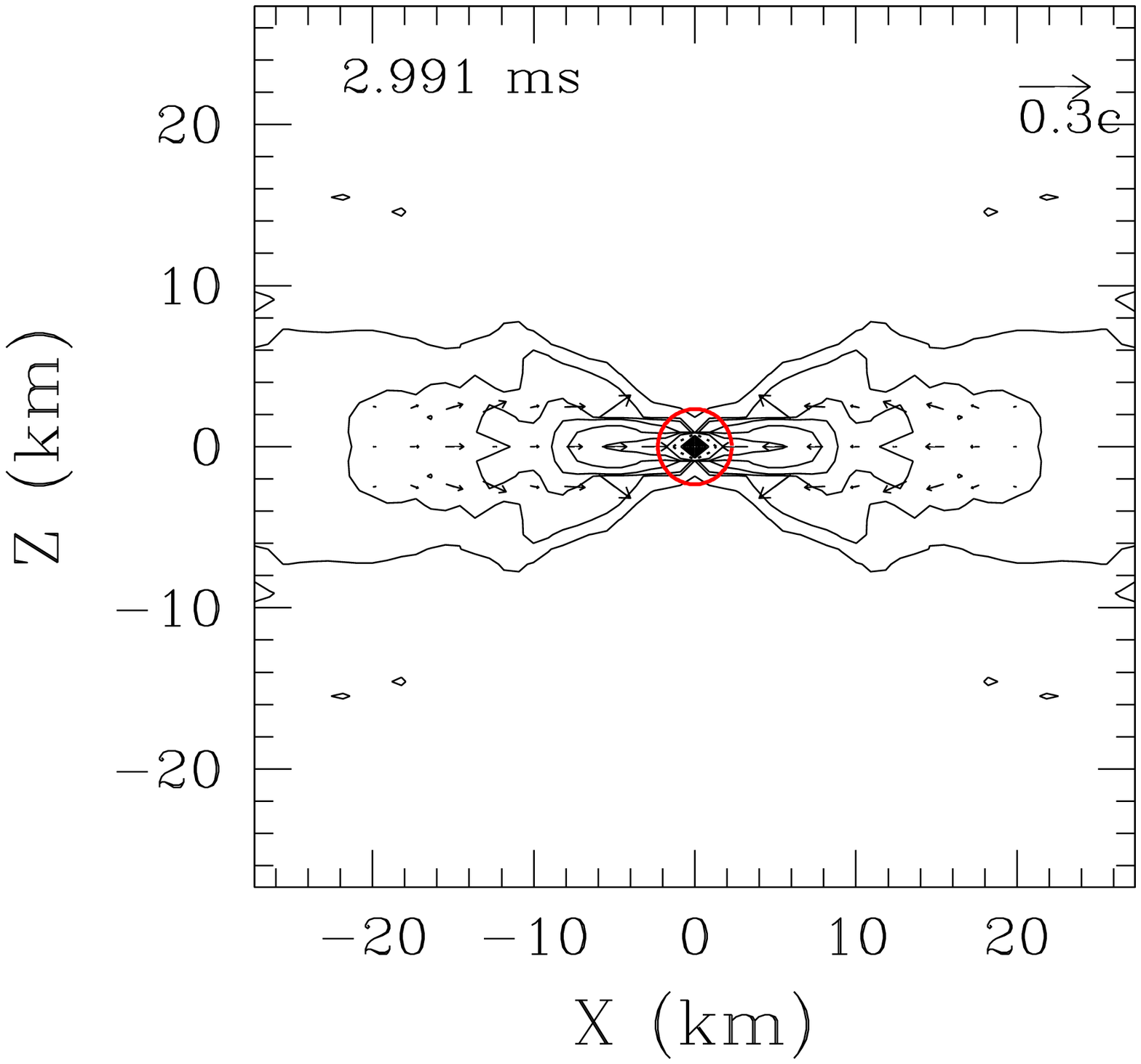}
  ~~(b)\includegraphics[width=2.7in]{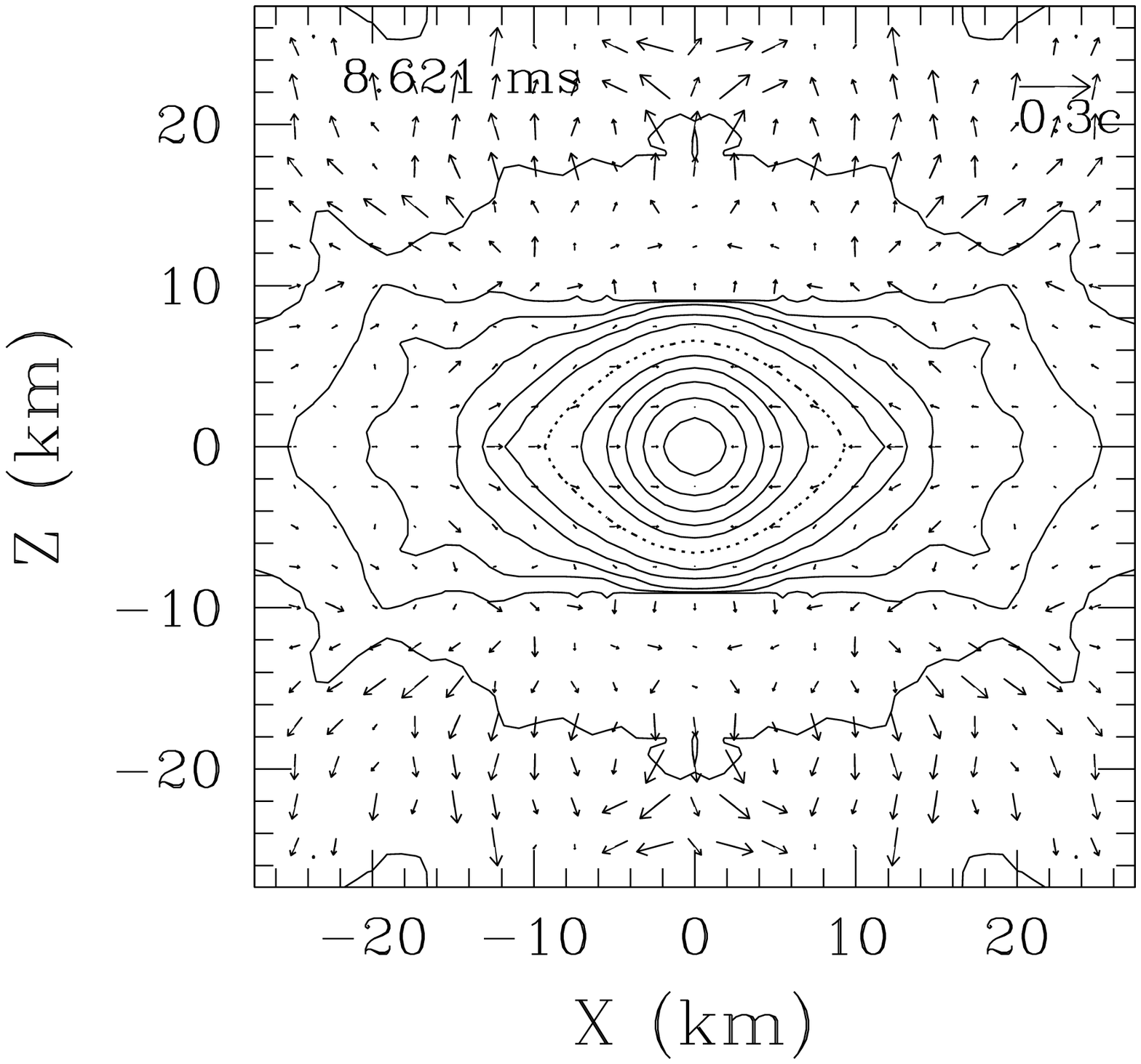}
\end{center}
\vspace{-4mm}
\caption{
Snapshots of the density contour curves for $\rho$ 
and the local velocity field $(v^x,v^z)$ in the $y=0$ plane
(a) at $t=2.991$ ms for model SLy1414a and 
(b) at $t=8.621$ ms for model SLy1313a.
The method for drawing the contour curves and
the velocity vectors is the same as that in Fig. \ref{FIG3}. 
\label{FIG6}}
\end{figure*}

Several quantities that characterize irrotational binary neutron stars
in quasiequilibrium circular orbits used as initial conditions for the
present simulations are summarized in Table II. We choose binaries of
an orbital separation which is slightly larger than that for an
innermost orbit.  Here, the innermost orbit is defined as a close
orbit for which Lagrange points appear at the inner edge of neutron
stars \cite{USE,GBM}. If the orbital separation becomes smaller than
that of the innermost orbit, mass transfer sets in and dumbbell-like
structure will be formed.  Until the innermost orbit is reached, the
circular orbit is stable, and hence, the innermost stable circular
orbit does not exist outside the innermost orbit for the present
cases. However, we should note that the innermost stable circular
orbit seems to be very close to the innermost orbit since the decrease
rates of the energy and the angular momentum as functions of the
orbital separation are very small near the innermost orbit.

The ADM mass of each neutron star, when it is in isolation (i.e., when
the orbital separation is infinity), is chosen in the range between
$1.2M_{\odot}$ and $1.45M_{\odot}$. Models SLy1212, SLy1313,
SLy135135, SLy1414, FPS1212, FPS125125, FPS1313, and FPS1414 are equal-mass
binaries, and SLy125135 and SLy135145 are unequal-mass ones.  For the
unequal-mass case, the mass ratio $Q_M$ is chosen to be $\agt 0.9$
since all the observed binary neutron stars for which each mass is
determined accurately indeed have such mass ratio \cite{Stairs}. Mass
of each neutron star in model SLy125135 is approximately the same as
that of PSRJ0737-3039 \cite{NEW}, while the mass in model SLy135145 is
similar to that of PSRB1913+16 \cite{HT}.  The total baryon rest-mass
for models SLy1313 and SLy125135 and for models SLy1414 and SLy135145
are approximately identical, respectively. For all these binaries, the
orbital period of the initial condition is about 2 ms. This implies
that the frequency of emitted gravitational waves is about 1 kHz.

The simulations were performed using a fixed uniform grid 
and assuming reflection symmetry with respect to the equatorial plane 
(here, the equatorial plane is chosen to be the orbital plane).
The detailed simulations were performed 
with the SLy equation of state. In this equation of state, 
the used grid size is (633, 633, 317) or (377, 377, 189) for $(x, y, z)$.
In the FPS equation of state, simulations were performed with the
(377, 377, 189) grid size to save the computational time. 
The grid covers the region $-L \leq x \leq L$, $-L \leq y \leq L$, and 
$0 \leq z \leq L$ where $L$ is a constant. 
The grid spacing is determined 
from the condition that the major diameter of each star is covered 
with about 50 grid points initially. We have shown that with this
grid spacing, a convergent numerical result is obtained \cite{STU}. 
The circumferential radius of spherical neutron stars 
with the SLy and FPS equations of state is about 11.6 and 10.7 km for 
$M=1.4M_{\odot}$, respectively (see Fig. \ref{FIG2}). Thus, the
grid spacing is $\sim 0.4$ km. 

Accuracy in the computation of gravitational waveforms and the
radiation reaction depends on the location of the outer boundaries
if the wavelength, $\lambda$, is larger than $L$ \cite{STU}.
For $L \alt 0.4 \lambda$, the amplitude and the 
radiation reaction of gravitational waves are significantly 
overestimated \cite{STU,SU01}. 
Due to the restriction of the computational power,
it is difficult to take a huge grid size in which $L$ is much larger
than $\lambda$. As a consequence of the
present restricted computational resources,
$L$ has to be chosen as $\sim 0.4 \lambda_0$
where $\lambda_0$ denotes $\lambda$ of
the $l=m=2$ mode at $t=0$. Hence, the error associated with
the small value of $L$ is inevitable, and thus, the amplitude and 
radiation reaction of gravitational waves are overestimated
in the early phase of the simulation. 
However, the typical wavelength of gravitational waves becomes
shorter and shorter in the late inspiral phase, and hence, 
the accuracy of the wave extraction is improved with the evolution
of the system. This point will be confirmed in Sec. III.

The wavelength of quasiperiodic gravitational waves emitted from the formed 
hypermassive neutron star (denoted by $\lambda_{\rm merger}$) 
is much shorter than $\lambda_0$ and as large as $L$ 
(see Table III), so that the waveforms 
in the merger stage are computed accurately (within $\sim 10\%$ error) 
in the case of neutron star formation irrespective of the grid size.
We performed simulations for
models SLy1313, SLy125135, SLy1414, and SLy135145 with the two grid sizes,
and confirmed that this is indeed the case. 
We demonstrate this fact in Sec. III by comparing the 
results for models SLy1313a and SLy1313b. 
From the numerical results for four models, we have also confirmed that 
the outcome in the merger does not depend on the grid size. 
Thus, when we are interested in the outcome or in gravitational waves 
emitted by the hypermassive neutron stars, 
simulations may be performed in a small grid size
such as (377, 377, 189). 

With the (633, 633, 317) grid size, about 240 GBytes 
computational memory is required. For the case of
the hypermassive neutron star formation, 
the simulations are performed for about 30,000 time steps
(until $t \sim 10$ ms) and then stopped to save the computational time.
The computational time for one model in 
such a simulation is about 180 CPU hours 
using 32 processors on FACOM VPP5000 in the data processing center of
National Astronomical Observatory of Japan (NAOJ). 
For the case of the black hole formation, the simulations 
crash soon after the formation of apparent horizon because of
the so-called grid stretching around the black hole formation region.
In this case, the computational time is about 60 CPU hours for
about 10,000 time steps. 

\subsection{Characteristics of the merger}

\begin{figure*}[thb]
\vspace{-4mm}
\begin{center}
  (a)\includegraphics[width=3.2in]{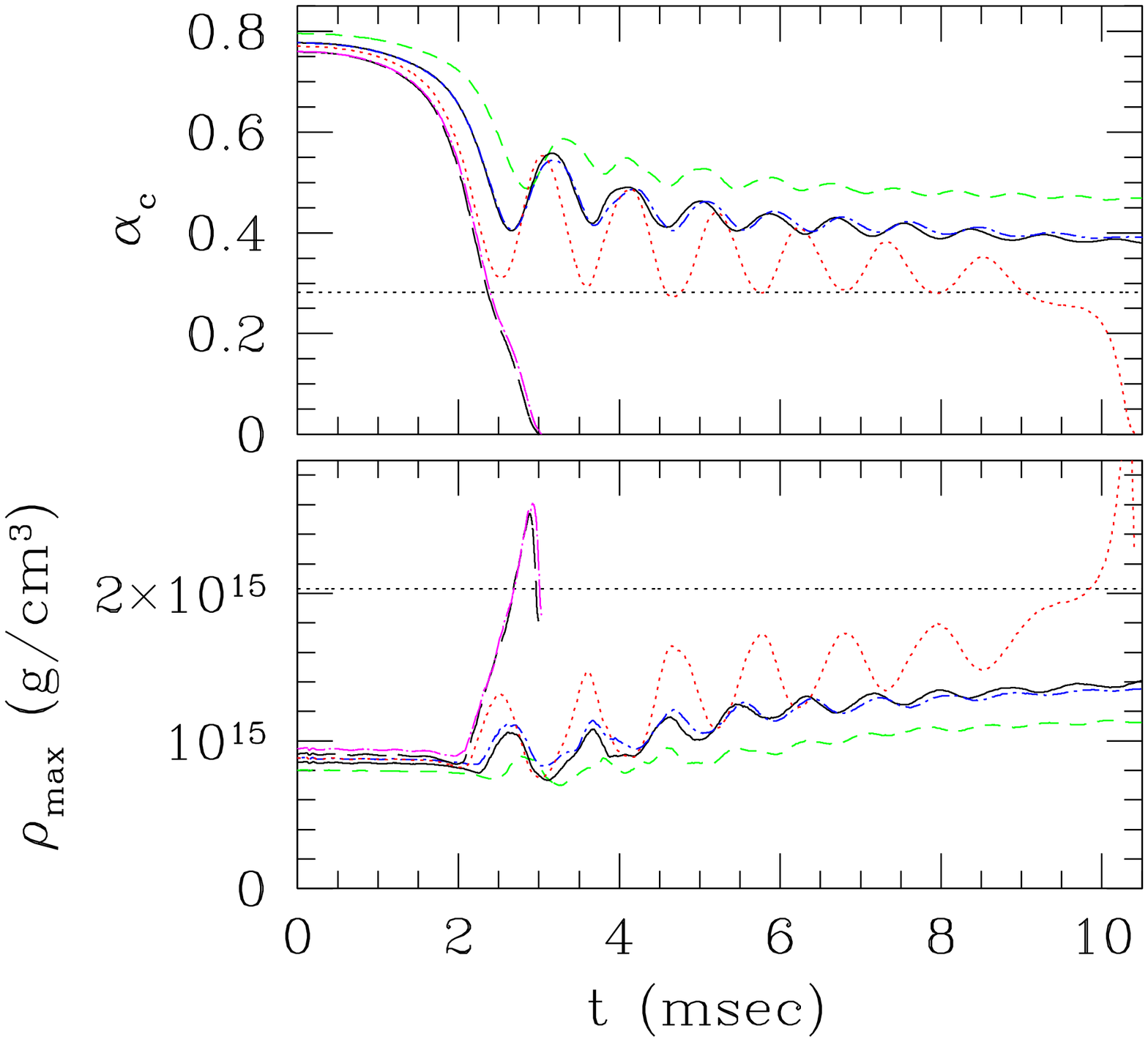}
  ~~~~(b)\includegraphics[width=3.2in]{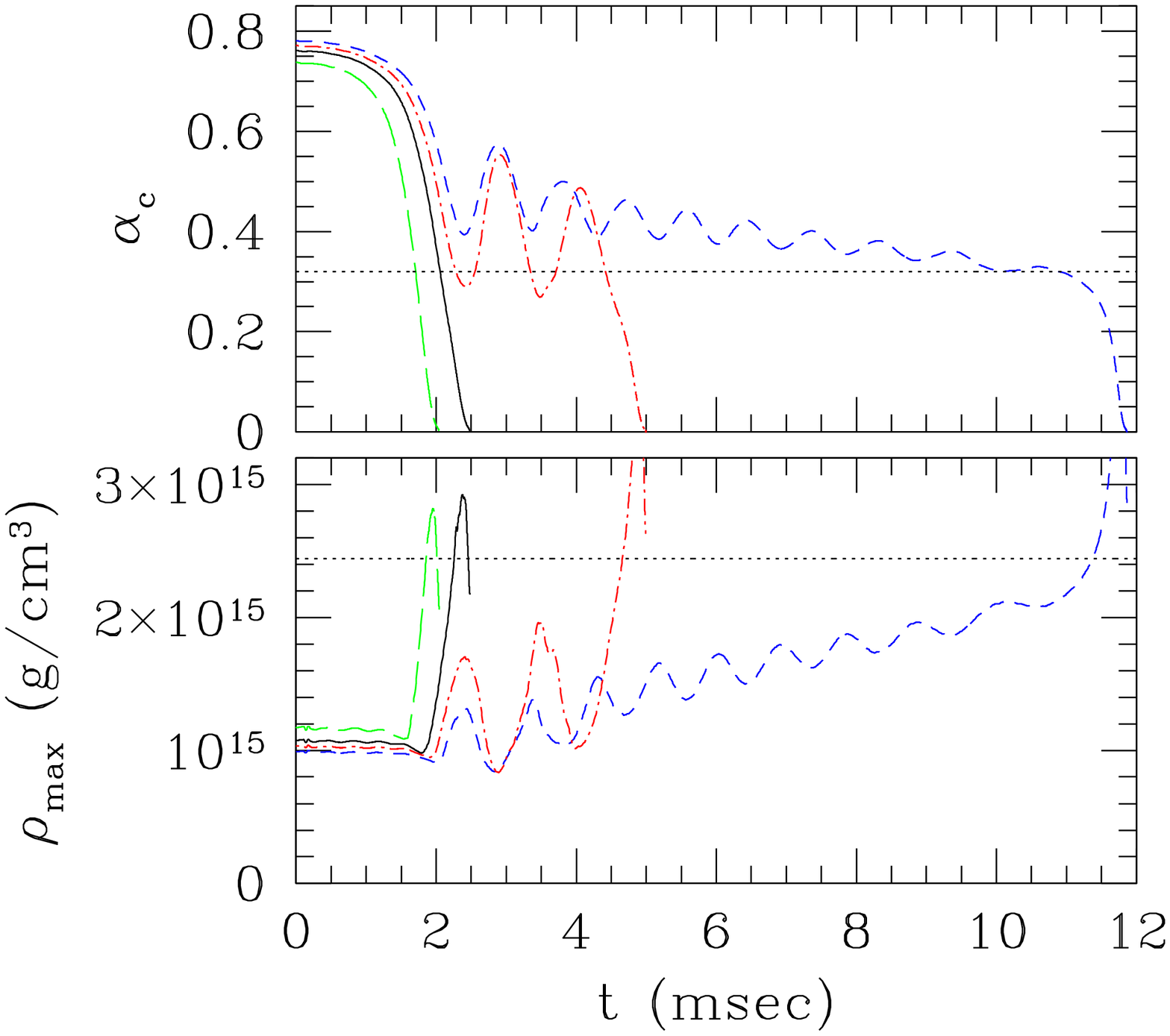}
\end{center}
\vspace{-2mm}
\caption{Evolution of the maximum values of $\rho$ and
the central value of $\alpha$
(a) for models SLy1414a (long dashed curves),
SLy135145a (dotted long-dashed curves which approximately coincide
with long dashed curve for $\alpha_c$), 
SLy135135b (dotted curves), 
SLy1313a (solid curves),
SLy125135a (dotted dashed curves), and SLy1212b (dashed curves), and 
(b) for models FPS1414b (long dashed curves), 
FPS1313b (solid curves), FPS125125 (dotted-dashed curves),
and FPS1212b (dashed curves). 
The dotted horizontal lines denote the central values of the lapse and
density of the marginally stable and spherical 
star in equilibrium for given cold equations of state. 
The reason that the maximum density decreases in the final stage
for the black hole formation case
is as follows: We choose $\rho_*$ as a fundamental variable
to be evolved and compute $\rho$ from $\rho_*/(\alpha u^t e^{6\phi})$.
In the final stage, $\phi$ is very large ($> 1$) and, hence, a 
small error in $\phi$ results in a large error in $\rho$. Note that 
the maximum value of $\rho_*$ increases monotonically
by many orders of magnitude. 
\label{FIG7}}
\end{figure*}

\begin{figure*}[thb]
\vspace{-4mm}
\begin{center}
  (a)\includegraphics[width=3.in]{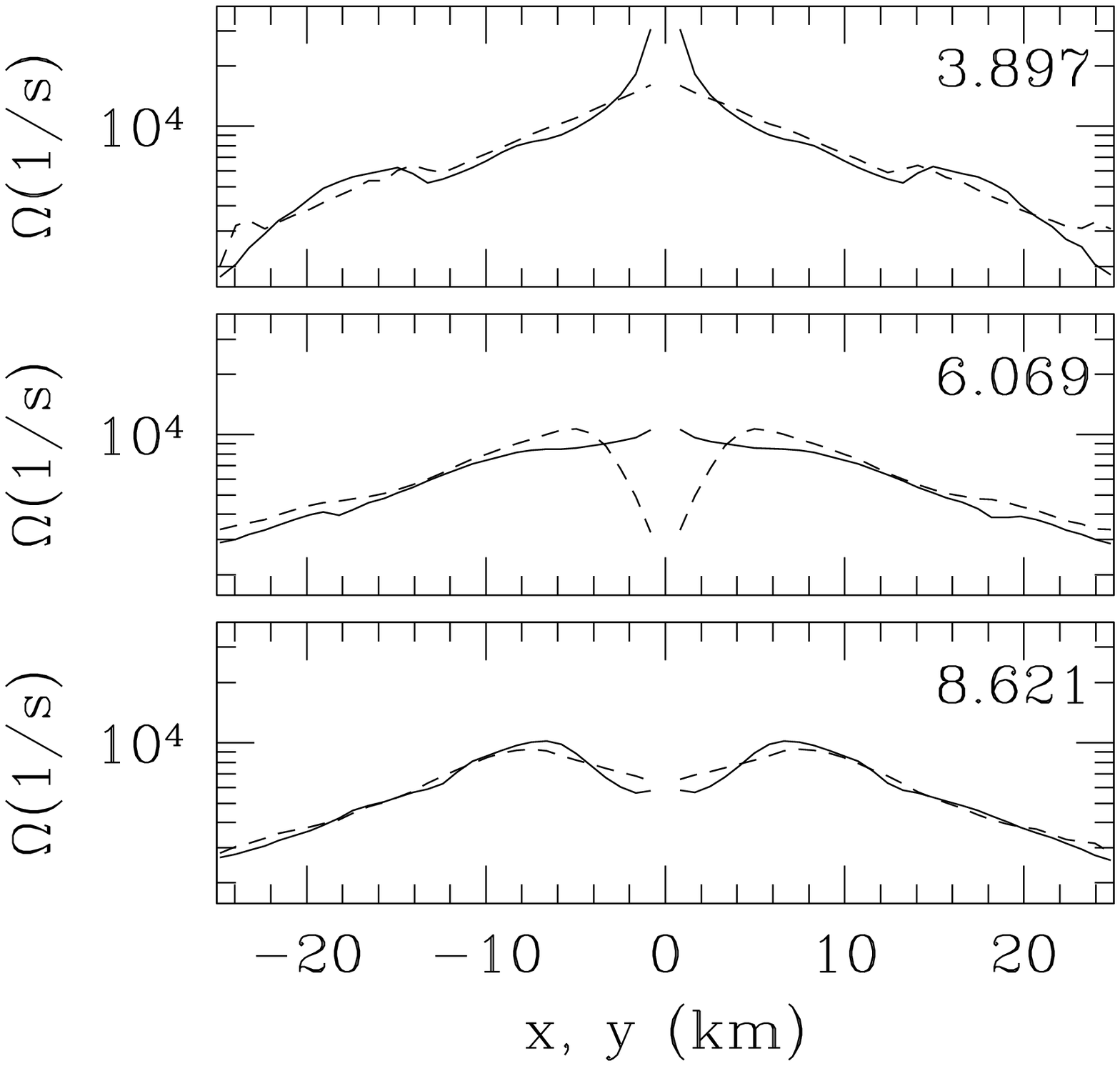}
  ~~~(b)\includegraphics[width=3.in]{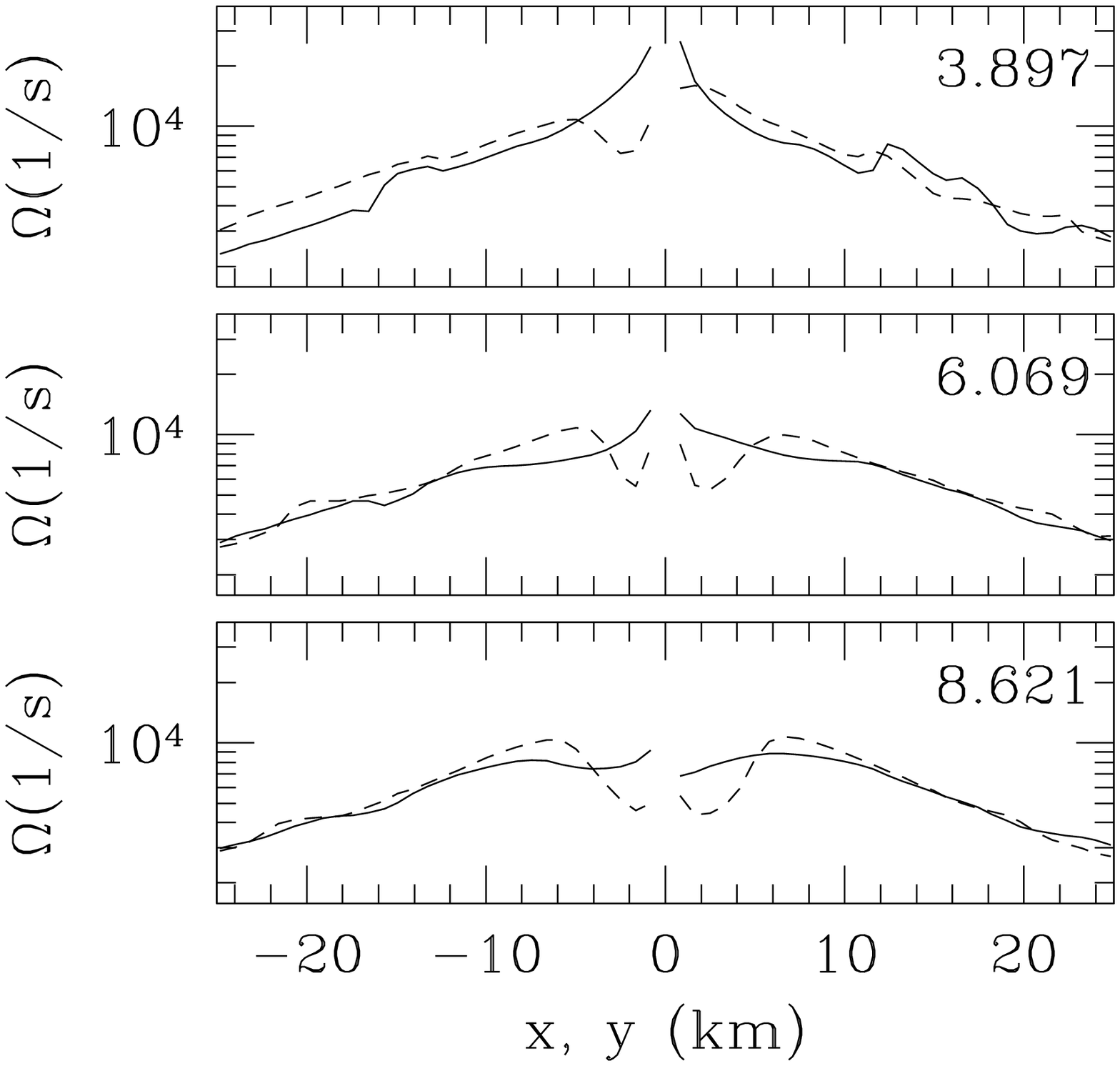}
\end{center}
\vspace{-2mm}
\caption{Evolution of the angular velocity $\Omega$
along $x$ (solid curves) and $y$ (dashed curves) axes 
(a) for models SLy1313a and (b) SLy125135a.
The time is shown in the upper right corner of
each panel in units of ms. 
$\Omega$ along $x$ and $y$ axes is computed by $v^y/x$ and
$v^x/y$, and hence, it is not determined at the origin. 
Since the formed hypermassive neutron star is not spheroidal,
$\Omega$ along two axes is significantly different. 
\label{FIG8} }
\end{figure*}

\subsubsection{General feature}\label{sec:gen}

In Figs. \ref{FIG3}--\ref{FIG5},
we display the snapshots of the density contour curves and 
the velocity vectors in the equatorial plane
at selected time steps for models SLy1414a, SLy1313a, and SLy125135a,
respectively. 
Figure \ref{FIG6} displays the density contour curves and 
the velocity vectors in the $y=0$ plane at a late time 
for SLy1414a and SLy1313a. 
Figures \ref{FIG3} and \ref{FIG6}(a) indicate typical evolution 
of the density contour curves in the case of prompt black hole 
formation. On the other hand, Figs. \ref{FIG4}, \ref{FIG5}, and 
\ref{FIG6}(b) show those in the formation of hypermassive neutron stars. 

Figure \ref{FIG7} displays the evolution of the maximum value of 
$\rho$ (hereafter $\rho_{\rm max}$) and the central value of $\alpha$
(hereafter $\alpha_c$) for models 
SLy1414a, SLy135145a, SLy135135b, SLy1313a, SLy125135a, and SLy1212b
(Fig. \ref{FIG7}(a)) and 
for models FPS1414b, FPS1313b, FPS125125b, and FPS1212b (Fig. \ref{FIG7}(b)).
This shows that in the prompt black hole formation,
$\alpha_{c}$ monotonically decreases toward zero. 
On the other hand, $\alpha_c$ and $\rho_{\rm max}$ settle down
to certain values in the hypermassive neutron star formation.
For model SLy135135b, a hypermassive neutron star is formed first,
but after several quasiradial oscillations of high amplitude, it
collapses to a black hole due to dissipation of the angular momentum
by gravitational radiation. The large oscillation amplitude results from
the fact that the selfgravity is large enough to deeply shrink 
surmounting the centrifugal force. These indicate that the total ADM 
mass of this model ($M \approx 2.7M_{\odot}$) is only slightly smaller than
the threshold value for the prompt black hole formation.
The quasiradial oscillation of the large amplitude induces a
characteristic feature in gravitational waveforms and the
Fourier spectrum (cf. Sec. \ref{sec:gw}). 

In the case of black hole formation 
(models SLy1414, SLy135145, SLy135135, FPS1414, FPS1313, FPS125125,
and FPS1212), 
the computation crashed soon after the formation of apparent horizons 
since the region around the apparent horizon of the
formed black hole was stretched significantly and 
the grid resolution became too poor to resolve such region. 
On the other hand, we stopped the simulations for other cases
to save the computational time, after the evolution of
the formed massive neutron stars was followed for a sufficiently long time. 
At the termination of these simulations, the averaged violation of the
Hamiltonian constraint $H$ remains of order 0.01 (cf. Fig. \ref{FIG18}).
We expect that the simulations could be continued for a much longer time 
than $10$ ms if we could have sufficient computational time. 

In every model, the binary orbit is stable at $t=0$
and the orbital separation gradually decreases 
due to the radiation reaction of gravitational waves
for which the emission time scale is longer than the orbital period.
If the orbital separation becomes sufficiently small,
each star is elongated by tidal effects. As a result,
the attraction force due to the tidal interaction between
two stars becomes strong enough to make the orbit unstable to merger. 
The merger starts after about one orbit at $t \sim 2$ ms irrespective
of models. Since the orbital separation at $t=0$ is very close to
that for a marginally stable orbit, a small decrease of the angular momentum
and energy is sufficient to induce the merger in the present simulations. 
If the total mass of the system is high enough, a black hole is directly 
formed within about 1 ms after the merger sets in.
On the other hand, for models with mass 
smaller than a threshold mass $M_{\rm thr}$, 
a hypermassive neutron star is formed at least temporarily. 
The hypermassive neutron star is stable against gravitational collapse 
for a while after its formation, but it will collapse to a black hole
eventually due to radiation reaction of gravitational waves
or due to outward angular momentum transfer (see discussion later). 

In the formation of the hypermassive neutron stars,
a double core structure is first formed, and then, it relaxes to
a highly nonaxisymmetric ellipsoid 
(cf. Figs. \ref{FIG4}, \ref{FIG5}, and \ref{FIG6}(b)). 
The contour plots drawn for a high-density region 
with $\rho \geq 4 \times 10^{14}~{\rm g/cm^3}$ show that 
the axial ratio of the bar measured in the equatorial plane is $\sim 0.5$;
the axial lengths of the semi major and minor axes are $\sim 20$ and 10 km,
respectively. Figure \ref{FIG6}(b) also shows 
that the axial length along the $z$ axis is about 10 km. 
Namely, a highly elliptical rotating ellipsoid is formed.
This outcome is significantly different from 
the previous ones found with the $\Gamma=2$ equation of state \cite{STU},
in which nearly axisymmetric spheroidal neutron
stars are formed. The reason is that the adiabatic index of the
realistic equations of state adopted in this paper is much larger 
than 2 that is adopted in the previous one. 
According to a Newtonian study \cite{james}, a uniformly
rotating ellipsoid (Jacobi-like ellipsoid) 
exists only for $\Gamma \agt 2.25$. 
This fact suggests that rapidly rotating stars with a large
adiabatic index are only subject to the ellipsoidal
deformation. Note that similar results have been already reported 
in Newtonian and post Newtonian simulations \cite{RS,C,FR}. 

The rotating hypermassive neutron stars also oscillate in a
quasiradial manner (cf. Fig. \ref{FIG7}).  Such oscillation is induced
by the approaching velocity at the collision of two stars. By the
radial motion, shocks are formed at the outer region of the
hypermassive neutron stars to convert the kinetic energy to the
thermal energy.  The shocks are also generated when the spiral arms
hit the oscillating hypermassive neutron stars. These shocks heat up
the outer region of the hypermassive neutron stars for many times, and 
as a result, the thermal energy of the envelope increases fairly
uniformly. The further detail of these thermal properties is discussed
in Sec. \ref{sec:therm}. 

Since the degree of the nonaxial symmetry is sufficiently large, the
hypermassive neutron star found in this paper is a 
stronger emitter of gravitational waves than that found in \cite{STU}. 
The significant radiation decreases the angular momentum of the 
hypermassive neutron stars. 
The nonaxisymmetric structure also induces the angular momentum 
transfer from the inner region to the outer one due to the hydrodynamic
interaction. As a result of these effects,
the rotational angular velocity $\Omega=v^{\varphi}$
decreases and its profile is modified.
In Fig. \ref{FIG8}, we show the evolution of $\Omega$ of
the hypermassive neutron stars along $x$ and $y$ axes at $t=3.897$,
6.069, and 8.621 ms for models SLy1313a and SLy125135a. At its
formation, the hypermassive neutron stars are strongly differentially
and rapidly rotating. The strong differential rotation yields the
strong centrifugal force, which plays an important role for sustaining
the large selfgravity of the hypermassive neutron stars
\cite{BSS,SBS}.  Since the angular momentum is dissipated by the 
gravitational radiation and redistributed by the hydrodynamic interaction,
$\Omega$ decreases significantly in the central region, and hence, 
the steepness of the differential rotation near the center decreases with
time. This effect eventually induces the collapse to a black hole. 

It should be also noted that 
$\Omega$ along two axes is significantly different near the origin.
The reason at $t=3.897$ ms is that the formed hypermassive neutron stars
have a double-core structure (cf. Figs. \ref{FIG4} and \ref{FIG5}) and
the angular velocity of the cores are larger than the
low density region surrounding them. 
The reason for $t \agt 6$ ms is that the hypermassive neutron star is not a
spheroid but an ellipsoid of high ellipticity and the
angular velocity depends strongly on the coordinate $\varphi$. 

Figures \ref{FIG4} and \ref{FIG5} show that 
even after the emission of gravitational waves for $\sim$ 10 ms, the
hypermassive neutron star is still highly nonaxisymmetric. This
indicates that gravitational waves will be emitted for much longer
time scale than 10 ms. Thus, the rotational kinetic energy and the
angular momentum will be subsequently dissipated by a large factor, 
eventually inducing the collapse to a black hole.

We here estimate the lifetime of the hypermassive neutron stars using
Fig. \ref{FIG7} which shows that the value of $\alpha_c$ decreases 
gradually with time. 
It is reasonable to expect that the collapse to a black hole sets in
when the value of $\alpha_c$ becomes 
smaller than a critical value. Since the angular momentum should be
sufficiently dissipated before the collapse sets in, 
the threshold value of $\alpha_c$ for the onset of the
collapse will be approximately equal to that of marginally stable 
spherical stars (i.e., the dotted horizontal line in Fig. \ref{FIG7}).
One should keep in mind that the threshold value depends on the
slicing condition, and thus, this method can work only when the same
slicing is used for computation of the spherical star and for simulation. 
In this paper, the (approximate) maximal slicing is adopted both in the
simulation and in computation of spherical equilibria so that this
method can be used. The results for models SLy135135b and FPS1212b
indeed illustrate that the prediction by this method is appropriate.

For models SLy1313a, SLy125135a and SLy1212b, the decrease rate of the
value of $\alpha_c$ estimated from the data for 5 ms $\alt t \alt 10$ ms
is $\sim 0.005$ ${\rm ms}^{-1}$. Extrapolating this
result suggests that the hypermassive neutron stars will collapse
to a black hole at $t \sim 30$ ms for models SLy1313a and SLy125135a
and at $t \sim 50$ ms for model SLy1212b. 
These time scales are much shorter than the dissipation time scale by
viscosities or the redistribution time scale of the angular momentum
by the magnetic field \cite{BSS}. Therefore, 
the gravitational radiation or the outward angular momentum transfer
by the hydrodynamic interaction plays the most important role in the 
evolution of the hypermassive neutron stars. 

In the prompt formation of a black hole, 
most of the fluid elements are swallowed into the black hole
in 1 ms after the merger sets in. Thus, 
the final outcome is a system of a rotating black hole
and a surrounding disk of very small mass (cf. Fig. \ref{FIG6}(a)). 
In Fig. \ref{FIG9}, we plot the evolution of the 
total baryon rest-mass located outside a radius $r$, $M_*(r)$, for models
SLy1414a and SLy135145a. $M_*(r)$ is defined by
\beqn
M_*(r)=\int_{r \leq r' \leq r_{\rm max}} \rho_* d^3x', \label{baryon}
\eeqn
where $r_{\rm max}$ is introduced to exclude the contribution from
the small-density atmosphere. 
In the present work we choose as $r_{\rm max}=7.5M_0 \approx 31$ km within
which the integrated mass of the atmosphere is negligible ($<$0.01\% of
the total mass). The results are plotted for $r/M_0=1.5$, 3, and 4.5. 
Note that the apparent horizon is located for $r \approx 0.5M_0$
at $t \approx 3.0$ ms for models SLy1414a and SLy135145a, and 
inside the horizon about 99\% and 98\% of the initial mass 
are enclosed for these cases, respectively.
Figure \ref{FIG9} indicates that the fluid elements 
still continue to fall into the black hole at the end of the
simulation. This suggests that the final disk mass will be smaller
than 1\% of the total baryon rest-mass. 

In \cite{STU}, we found that even the small mass difference with
$Q_M \sim 0.9$ increases the fraction of disk mass around the black
hole significantly. However, in the present equations of state,
$Q_M \sim 0.9$ is not small enough to significantly increase the disk mass.
This results from the difference in the equations of state. 
The detailed reason is discussed in Sec. \ref{sec:massdif}.

The area of the apparent horizons $A$ is determined in the black hole
formation cases. We find that
\beqn
{A \over 16\pi M_0^2} \sim 0.85,
\eeqn
for models SLy1414a and SLy135145a. Since most of the fluid elements
are swallowed into the black hole and also the energy
carried out by gravitational radiation is at most $\sim 0.01M_0$
(cf. Fig. \ref{FIG18}), 
the mass of the formed black hole is approximately $\sim 0.99M_0$.
Assuming that the area of the apparent horizon is equal to that 
of the event horizon, the nondimensional spin parameter of 
the black hole defined by $q\equiv J_{\rm BH}/M_{\rm BH}^2$, where 
$J_{\rm BH}$ and $M_{\rm BH}$ are the angular momentum and the mass
of the black hole, are computed from
\beqn
{A \over 16\pi M_{\rm BH}^2}={1\over 2}\Bigl[1+(1-q^2)^{1/2}\Bigr].
\label{aaa}
\eeqn
Equation (\ref{aaa}) implies that 
for $A/16\pi M_{\rm BH}^2 \sim 0.85$, $q \sim 0.7$.

On the other hand, we can estimate the value of $q$ in the following manner. 
As shown in Sec. \ref{sec:gw}, the angular momentum is dissipated
by $\sim 10$--15\% by gravitational radiation, while the ADM mass
decreases by $\sim 1\%$.
As listed in Table II, the initial value of $q$ is $\sim 0.9$.
Therefore, the value of $q$ in the final stage should be $\sim 0.75$--0.8. 
The values of $q$ derived by two independent methods agree with each other
within $\sim 10\%$ error. This indicates that the location and the area of the
black holes are determined within $\sim 10\%$ error. 

For $q=0.7$--0.8 and $M_{\rm BH} \approx 2.8M_{\odot}$, the frequency of 
the quasinormal mode for the black hole oscillation is about
6.5--7$(2.8M_{\odot}/M_{\rm BH})$ kHz 
\cite{Leaver}. This value is rather high and far out of the
best sensitive frequency range 
of the laser interferometric gravitational wave detectors \cite{KIP}. 
Thus, in the following, we do not touch on gravitational waveforms 
in the prompt black hole formation. 

\subsubsection{Threshold mass for black hole formation}

The threshold value of the
total ADM mass above which a black hole is promptly formed is 
$M_{\rm thr} \sim 2.7M_{\odot}$ for the SLy equation of state and
$M_{\rm thr} \sim 2.5M_{\odot}$ for the FPS one with
$\Gamma_{\rm th}=2$. For the SLy case, we find that
the value does not depend on the mass ratio for 
$0.9 \alt Q_M \leq 1$. The maximum allowed mass for
the stable and spherical neutron stars 
is $2.04M_{\odot}$ and $1.80M_{\odot}$ for the SLy and FPS 
equations of state, respectively.
This implies that if the total mass is by $\sim 30$--40\% larger
than the maximum allowed mass for stable and spherical stars,
a black hole is promptly formed. 
In a previous study with the $\Gamma=2$ equation of state \cite{STU},
we found that threshold mass is by about 70\% larger than the
maximum allowed mass for stable and spherical neutron stars.
Thus, comparing the threshold value of the total ADM mass, 
we can say that a black hole is more subject to be formed promptly 
with the realistic equations of state.

In \cite{LBS,MBS}, the maximum mass of differentially rotating 
stars in axisymmetric equilibrium (hereafter $M_{\rm max:dif}$)  
is studied for various equations of state. The authors compare 
$M_{\rm max:dif}$ with the maximum mass of spherical stars 
(hereafter $M_{\rm max:sph}$) for given equations of state. 
They find that the ratio $M_{\rm max:dif}/M_{\rm max:sph}$ 
for FPS and APR equations of state (APR is similar to SLy equation 
of state) is much smaller than that for $\Gamma=2$ equation of state. 
Their study is carried out for axisymmetric rotating stars in 
equilibrium and with a particular rotational law, 
and hence, their results cannot be simply compared with our results 
obtained for dynamical and nonaxisymmetric spacetime. 
However, their results suggest that the merged object may be 
more susceptible to collapse to a black hole with the
realistic equations of state. This tendency agrees with our conclusion. 

The compactness in each neutron star of no rotation 
in isolation is defined by $C=GM_{\rm sph}/R_{\rm sph} c^2$
where $M_{\rm sph}$
and $R_{\rm sph}$ denote the ADM mass and the circumferential radius of the
spherical stars. For the SLy equation of state, $C \approx 0.151$,
0.165, 0.172, and 0.178 for $M_{\rm sph}=1.2$, 1.3, 1.35,
and $1.4M_{\odot}$, respectively.
For FPS one, $C \approx 0.162$, 0.169, 0.177, and 0.192 
for $M_{\rm sph}=1.2$, 1.25, 1.3, and $1.4M_{\odot}$, 
respectively. This indicates that a black hole 
is promptly formed for $C \agt 0.17$ after merger of
two (nearly) identical neutron stars. 
In the $\Gamma=2$ equation of state, 
the threshold value of $C$ is $\sim 0.15$--0.16 \cite{STU}.
Thus, comparing the threshold value of the compactness,
we can say that a black hole is less subject to be formed
with the realistic equations of state. 

The reason that the threshold mass for the prompt black hole formation
is smaller with the realistic equations of state may be mentioned
in the following manner: In the realistic equations of state,
the compactness of each neutron star 
is larger than that with the $\Gamma=2$ equation of state
for a given mass. Accordingly, for a given total mass, 
the binary system at the onset of the merger 
is more compact. This implies that the angular momentum is dissipated more 
before the merger sets in with the realistic equations of state.
In the case of the hypermassive neutron star formation,
the centrifugal force plays the most important role for
sustaining the large selfgravity. Thus, the large dissipation of the
angular momentum before the merger 
helps the prompt black hole formation. Therefore, 
a black hole is more subject to be formed in the realistic
equations of state. 

\subsubsection{Thermal properties}\label{sec:therm}. 

In Fig. \ref{FIG10}(a)--(c), we show profiles of $\varep$ and
$\varep_{\rm th}$ as well as that of $\rho$ along $x$ and $y$ axes
at $t=2.991$ ms for model SLy1414a, 
at $t=8.621$ ms for model SLy1313a, and 
at $t=8.621$ ms for model SLy125135a, respectively. 
The density contour curves at the corresponding
time steps are displayed in the last panel of Figs. \ref{FIG3}--\ref{FIG5}.
Figure \ref{FIG10}(d) shows the evolution of the total internal
energy and thermal energy defined by 
\beqn
&&U \equiv \int \rho_* \varep d^3x,\\
&&U_{\rm th} \equiv \int \rho_* \varep_{\rm th} d^3x,
\eeqn
for models SLy1313a, SLy125135a, and SLy1212b. 
Note that in the absence of shock heating, $\varep$ should be 
equal to $\varep_{\rm cold}$. Thus, $\varep_{\rm th}$ 
denotes the specific thermal energy generated by the 
shock heating.

\begin{figure}[thb]
\vspace{-4mm}
\begin{center}
  \includegraphics[width=3.in]{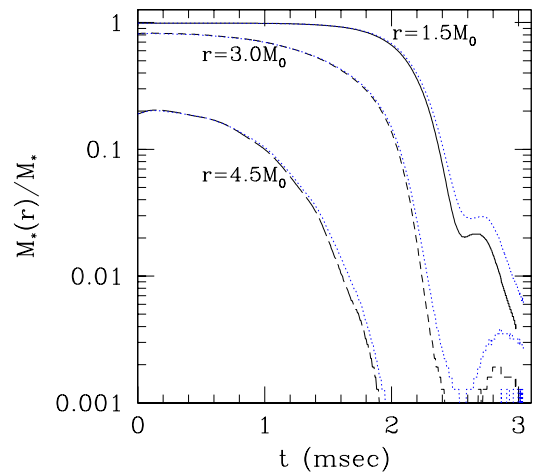}
\end{center}
\vspace{-8mm}
\caption{Evolution of $M_*(r)/M_*$ for $r=1.5$, 3, and $4.5M_0$
for models SLy1414a (curves except for the dotted curve)
and SLy135145a (dotted curves). $M_0$ and $M_*$ denote the ADM mass 
and total baryon rest-mass of binary neutron stars at $t=0$.
Definition for $M_*(r)$ is shown in Eq. (\ref{baryon}). 
\label{FIG9} }
\end{figure}

\begin{figure*}[thb]
\vspace{-4mm}
\begin{center}
  (a)\includegraphics[width=2.8in]{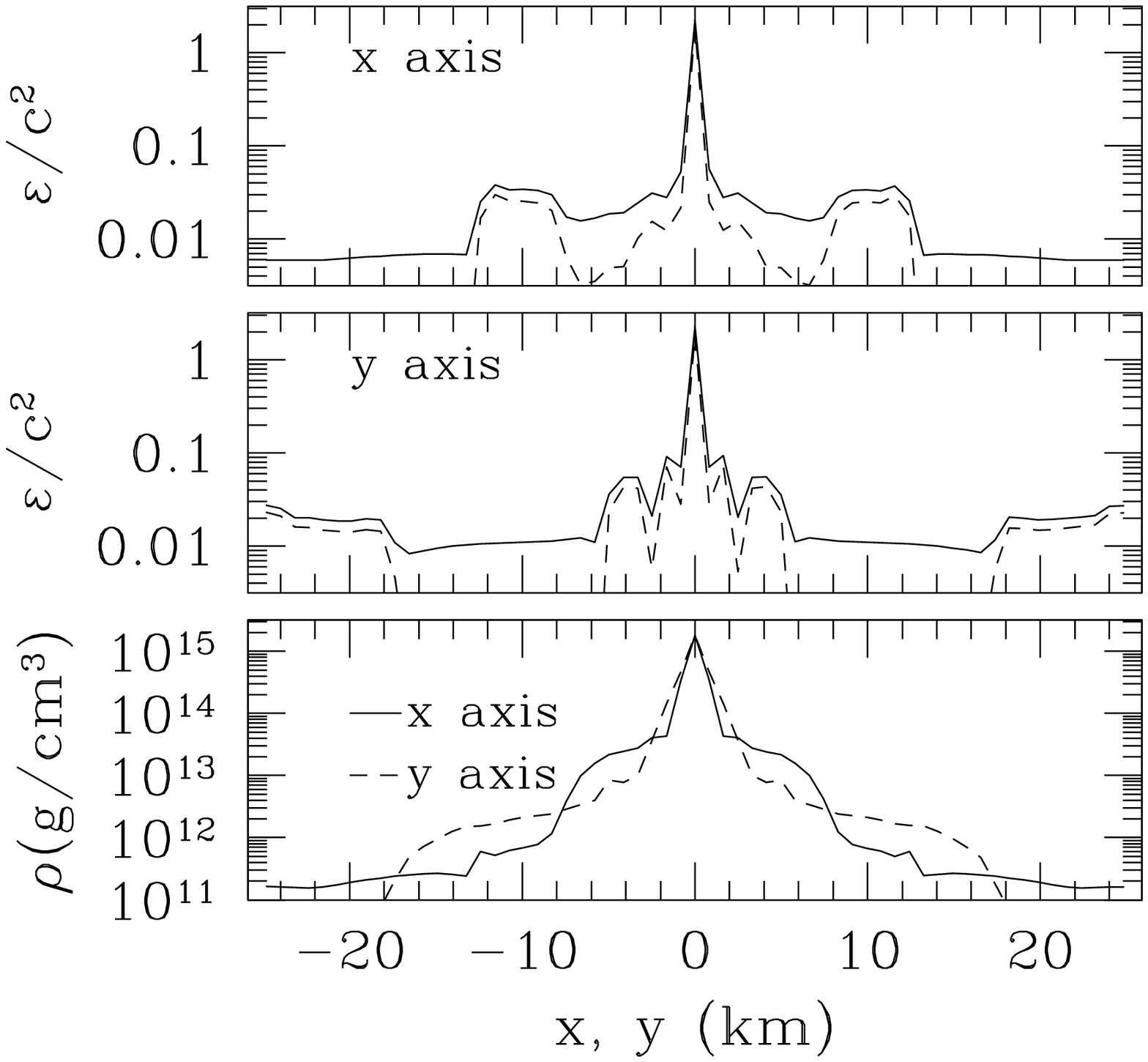}
  ~~~(b)\includegraphics[width=2.8in]{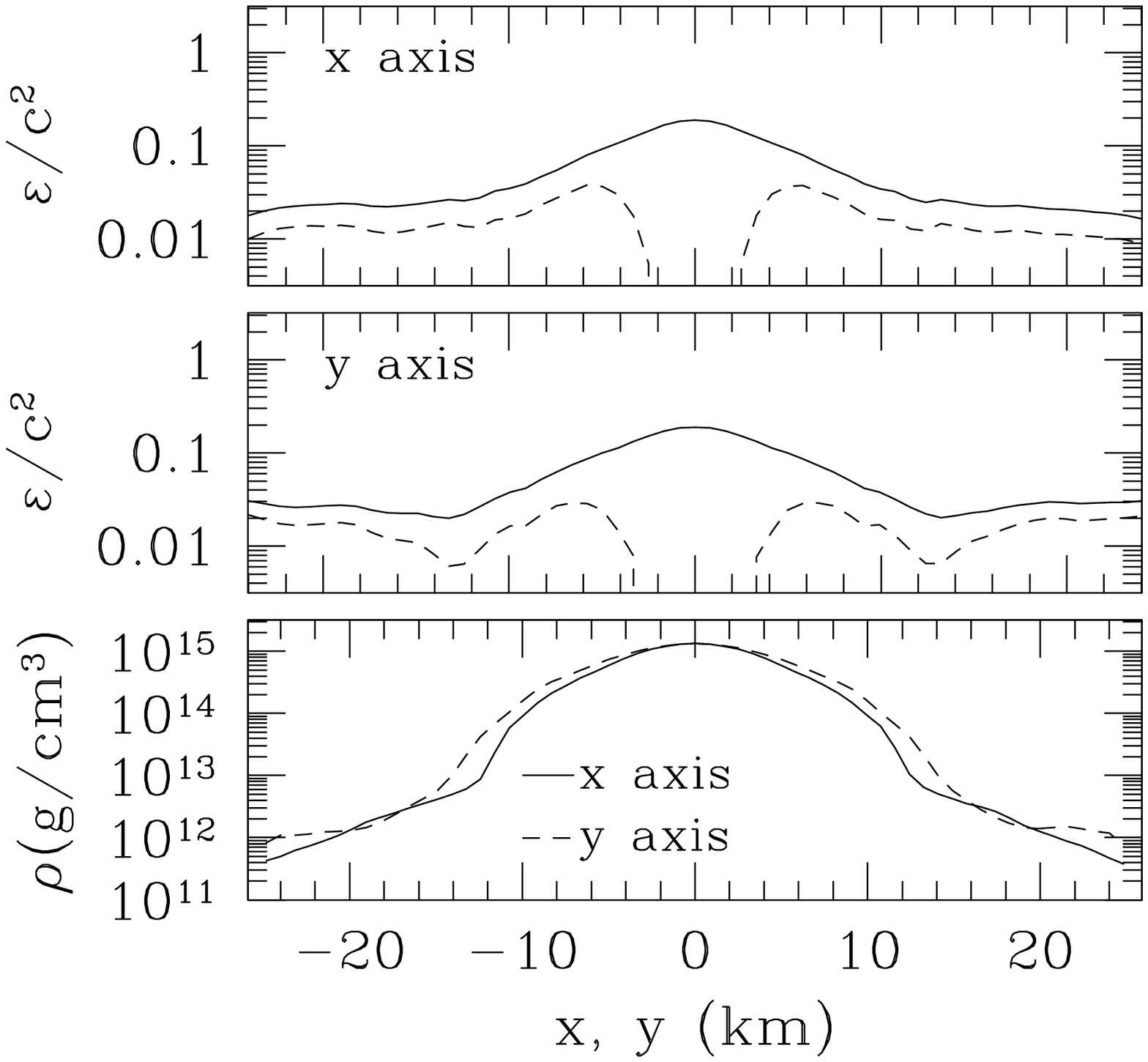} \\
\vspace{-4mm}
  (c)\includegraphics[width=2.8in]{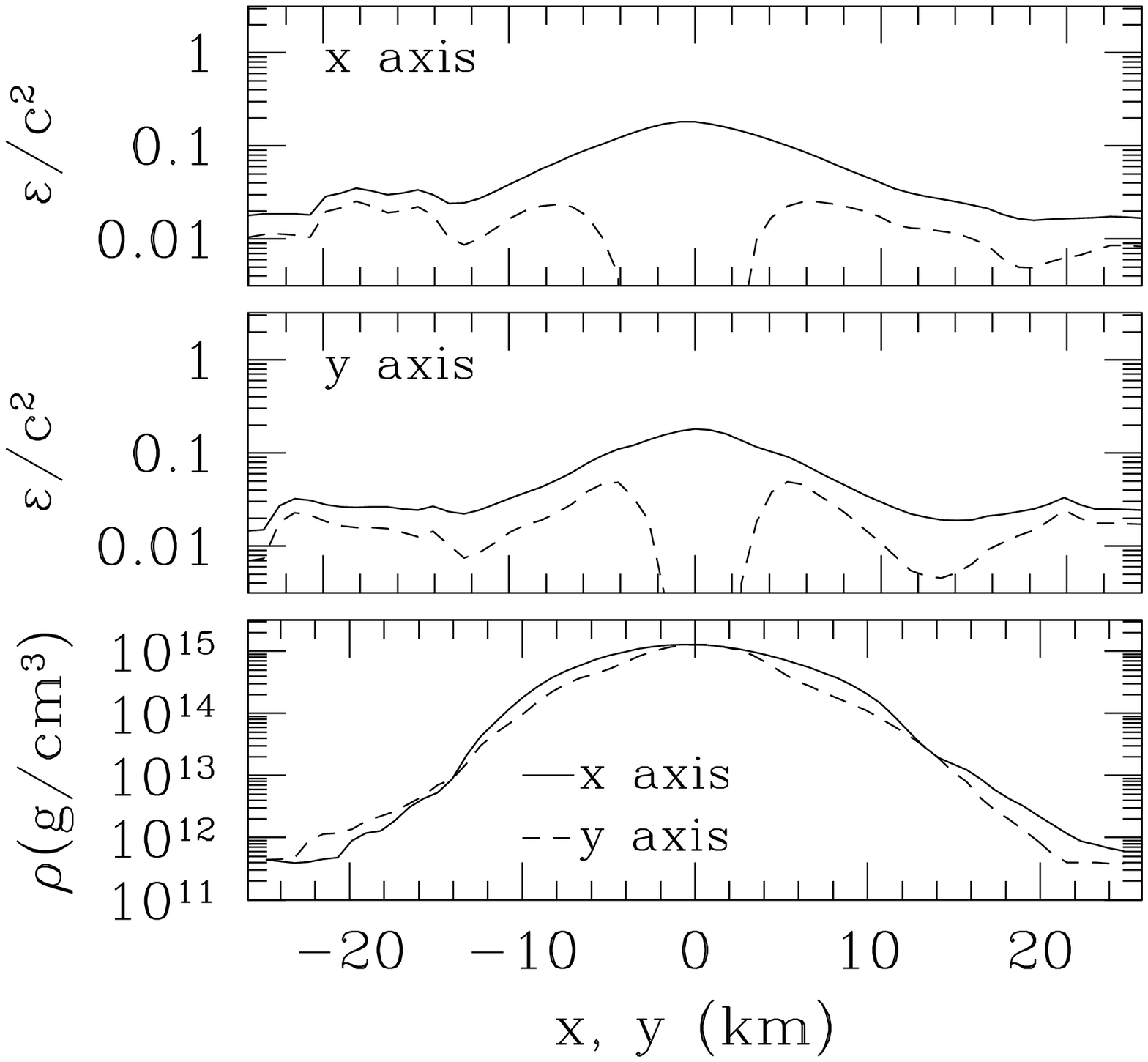}
  ~~~(d)\includegraphics[width=2.8in]{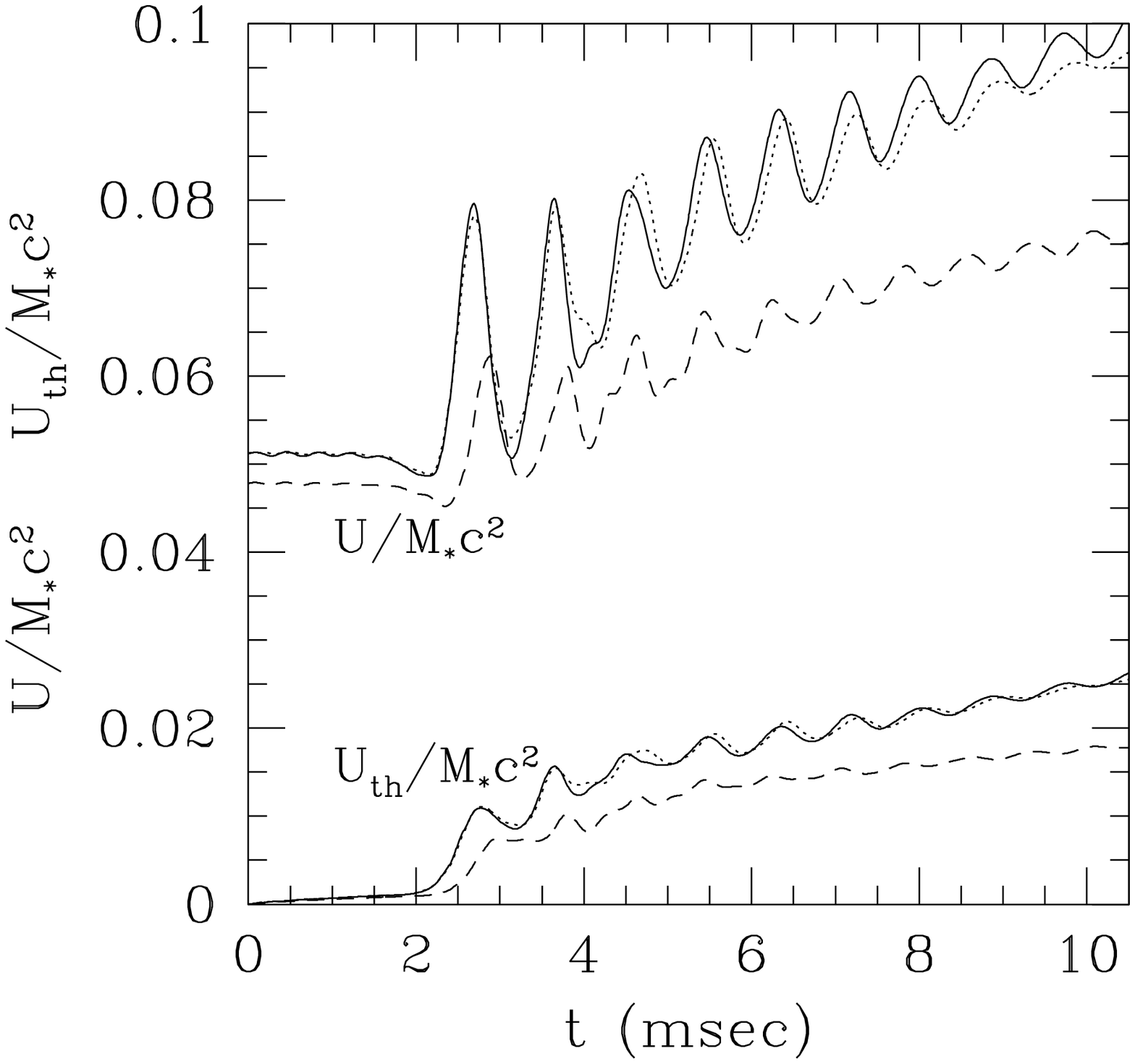}
\end{center}
\vspace{-2mm}
\caption{
Profiles of $\varep$ (solid curves) and
$\varep_{\rm th}=\varep-\varep_{\rm cold}$ (dashed curves) 
as well as that of $\rho$ along $x$ and $y$ axes
(a) at $t=2.991$ ms for model SLy1414a, 
(b) at $t=8.621$ ms for model SLy1313a, and 
(c) at $t=8.621$ ms for model SLy125135a.
Note that the region of $r \alt 2$ km for panel (a)
is inside the apparent horizon (see Figs. \ref{FIG3} and \ref{FIG6}(a)).
(d) Evolution of $U$ and $U_{\rm th}$ in unit of the rest-mass
energy $M_*c^2$ for models SLy1313a (solid curves), 
SLy125135a (dotted curves), and SLy1212b (dashed curves). 
\label{FIG10} }
\end{figure*}

First, we focus on the thermal property for models SLy1313a and SLy125135a 
which are the representative models of hypermassive neutron star formation. 
In these cases, the heating is not very important in the central 
region. This is reasonable because the shocks generated
at the collision of two stars are not very strong, and thus, the central
part of the hypermassive neutron stars 
is formed without experiencing the strong shock heating.
On the other hand, the shock heating plays an important role 
in the outer region of the hypermassive neutron stars and
in the surrounding accretion disk since 
the spiral arms hit the hypermassive neutron stars for many times. 

The typical value of $\varep_{\rm th}$ is $0.01c^2$--$0.02c^2$.
Here, we recover $c$ for making the unit clear. In the following, we 
assume that the components of the hypermassive neutron stars and surrounding 
disks are neutrons. Then, the value of $\varep_{\rm th}=0.01c^2$ 
implies that the thermal energy per nucleon is 
\beqn
9.37 \biggl({\varep_{\rm th} \over 0.01c^2}\biggr)~{\rm MeV/nucleon}.  
\eeqn
Since the typical value of $\varep_{\rm th}$ is $0.01c^2$--$0.02c^2$,
the typical thermal energy is 10--20 MeV. This value agrees approximately
with that computed in \cite{RJ,RJ2}.

Figure \ref{FIG10}(d) shows that the total internal energy and
thermal energy are relaxed to be 
\beqn
&&U \sim 0.1 M_* c^2 \sim 5 \times 10^{53}~{\rm erg},\\
&&U_{\rm th} \sim 0.025 M_* c^2 \sim 1 \times 10^{53}~{\rm erg},
\eeqn
for both models SLy1313a and SLy125135a.
Thus, the thermal energy increases up to $\sim 25\%$ of the total
internal energy. 
We note that these values are approximately identical between
models SLy1313a and SLy125135a. This implies that the mass ratio
of $Q_M \sim 0.9$ does not significantly modify the thermal properties
of the hypermassive neutron stars in the realistic equations of state. 

The region of $\varep_{\rm th} \sim 0.01c^2$ 
will be cooled via the emission of neutrinos \cite{RJ,RJ2}.
According to \cite{RJ,RJ2}, the emission rate in the
hypermassive neutron star with the averaged value of
$\varep_{\rm th} \sim 10$--30 MeV is $10^{52}$--$10^{53}$ erg/s.
Thus, if all the amounts of the thermal energy are assumed to be
dissipated by the
neutrino cooling, the time scale for the emission of the neutrinos
will be 1--10 s. This is much longer than the lifetime of the
hypermassive neutron stars $\alt 100$ ms. Therefore,
the cooling does not play an important role in their evolution. 

Since the lifetime of the hypermassive neutron stars $\alt 100$ ms
is nearly equal to the time duration of the short gamma-ray bursts 
\cite{GRB}, it is interesting to ask if 
they could generate the typical energy of the bursts. 
In a model for central engines of the gamma-ray bursts,
a fireball of the electron-positron pair and photon is produced
by the pair annihilation of the neutrino and antineutrino \cite{RJ2,GRB}.
In \cite{RJ2}, Janka and Ruffert estimate the efficiency of the
annihilation as several$\times 10^{-3}$ for the neutrino luminosity
$\sim 10^{52}$ erg/s, the mean energy of neutrino $\sim 10$ MeV,
and the radius of the hypermassive neutron star $\sim 10$ km (see Eq. (1)
of \cite{RJ2}). This suggests that the energy generation rate of the 
electron-positron pair is $\sim 10^{50}$ erg/s. Since the
lifetime of the hypermassive neutron stars is $\alt 100$ ms, 
the energy available for the fireball will be at most $\sim 10^{49}$ erg. 
This value is not large enough to explain typical
cosmological gamma-ray bursts. Furthermore, as Janka and Ruffert found
\cite{RJ2}, the pair annihilation of the neutrino and antineutrino is most
efficient in a region near the hypermassive neutron star, for which
the baryon density is large enough (cf. Fig. \ref{FIG6}(b))
to convert the energy of the
fireball to the kinetic energy of the baryon. Therefore, it is
not very likely that the hypermassive neutron stars are the central
engines of the typical short gamma-ray bursts. 


Now, we focus on model SLy1414a in which a black hole is promptly
formed after the merger. 
Comparing Fig. \ref{FIG10}(a) with the last panel of Fig. \ref{FIG3},
the region of high thermal energy is
located along the spiral arms of the accretion disk surrounding
the central black hole. 
(Note that the region of $r \alt 2$ km is inside the
apparent horizon, and hence, we do not consider such region.)
The part of the matter in the spiral arms with small orbital radius
$r \alt 5$ km is likely to be inside the radius of
an innermost stable circular orbit around the black hole, and
hence, be swallowed into the black hole. Otherwise, 
the matter in the spiral arms will form an accretion
disk surrounding the black hole. Thus, eventually a hot
accretion disk will be formed. 
However, the region of high thermal energy for $r \agt 10$ km
is of low density with $\rho \alt 10^{12}~{\rm g/cm^3}$, 
and the total mass of the disk will be 
$10^{-3} M_{\odot}\alt M_* \alt 0.01M_{\odot}$ (see Fig. \ref{FIG9}). 
The total thermal energy of the accretion disk is estimated as 
\beqn
U_{\rm th} &&\sim M_{\rm disk}\bar \varep_{\rm th} \nonumber \\
&&=1.8 \times 10^{50} 
\biggl({M_{\rm disk} \over 0.01M_{\odot}}\biggr)
\biggl({\bar \varep_{\rm th} \over 0.01c^2}\biggr) ~{\rm erg}. 
\eeqn
where $M_{\rm disk}$ and $\bar \varep_{\rm th}$ denote
the mass of the accretion disk and the averaged value of the 
specific thermal energy. Hence, even if all the amounts of
the thermal energy are dissipated by the emission of neutrinos, 
the total energy of the radiated neutrinos will be at most 
several $\times 10^{50}$ erg. According to \cite{RJ2}, 
the efficiency of the annihilation of the neutrino and antineutrino
is several $\times 10^{-5}$ for the neutrino luminosity $\sim 10^{50}$ erg/s,
the mean energy of neutrino $\sim 10$ MeV, and the disk radius $\sim 10$ km.  
This indicates that the energy of the fireball is at most $\sim 10^{46}$ erg.
Although the density of the baryon at the region that the
pair annihilation is likely to happen is small enough to avoid
the baryon loading problem, 
this energy is too small to explain cosmological gamma-ray bursts. 

\subsubsection{Effects of mass difference}\label{sec:massdif}

Comparing the evolution of the contour curves, the maximum density, and
the central value of the lapse function for models SLy1313a and SLy125135a
(see Figs. \ref{FIG4}, \ref{FIG5}, and \ref{FIG7}(a)), it is found that
the mass difference plays a minor role in the formation of a hypermassive
neutron star as far as $Q_M$ is in the range between 0.9 and 1. 
Figures \ref{FIG7}(a) and \ref{FIG9} also illustrate that the evolution of
the system to a black hole is very similar for models SLy1414a and
SLy135145a. In \cite{STU} in which simulations were 
performed using the $\Gamma=2$ equation of state, 
we found that the mass difference 
with $Q_M \sim 0.9$ significantly induces an asymmetry in
the merger which contributes to formation of large spiral arms and 
the outward angular momentum transfer, which are not 
very outstanding in the present results.
The reason seems to be as follows. In the previous equation of state,
the mass difference with $Q_M \sim 0.9$ results in a relatively
large ($\sim 15\%$) difference of the compactness between two stars.
On the other hand, the difference in the compactness between two stars 
with the present equations of state is $\sim 10\%$ for $Q_M \sim 0.9$.
This is due to the fact that the stellar radius depends weakly on
the mass in the range $0.8M_{\odot} \alt M \alt 1.5M_{\odot}$
(see Fig. \ref{FIG2}). 
As a result of the smaller difference in the compactness,
the tidal effect from the more massive star to the companion
becomes smaller, and therefore, the asymmetry is suppressed.
To yield a system of a black hole and a massive disk,
smaller mass ratio with $Q_M < 0.9$ will be necessary in the
realistic equations of state. 

Another possible reason is that neutron stars in the realistic equations
of state are more compact than those in the $\Gamma=2$ 
equation of state. Due to this fact, at the merger, the system
is more compact, and hence, even in the formation of the asymmetric 
spiral arms, they cannot spread outward extensively but wind 
around the formed neutron star quickly. 
Consequently, the mass of the disk around the central object
is suppressed to be small and also the asymmetric density configuration
does not become very outstanding.

\subsubsection{Dependence of dynamics on the grid size and
$\Gamma_{\rm th}$}\label{sec:gamma}

\begin{figure}[thb]
\vspace{-4mm}
\begin{center}
  \includegraphics[width=3.2in]{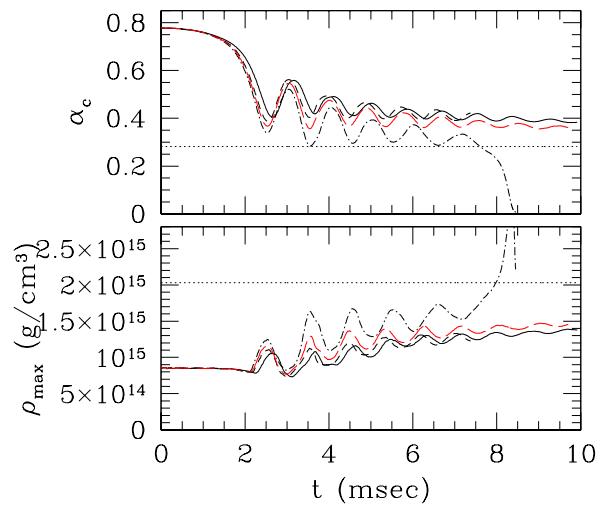}
\end{center}
\vspace{-8mm}
\caption{The same as Fig. \ref{FIG7} but 
for models SLy1313a (solid curves), SLy1313b (dashed curves),
SLy1313c (dotted dashed curves), and SLy1313d (long dashed curves).
Note that the simulation for SLy1313b was stopped at 
$t \sim 7.5$ ms to save computational time. 
\label{FIG11}
}
\end{figure}

For model SLy1313, we performed simulations changing the value of
$\Gamma_{\rm th}$ with the grid size (377, 377, 189).
In Fig. \ref{FIG11}, evolution of $\alpha_c$ and $\rho_{\rm max}$ 
is shown for models SLy1313b--SLy1313d. 
Note that the grid size and grid spacing are identical
for these models. The results for model SLy1313a are shown together
for comparison with those for model SLy1313b for which 
the parameters are identical but for the grid size. 

By comparing the results for models SLy1313a and SLy1313b, 
the magnitude of the error associated with the small size of $L$
is investigated. Figure \ref{FIG11} shows that two results 
are approximately identical besides a systematic phase shift
of the oscillation. This shift is caused by the inappropriate
computation of the radiation reaction in the late inspiral stage
for $t \alt 2$ ms: For model SLy1313b, 
$L$ is smaller, and hence, the radiation reaction in the
inspiral stage is significantly overestimated to spuriously
accelerate the merger resulting in the phase shift. 
However, besides the phase shift, the results are approximately 
identical. In particular, the results agree well in 
the merging phase. This indicates that even with the smaller grid size
(377, 377, 189), the formation and evolution of the hypermassive neutron
star can be followed within a small error. 

Comparison of the results among models SLy1313b--SLy1313d tells 
that for the smaller value of $\Gamma_{\rm th}$,
the maximum density (central lapse) of a hypermassive neutron star
formed during the merger is larger (smaller). 
This is due to the fact that the strength of the shock formed at
the collision of two stars, which provides the thermal energy
in the outer region of the formed hypermassive neutron stars to expand,
is proportional to the value of $\Gamma_{\rm th}-1$. 
Figure \ref{FIG11} also indicates that the evolution of the system
does not depend strongly on the value of $\Gamma_{\rm th}$ for 
$\Gamma_{\rm th} \agt 1.65$. However, 
for $\Gamma_{\rm th}=1.3$, the formed hypermassive
neutron star is very compact at its birth, and hence, 
collapses to a black hole in a short time scale (at $t \sim 8$ ms) 
after the angular momentum is dissipated by gravitational radiation.
This time scale for black hole formation is much shorter than that
for models SLy1313b and SLy1313d for which it would be $\sim 30$ ms.
This indicates that for small values of $\Gamma_{\rm th}-1 \alt 0.3$,
the collapse to a black hole is significantly enhanced. 

In reality, in the presence of cooling processes such as neutrino cooling, 
the adiabatic index $\Gamma_{\rm th}$ decreases effectively. 
Such cooling mechanism may accelerate the formation of a black hole. 
However, the emission time scale of the neutrino is $\sim 1$--10 s as 
mentioned in Sec. \ref{sec:therm}. Thus, the effect does not seem to be 
very strong. 

\subsection{Gravitational waveforms}\label{sec:gw}

\subsubsection{Waveforms in the formation of hypermassive neutron stars}

\begin{figure*}[thb]
\vspace{-4mm}
\begin{center}
  (a)\includegraphics[width=3.in]{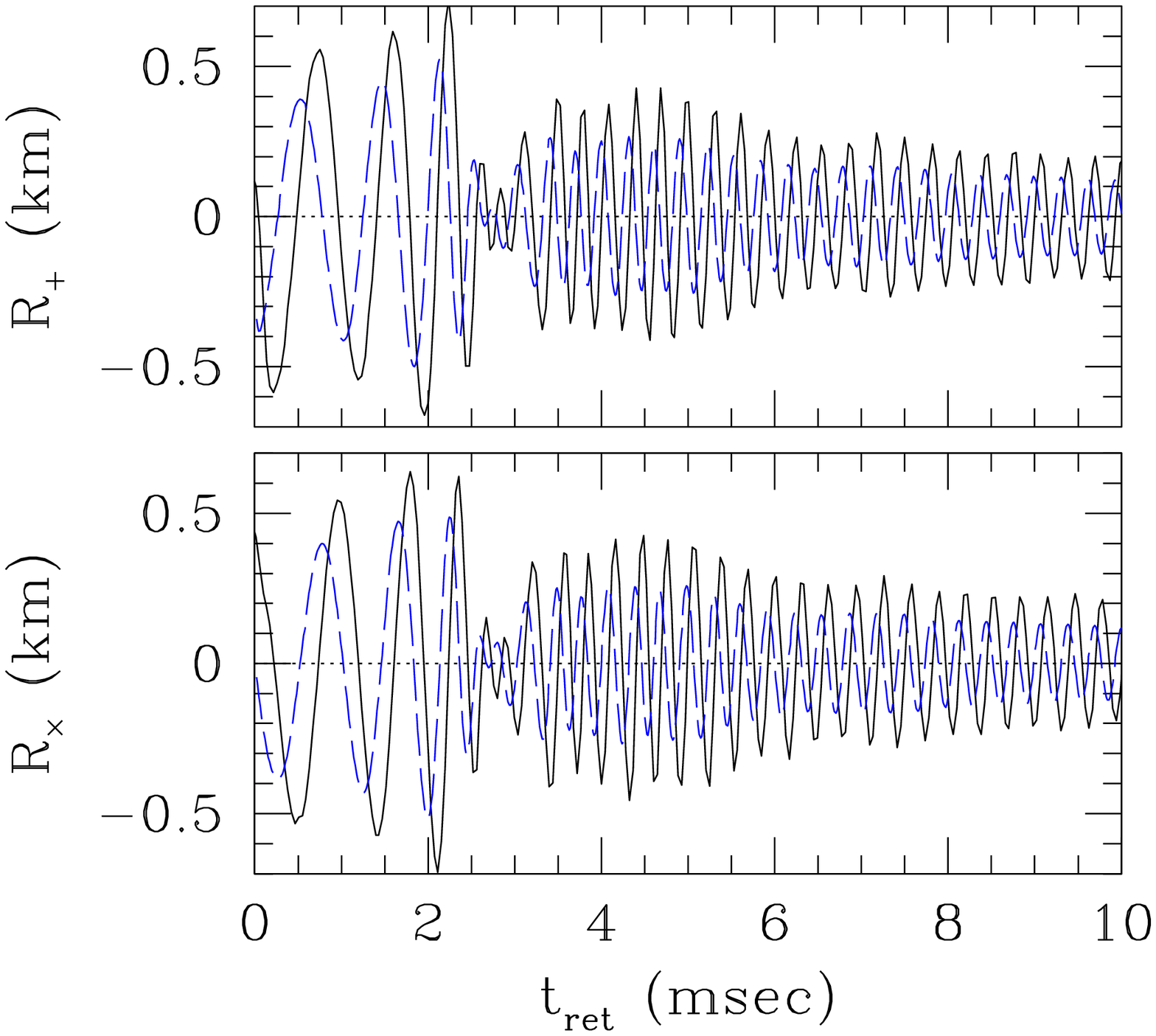}
  ~~~(b)\includegraphics[width=3.in]{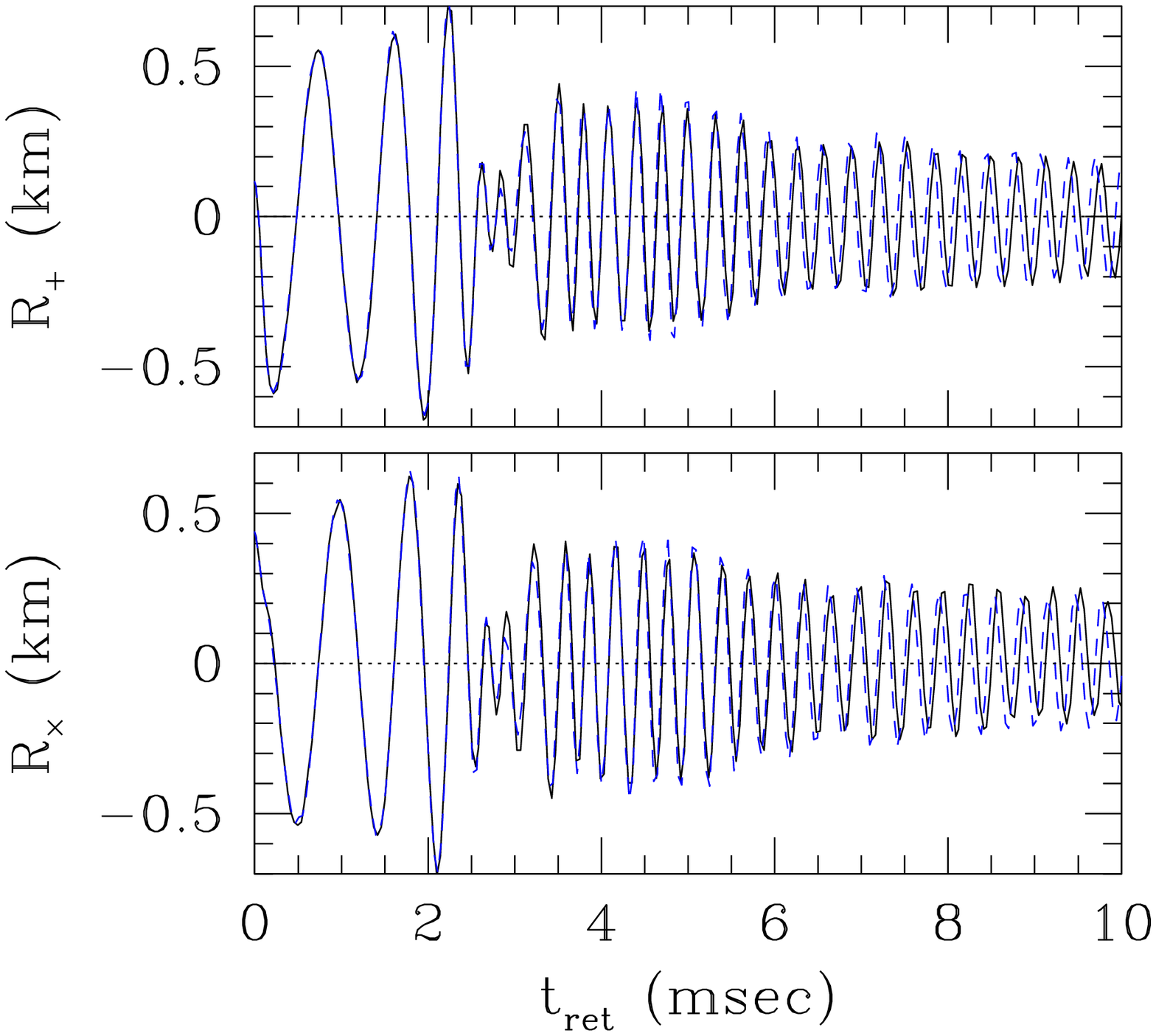}
\end{center}
\vspace{-3mm}
\caption{(a) Gravitational waves for model SLy1313a. 
$R_{+}$ and $R_{\times}$ (solid curves) and 
$A_{+}$ and $A_{\times}$ (dashed curves) 
as functions of the retarded time are shown.
(b) $R_{+}$ and $R_{\times}$ 
as functions of the retarded time for model SLy125135a (solid curves). 
For comparison, those for SLy1313a are shown by the dashed curves. 
\label{FIG12}
}
\end{figure*}

\begin{figure}[thb]
\vspace{-4mm}
\begin{center}
  \includegraphics[width=3.in]{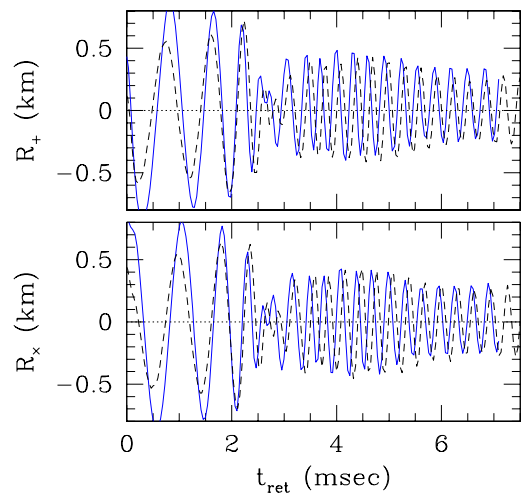}
\end{center}
\vspace{-8mm}
\caption{$R_{+}$ and $R_{\times}$ 
for models SLy1313a (dashed curves) and SLy1313b (solid curves) 
as functions of the retarded time. 
It is found that gravitational waveforms in the merger stage 
agree well. Note that the simulation for SLy1313b was stopped at
$t \sim 7.5$ ms to save computational time. 
\label{FIG13}
}
\end{figure}

\begin{figure}[thb]
\vspace{-4mm}
\begin{center}
  \includegraphics[width=3.in]{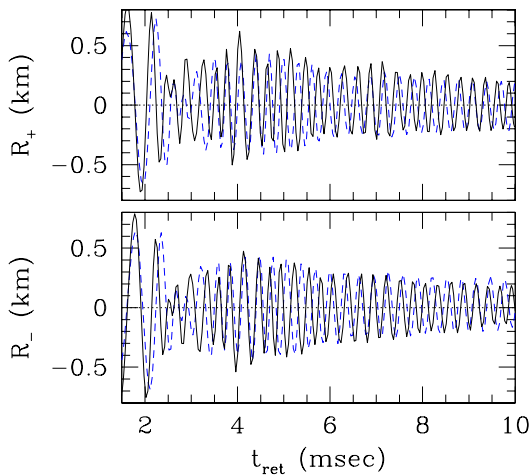}
\end{center}
\vspace{-8mm}
\caption{$R_{+}$ and $R_{\times}$ 
for models SLy135135b (solid curves) and SLy1313a (dashed curves) 
as functions of the retarded time. 
\label{FIG14}
}
\end{figure}

\begin{figure*}[thb]
\vspace{-4mm}
\begin{center}
  (a)\includegraphics[width=3.in]{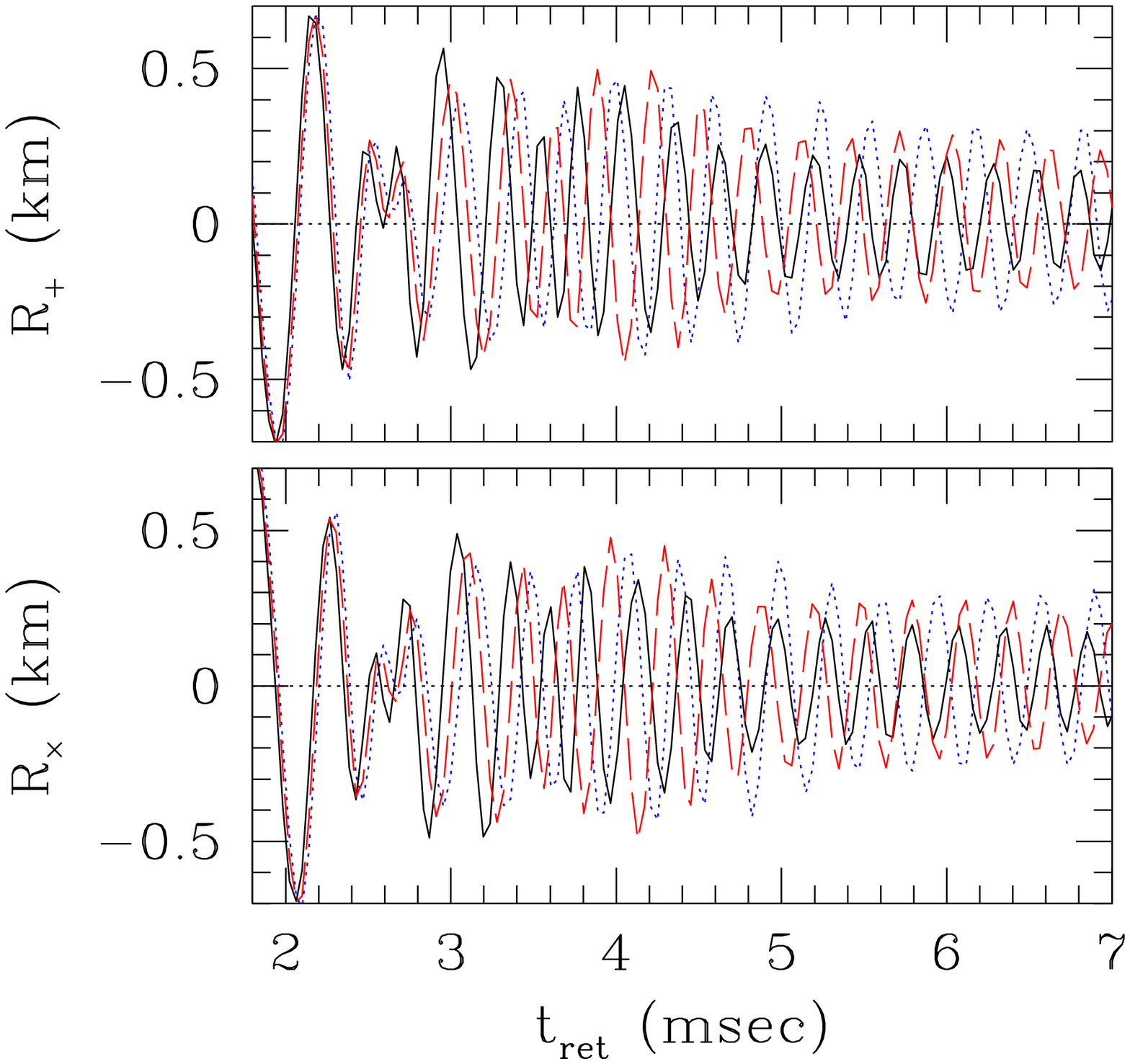}
  ~~~(b)\includegraphics[width=3.in]{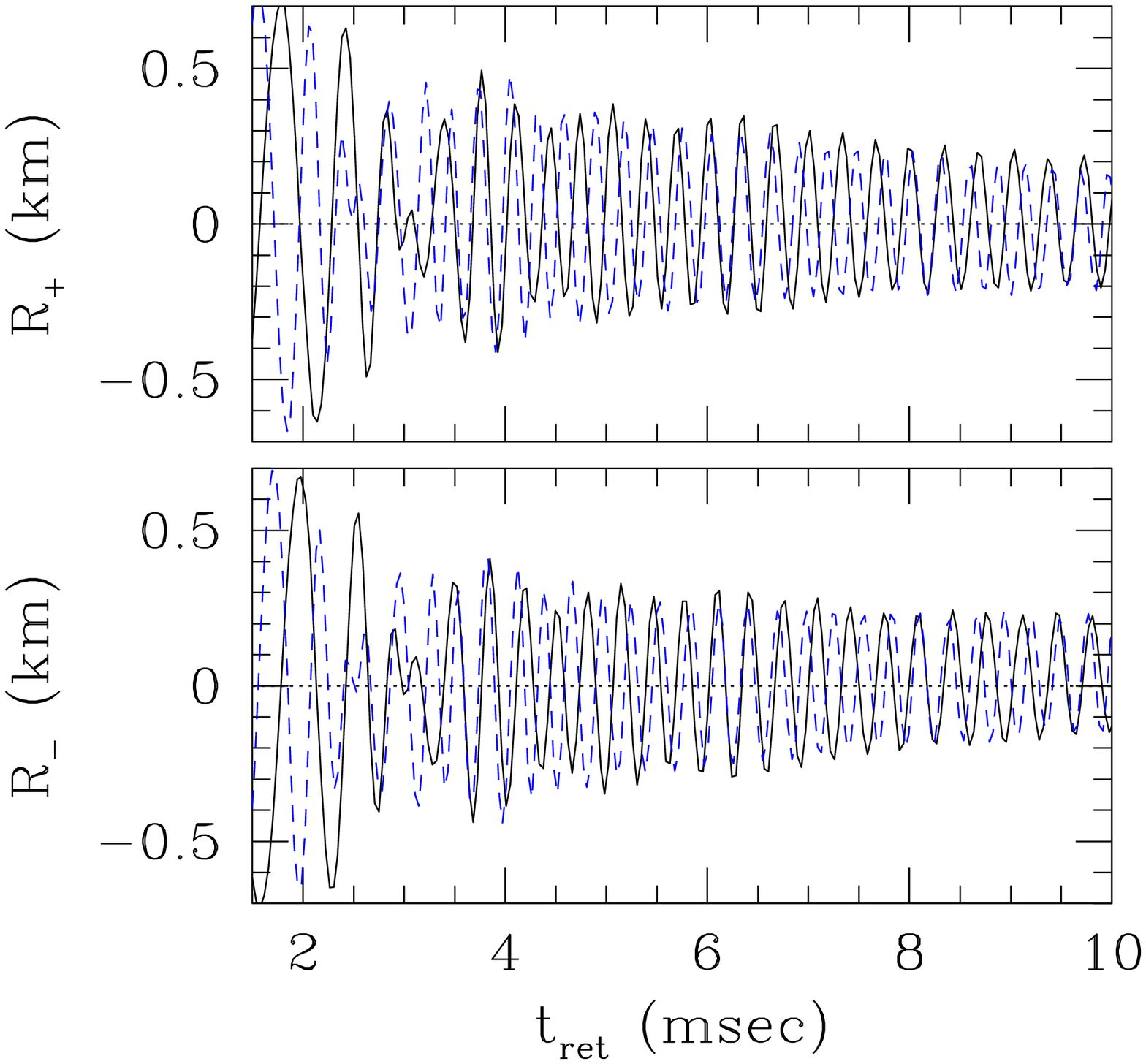}
\end{center}
\vspace{-2mm}
\caption{$R_{+}$ and $R_{\times}$ 
(a) for models SLy1313b (dotted curves), SLy1313c (solid curves),
and SLy1313d (dashed curves), and 
(b) for models SLy1212b (solid curves) 
and FPS1212b (dashed curves) as functions of the retarded time.
\label{FIG15}
}
\end{figure*}

In Figs. \ref{FIG12}--\ref{FIG15}, we present gravitational waveforms
in the formation of hypermassive neutron stars 
for several models. Figure \ref{FIG12}(a) shows $R_+$ and $R_{\times}$
for model SLy1313a. For comparison, gravitational waves computed in
terms of a quadrupole formula ($A_+$ and $A_{\times}$)
defined in Sec. II B are shown together
by the dashed curves. The amplitude of gravitational waves, $h$, observed 
at a distance of $r$ along the optimistic direction ($\theta=0$)
is written as 
\beqn
h \approx 10^{-22} \biggl( {R_{+,\times} \over 0.31~{\rm km}}\biggr)
\biggl({100~{\rm Mpc} \over r}\biggr). \label{hamp}
\eeqn
Thus, the maximum amplitude observed along the most optimistic
direction is $\sim 2 \times 10^{-22}$ at a distance of 100 Mpc.

In the real data analysis of gravitational waves, 
a matched filtering technique \cite{KIP} is employed.
In this method, the signal of the identical frequency can
be accumulated using appropriate templates, and 
as a result, the effective amplitude increases by a factor of $N^{1/2}$
where $N$ denotes the number of the cycle of gravitational
waves for a given frequency.
We determine such effective amplitude in Sec. \ref{sec:fourier}
(cf. Eq. (\ref{heff})).

The waveforms shown in Fig. \ref{FIG12}(a) are typical ones in the
formation of a hypermassive neutron star.  In the early phase
($t_{\rm ret} \alt 2$ ms), gravitational waves associated with the
inspiral motion are emitted, while for $t_{\rm ret} \agt 2$ ms,
those by the rotating and oscillating hypermassive neutron star are
emitted. In the following, we focus only on the waveforms for
$t_{\rm ret} \agt 2$ ms. 

Gravitational waves from the hypermassive neutron stars are
characterized by quasiperiodic waves for which the amplitude 
and the frequency decrease slowly. 
The amplitude decreases with the ellipticity, which is decreased by 
the effects that the angular momentum decreases due to the radiation
reaction and is transferred from the inner region to the outer one
by the hydrodynamic interaction associated with the nonaxisymmetric structure. 
However, the time scale for the decrease appears to be much longer 
than $10$ ms as illustrated in Figs. \ref{FIG12}--\ref{FIG15}. 
The oscillation frequency varies even more slowly. The reason seems to
be that the following two effects approximately cancel each other:
(i) with the decrease of the angular momentum of the hypermassive
neutron stars due to the radiation reaction as well as the angular momentum
transfer by the hydrodynamic interaction with outer envelope, 
the characteristic frequency of the figure rotation decreases, 
while (ii) with the decrease of the angular momentum, the centrifugal
force is weakened to reduce the characteristic radius for a spin-up. 
(We note that the radiation reaction alone may increase the
frequency of the figure rotation \cite{LS}. In the hypermassive neutron
stars formed after the merger, the angular momentum transfer due to
the hydrodynamic interaction is likely to 
play an important role for the decrease of the frequency.)

In gravitational waveforms computed in terms of the quadrupole formula
(the dashed curves in Fig. \ref{FIG12}), 
the amplitude is systematically underestimated by a factor
of 30--40\%. This value is nearly equal to the magnitude of the compactness
of the hypermassive neutron star, $GM/Rc^2$, where $M$ and $R$ denote
the characteristic mass and radius. Since the quadrupole formula
is derived ignoring the terms of order $GM/Rc^2$, this magnitude for
the error is quite reasonable. In simulations with 
Newtonian, post Newtonian, and approximately relativistic frameworks, 
gravitational waves are computed in the quadrupole formula (e.g.,
\cite{C,FR,OUPT}). 
The results here indicate that the amplitudes for quasiperiodic
gravitational waves from hypermassive neutron stars presented
in those simulations are significantly underestimated 
\footnote{Besides systematic underestimation of the wave amplitude,
rather quick damping of quasiperiodic gravitational waves is seen 
in the results of these references. This quick damping also seems to be
due to a systematic error, which is likely to result from a relatively 
dissipative numerical method (SPH method) used in these references.}. 
Although the error in the amplitude is large, 
the wave phase is computed accurately except for a slight 
systematic phase shift. From the point of view of the data 
analysis, the wave phase is the most important information 
on gravitational waves. This suggests that a quadrupole formula 
may be a useful tool for computing approximate gravitational waves. 
We note that these features have been already found for
oscillating neutron stars \cite{SS1} and for nonaxisymmetric
protoneutron stars formed after stellar core collapse \cite{SS3}. 
Here, we reconfirm that the same feature holds for the merger of binary
neutron stars. 

In Fig. \ref{FIG12}(b), we display gravitational waveforms for
model SLy125135a. For comparison, those for SLy1313a are shown
together (dashed curves). It is found that two waveforms coincide
each other very well. As mentioned in Sec. \ref{sec:massdif}, the
mass difference with $Q_M \sim 0.9$ does not induce any outstanding
change in the merger dynamics. This fact is also reflected in the 
gravitational waveforms. 

In Fig. \ref{FIG13}, we compare gravitational waveforms for models 
SLy1313b (solid curves) and SLy1313a (dashed curves).  For SLy1313b,
the simulation was performed with a smaller grid size and 
gravitational waves were extracted in a near zone with $L/\lambda_0 
\approx 0.25$ and $L/\lambda_{\rm merger}\approx 0.83$
(cf. for model SLy1313a, $L/\lambda_0 \approx 0.41$ and
$L/\lambda_{\rm merger} \approx 1.39$). This implies that 
the waveforms for model SLy1313b are less accurately computed than 
those for SLy1313a.  Indeed, the wave amplitude for $t_{\rm ret} \alt 2$ ms is 
badly overestimated.  However, the waveforms from the formed 
hypermassive neutron stars for two models agree very well except 
for a systematic phase shift, which is caused by the overestimation for 
the radiation reaction in the early phase ($t_{\rm ret} \alt 2$ ms).
Thus, for computation of gravitational waves from 
the hypermassive neutron stars, we may choose the small grid size.
Making use of this fact, 
we compare gravitational waveforms among several models computed
with the small grid size in the following. 

In Fig. \ref{FIG14}, we compare gravitational waves from the
hypermassive neutron stars for models SLy1313a (dashed curves) and
SLy135135b (solid curves). As shown in Figs. \ref{FIG12} and
\ref{FIG13}, quasiperiodic waves for which the frequency is
approximately constant are emitted for model SLy1313a.  On the other
hand, the frequency is not constant but modulates with time for model
SLy135135b (e.g., see the waveforms at $t_{\rm ret}\approx 3.6$, 4.6,
5.6, and 6.6 ms for which the wavelength is relatively short). 
The reason is that the formed hypermassive 
neutron star quasiradially oscillates with a large amplitude and the 
frequency of gravitational waves varies with the change of the
characteristic radius. Due to this, the Fourier spectra for models
SLy1313 and SLy135135 are significantly different although the
difference of the total mass is very small (cf. Fig. \ref{FIG17}(b)).

In Fig. \ref{FIG15}(a), we compare gravitational waveforms for
models SLy1313b (dotted curves), SLy1313c (solid curves), and
SLy1313d (dashed curves). For these models,
the cold part of the equation of state is identical
but the value of $\Gamma_{\rm th}$ is different.
As mentioned in Sec. \ref{sec:gamma}, 
with the smaller values of $\Gamma_{\rm th}$, the shock heating is
less efficient, and as a result, the formed hypermassive neutron
star becomes more compact. Since the characteristic radius
decreases, the amplitude of gravitational waves decreases and
the frequency increases. 
This shows that the strength of the shock heating affects the
amplitude and the characteristic frequency of gravitational waves. 

In Fig. \ref{FIG15}(b), we compare gravitational waveforms for
models SLy1212b (solid curves) and FPS1212b (dashed curves).
For these models, the equations of state 
are different, but the total ADM mass is 
approximately identical. Since the FPS equation of state
is slightly softer than the SLy one, the compactness of
each neutron star is larger by a factor of 5--10\% (cf. Fig. \ref{FIG1}) 
and so is for the formed hypermassive neutron star. 
As a result, the frequency of gravitational waves for
the FPS equation of state is slightly ($\sim 15\%$) higher
(cf. Fig. \ref{FIG17}(d)). On the other hand, the 
amplitude of gravitational waves is not very different.
This is due to the fact that with increasing the compactness,
the radius of the hypermassive neutron star decreases while
the angular velocity increases. 
These two effects approximately cancel each other, and as a result,
dependence of the amplitude is not remarkable between two models.

\subsubsection{Emission rate of the energy and the angular momentum}
\label{sec:dedt}

\begin{figure*}[thb]
\vspace{-4mm}
\begin{center}
  (a)\includegraphics[width=3.in]{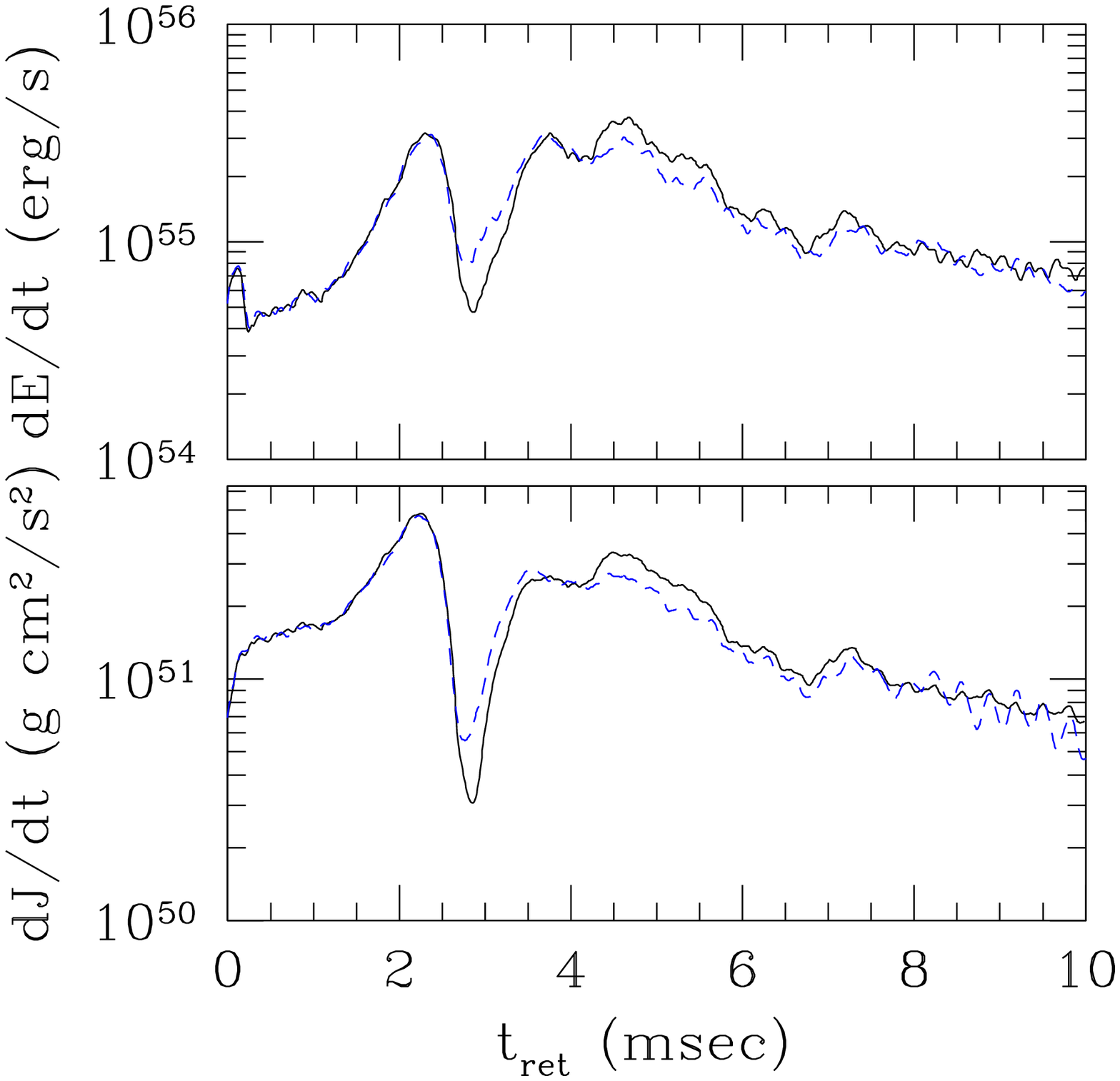}
  ~~~(b)\includegraphics[width=3.in]{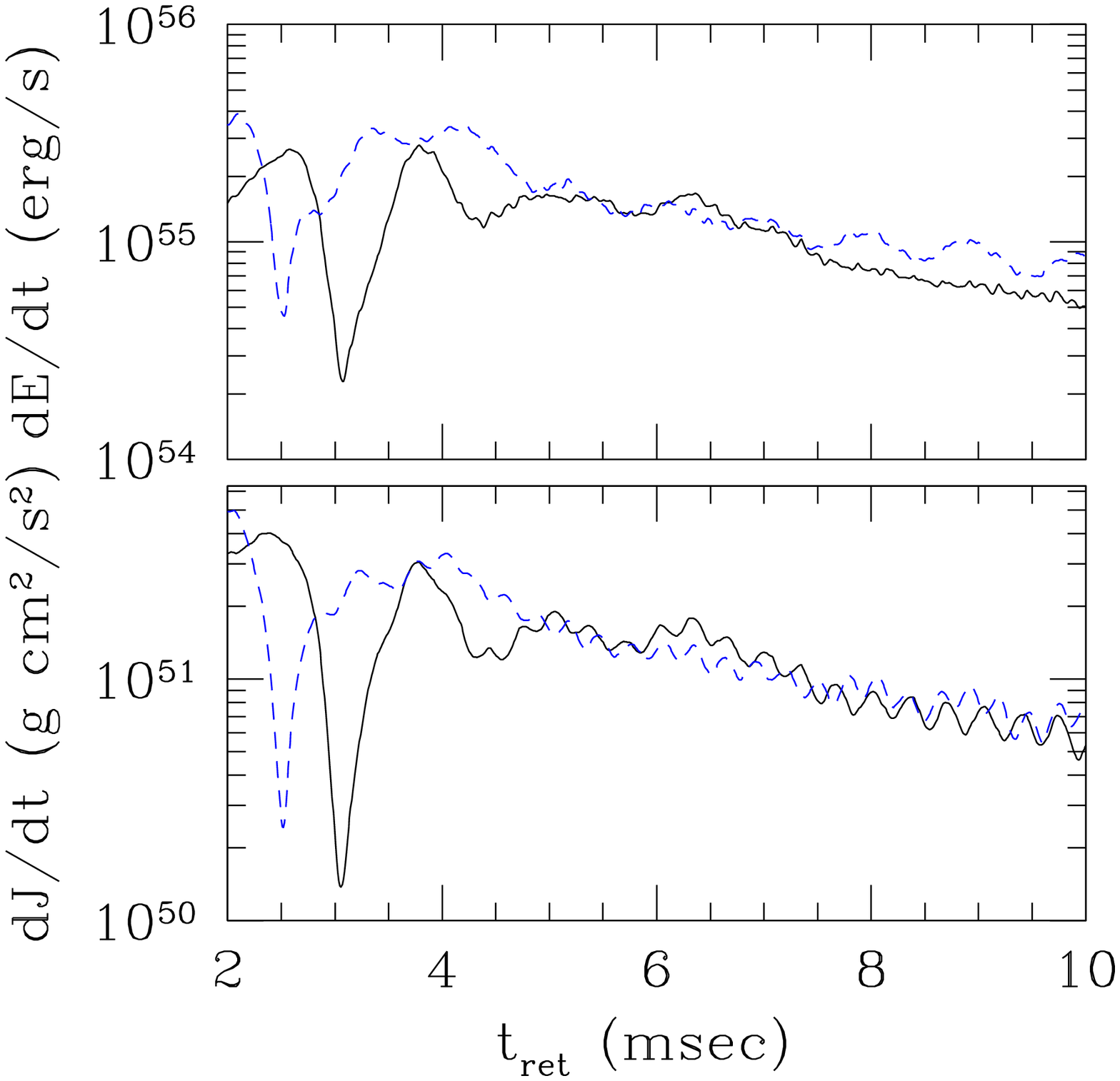}
\end{center}
\vspace{-3mm}
\caption{$dE/dt$ and $dJ/dt$ of gravitational radiation 
(a) for models SLy1313a (solid curves) and SLy125135a (dashed curves), and  
(b) for models SLy1212b (solid curves) and FPS1212b (dashed curves). 
\label{FIG16}
}
\end{figure*}

In Fig. \ref{FIG16}(a), the emission rates of the energy and the angular 
momentum by gravitational radiation are shown
for models SLy1313a (solid curves) and SLy125135a (dashed curves). 
In the inspiral phase for $t_{\rm ret} \alt 2$ ms, 
they increase with time since the amplitude and the frequency of the
chirp signal increase. After the peak is reached,
the emission rates quickly decrease by about one order of magnitude since
the merged object becomes a fairly axisymmetric transient object.
However, because of its large angular momentum, 
the formed hypermassive neutron star soon changes to a highly
ellipsoidal object which emits 
gravitational waves significantly. The luminosity from the
ellipsoidal neutron star is as high as the first
peak at $t_{\rm ret} \sim 2.2$ ms.
This is in contrast with the results obtained with the $\Gamma=2$ 
equation of state in which the magnitude of the 
second peak is 30--50\% as large as that of the first peak \cite{bina2}.
This reflects the fact that the degree of the
ellipticity of the formed hypermassive neutron star is much 
higher than that found in \cite{bina2} because of the large 
adiabatic index for the realistic equations of state. 

The emission rates of the energy and the angular momentum via
gravitational waves gradually decrease with time, since the degree
of the nonaxial symmetry decreases. However, the decrease rates are
not very large and the emission rates at $t_{\rm ret} \sim 10$ 
ms remain to be as high as that in the late inspiral phase as 
$dE/dt \sim 7 \times 10^{54}$ erg/s and
$dJ/dt \sim 7 \times 10^{50}~{\rm g~cm^2/s^2}$. 
The angular momentum at
$t \sim 10$ ms is $J \sim 0.7J_0 \sim 4 \times 10^{49}~{\rm g~cm^2/s}$.
Assuming that the emission rate of the angular momentum does not
change and remains $\sim 7 \times 10^{50}~{\rm g~cm^2/s}$, 
the emission time scale is evaluated as $J/(dJ/dt) \sim 50$ ms.
For more accurate estimation, we should compute
$(J-J_{\rm min})/(dJ/dt)$ where $J_{\rm min}$ denotes the minimum
allowed angular momentum for sustaining the hypermassive neutron star.
Since $J_{\rm min}$ is not clear, we set $J_{\rm min}=0$. 
Thus, the estimated value presented here is an approximate upper limit for the
emission time scale (see discussion below), and hence, the hypermassive
neutron star will collapse to a black hole within 50 ms. 
This estimate agrees with the value $\sim 30$ ms
obtained in terms of the change rate of $\alpha_c$ (cf. Sec. \ref{sec:gen}). 
Therefore, we conclude that the lifetime of the hypermassive neutron stars
and hence the time duration of the emission of quasiperiodic
gravitational waves are as short as $\sim 30$--50 ms
for models SLy1313a and SLy125135a. 

Figure \ref{FIG16}(b) displays
the emission rates of the energy and the angular 
momentum for models SLy1212b (solid curves) and FPS1212b (dashed curves).
For these models,
the value of $L$ is not large enough to accurately compute gravitational
waves in the inspiral phase for $t_{\rm ret} \alt 2$ ms.
Thus, we only present the results for the merger phase. 
The emission rates for SLy1212b are slightly 
smaller than those for model SLy1313a. 
This results from the fact that the total mass of the
system is smaller. On the other hand, 
the emission rates for FPS1212b is slightly larger than that for
SLy1212b. The reason is that the FPS equation of state is softer than 
the SLy one, and as a result, the formed hypermassive neutron star is
more compact and the rotational angular velocity is larger.
The hypermassive neutron star formed for FPS1212b collapses to a 
black hole at $t \sim 10$ ms. This is induced by the emission 
of the angular momentum by gravitational waves.
However, the collapse time is shorter than the emission time scale
evaluated by $J/(dJ/dt)$. This is reasonable because the
hypermassive neutron star formed for model FPS1212 is close to the
marginally stable configuration, and hence, a small amount of the
dissipation leads to the collapse. This illustrates that the
time scale $J/(dJ/dt)$ should be regarded as
the approximate upper limit for the collapse time scale. 

\subsubsection{Fourier power spectrum}\label{sec:fourier}

\begin{figure*}[thb]
\vspace{-4mm}
\begin{center}
  (a)\includegraphics[width=2.8in]{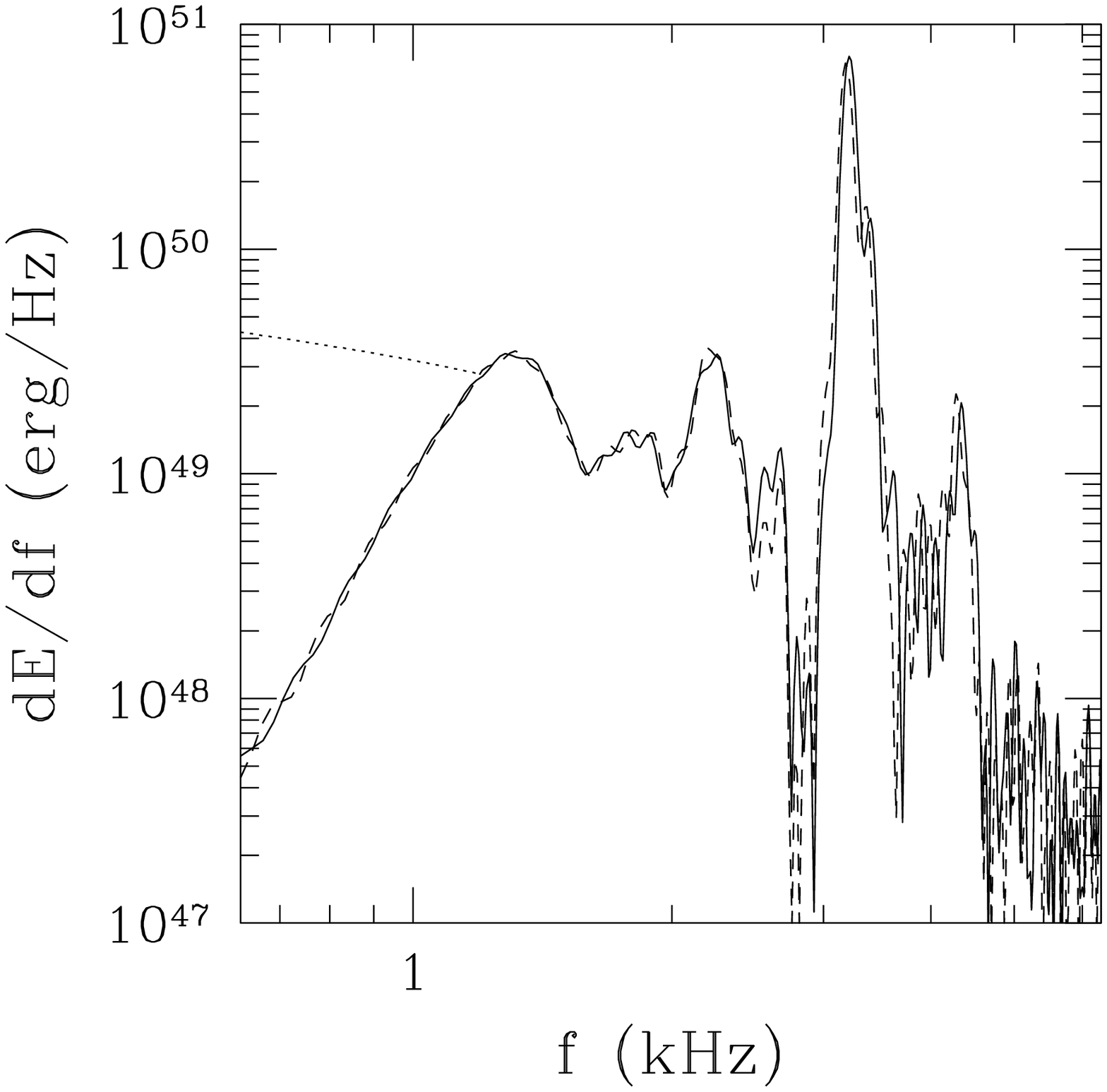}
  ~~~(b)\includegraphics[width=2.8in]{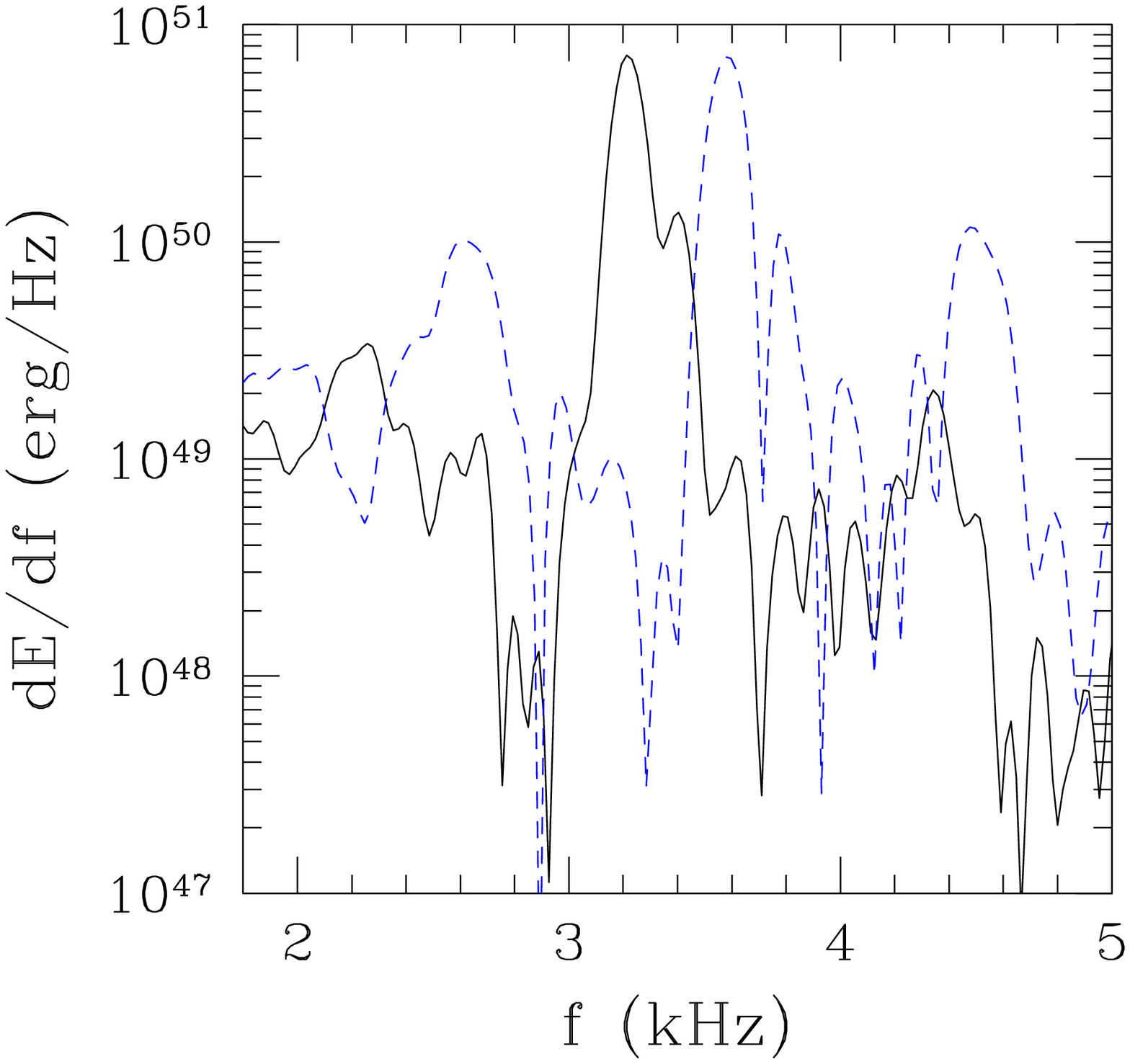}\\
\vspace{-4mm}
  (c)\includegraphics[width=2.8in]{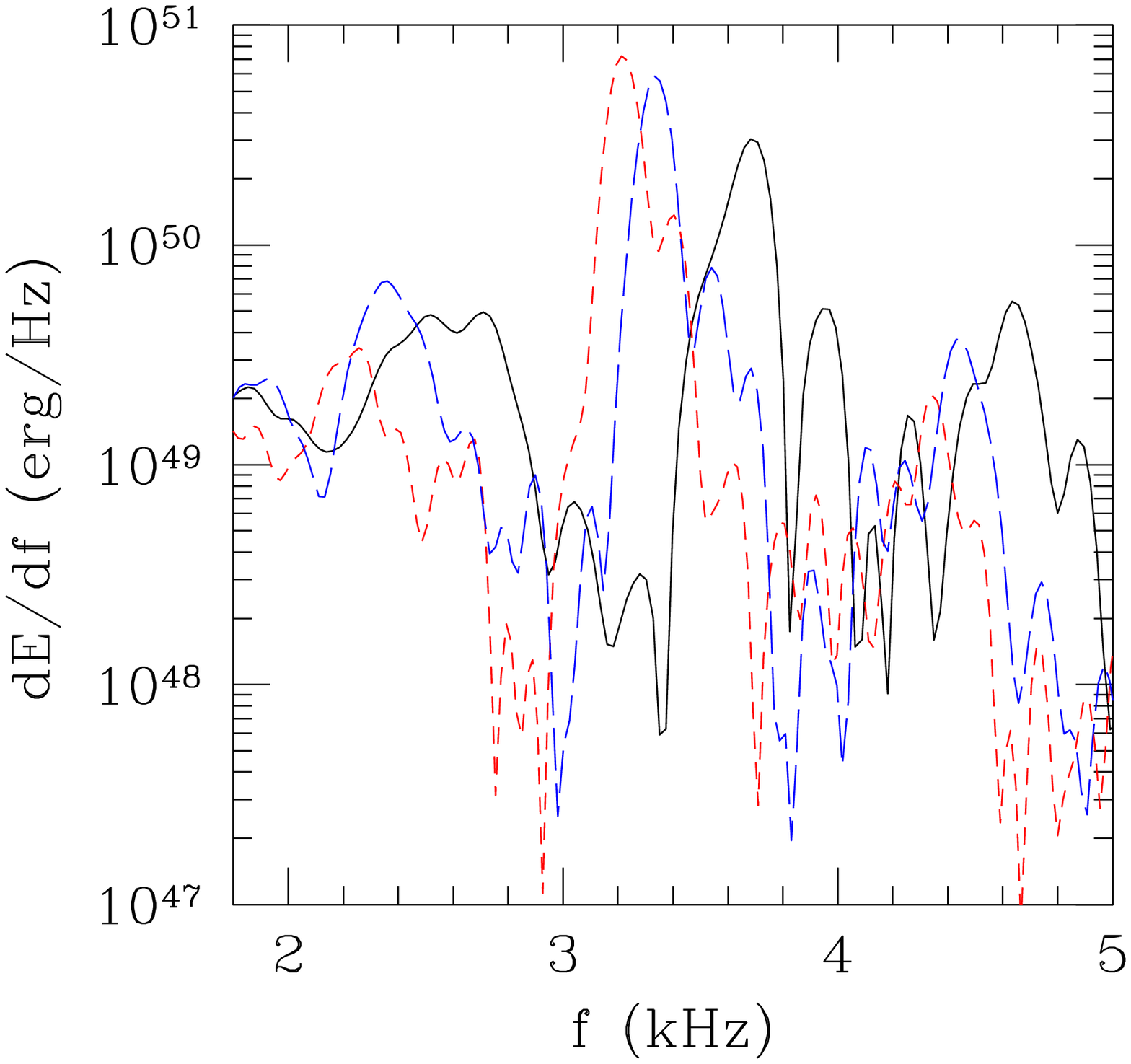}
  ~~~(d)\includegraphics[width=2.8in]{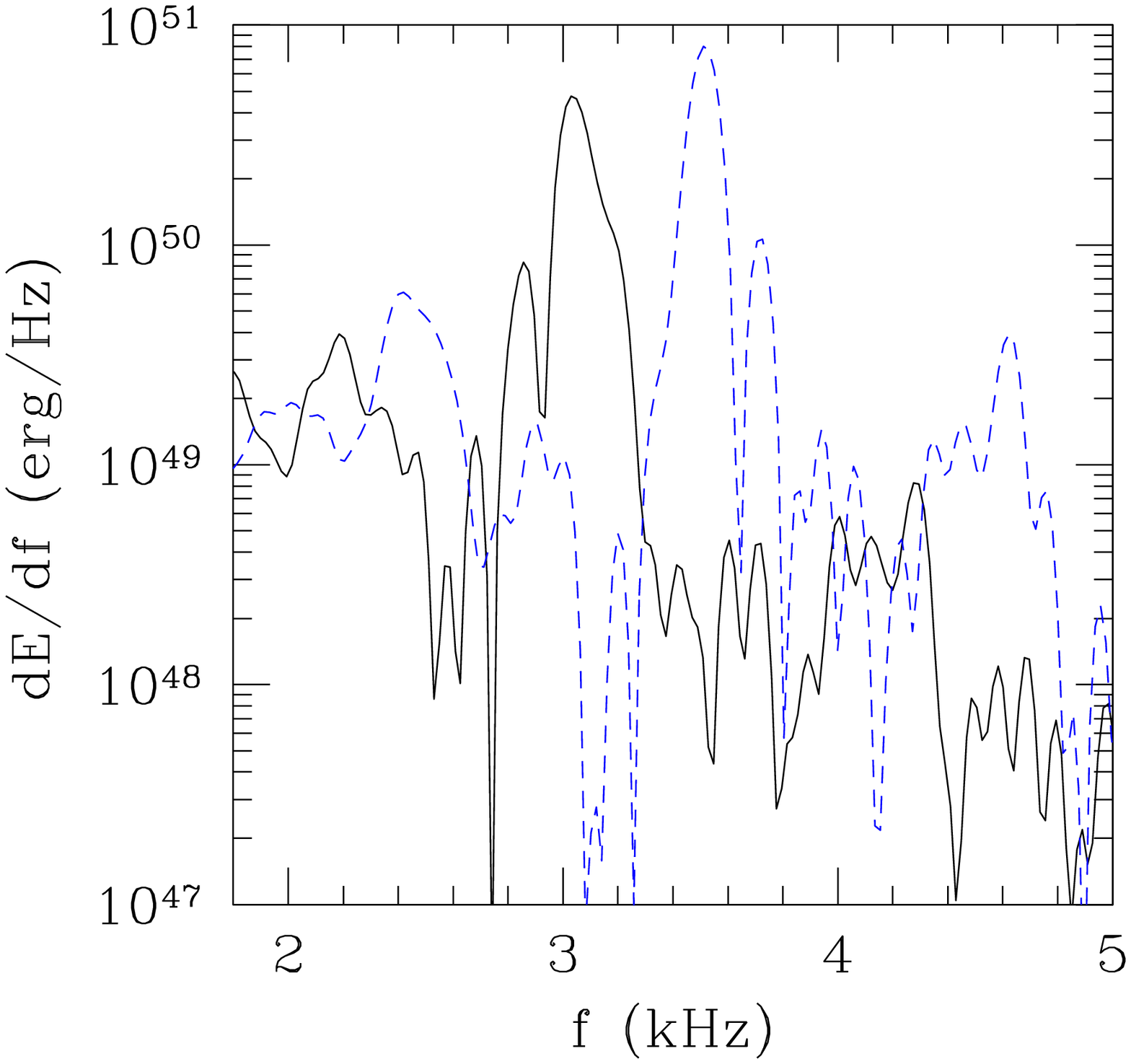}
\end{center}
\vspace{-3mm}
\caption{Fourier power spectrum of gravitational waves $dE/df$
(a) for models SLy1313a (solid curve) and SLy125135a (dashed curve),
(b) for models SLy1313a (solid curve) and SLy135135b (dashed curve), 
(c) for models SLy1313a (dashed curve), SLy1313c (solid curve), and
SLy1313d (long-dashed curve), and
(d) for models SLy1212b (solid curve) and FPS1212b (dashed curve). 
Since the simulations are started when the frequency of
gravitational waves is $\sim 1$ kHz, the spectrum
for $f < 1$ kHz is not correct. 
The dotted curve in the panel (a) denotes the analytical result 
of $dE/df$ in the second post Newtonian and point-particle approximation.
The real spectrum for $f \alt 1$ kHz is approximated by the dotted curves. 
\label{FIG17}
}
\end{figure*}

\begin{figure*}[thb]
\vspace{-4mm}
\begin{center}
  (a)\includegraphics[width=3.in]{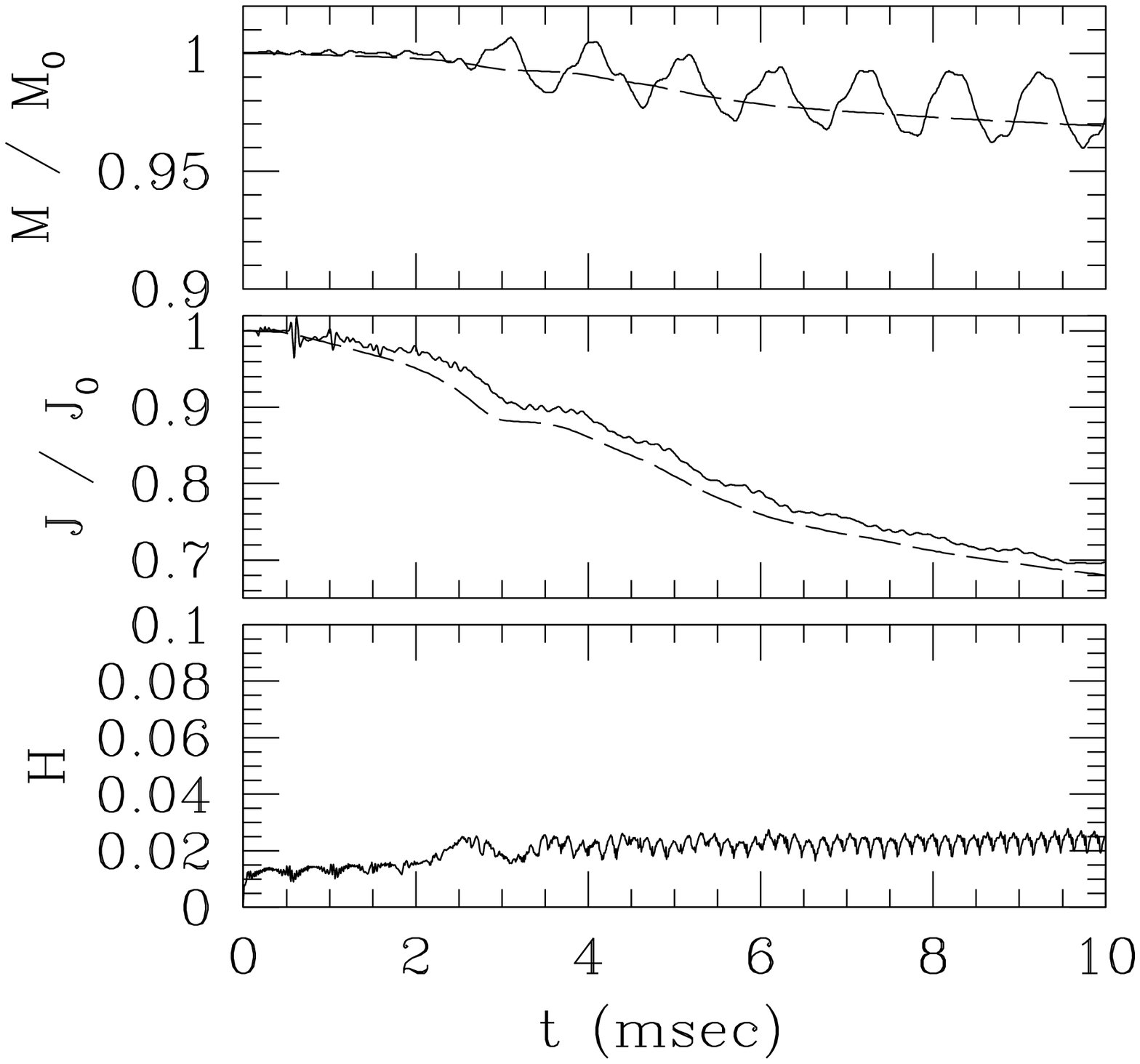}
  ~~~(b)\includegraphics[width=3.in]{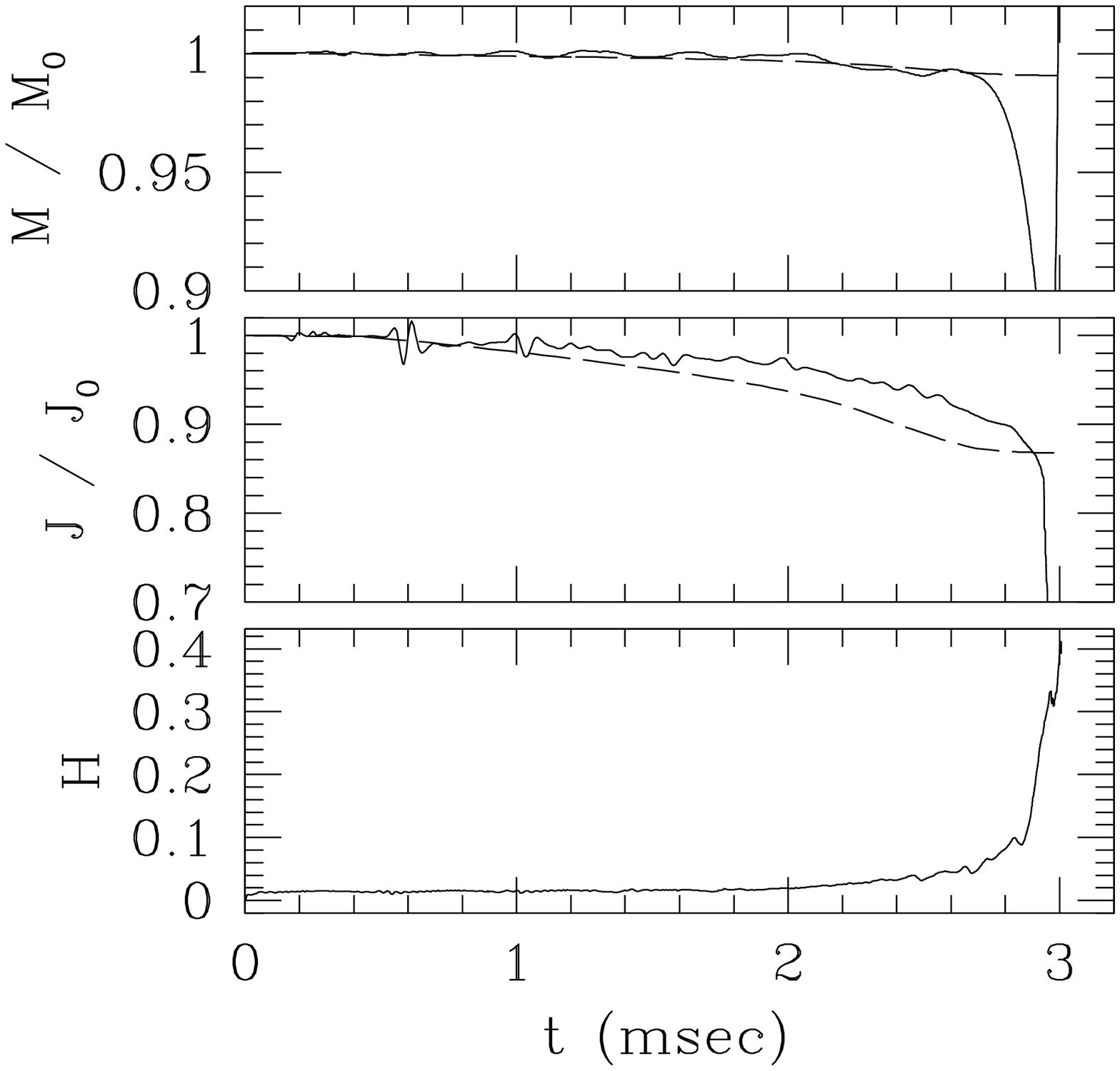}
\end{center}
\vspace{-3mm}
\caption{Evolution of ADM mass and angular momentum in units of
their initial values $M_0$ and $J_0$, and
violation of the Hamiltonian constraint (a) for model SLy1313a and
(b) for model SLy1414a. 
In the upper two panels, the solid curves denote $M/M_0$ and $J/J_0$
computed from Eqs. (\ref{eqm00}) and (\ref{eqj00}), 
while the long dashed curves denote $1-\Delta E(t)/M_0$ and
$1-\Delta J(t)/J_0$, respectively (see Eqs. (\ref{eqm01}) and (\ref{eqj01})). 
\label{FIG18}
}
\end{figure*}

To determine the characteristic frequency of gravitational waves,
we carried out the Fourier analysis. In Fig. \ref{FIG17}, 
the power spectrum $dE/df$ is presented 
(a) for models SLy1313a and SLy125135a, 
(b) for SLy1313a and SLy135135b, 
(c) for SLy1313a, SLy1313c, SLy1313d, and (d) for SLy1212b and FPS1212b,
respectively. Since the simulations were started with the initial 
condition of the orbital period $\sim 2$ ms 
(i.e., frequency of gravitational waves is $\sim 1$ kHz),
the spectrum of inspiraling binary neutron stars
for $f < 1$ kHz cannot be taken into account. Thus, 
only the spectrum for $f \agt 1$ kHz should be paid attention. 
In the panel (a), we plot the following Fourier power spectrum 
of two point particles in circular orbits in 
the second post Newtonian approximation \cite{blanchet}: 
\beqn
{dE \over df}&&={\mu M\over 3f}x\biggl[1
-\biggl({3\over 2}+{\mu \over 6M}\biggr)x \nonumber \\
&&~~~~~~
+3\biggl(-{27\over 8}+{19\mu \over 8M}-{\mu^2 \over 24M^2}\biggr)x^2
\biggr]. 
\eeqn
Here, $\mu$ and $M$ denote the reduced mass and the total mass of
the binary, and $x \equiv (M\pi f)^{2/3}$. We note that 
the third post Newtonian terms does not significantly modify 
the spectrum since their magnitude is $\sim 0.01$ 
of the leading-order term. 
Thus, the dotted curve should be regarded as the plausible
Fourier power spectrum for $f \alt 1$ kHz. 

Figure \ref{FIG17} shows that a sharp characteristic peak is present at 
$f=3$--4 kHz irrespective of models in which hypermassive neutron stars
with a long lifetime ($\agt 10$ ms) are formed (see also Table III for the
list of the characteristic frequency). This is associated with quasiperiodic 
gravitational waves emitted by the formed hypermassive neutron stars.
The amplitude of the peak is much higher than that in the $\Gamma=2$ 
equation of state \cite{STU}. The reason is that
with the realistic equations of state, 
the ellipticity of the formed hypermassive neutron stars is
much larger, and as a result, quasiperiodic 
gravitational waves of a higher amplitude are emitted. 
Also, the elliptic structure of the hypermassive 
neutron stars is preserved for a long time duration. 
These effects amplify the peak amplitude in the Fourier power spectrum.

The energy power spectra for models SLy1313a and SLy125135a
are very similar reflecting the fact that 
the waveforms for these two models are very similar (Fig. \ref{FIG17}(a)). 
This indicates that the spectral shape depends very weakly on
the mass ratio $Q_M$ as far as it is in the range between 0.9 and 1.
On the other hand, three peaks are present
at $f \approx 2.6$, 3.6, and 4.5 kHz in 
the energy power spectrum for model SLy135135b (Figs. \ref{FIG17}(b)).
Thus, the spectral shape is quite different from that for model SLy1313a 
although the total mass is only slightly different between two models.
The reason is that the amplitude of the quasiradial oscillation
of the hypermassive neutron star is very large and the characteristic
radius varies for a wide range for model SLy135135b, 
inducing the modulation of the wave frequency. Indeed, the difference
of the frequencies for the peaks is approximately 
equal to that of the quasiradial oscillation $\sim 1$ kHz. 
As a result, the intensity of the power spectrum is dispersively
distributed to multi peaks in this case, and 
the amplitude for the major peak at $f \sim 3.6$ kHz is suppressed.
The similar feature is also found for models SLy1313c and FPS1212b
for which the hypermassive neutron stars collapse to a black hole
within $\sim 10$ ms.


Figures \ref{FIG17}(c) and (d) illustrate that 
the amplitude and the frequency for the peak
around $f \sim 3$--4 kHz depend on the total mass, 
the value of $\Gamma_{\rm th}$, and the equations of state as in the case 
of gravitational waveforms. Figure \ref{FIG17}(c) indicates that 
for the larger total mass (but with
$M < M_{\rm thr}$), the peak frequency becomes higher.
Also, with the increase of the value of $\Gamma_{\rm th}$,
the peak frequency is decreased since the formed hypermassive neutron
star becomes less compact. As Fig. \ref{FIG17}(c) shows, 
the peak frequency is larger for the FPS equation of state than
for the SLy one for the same value of the total mass. 
This is also due to the fact that the hypermassive neutron star
in the FPS equation of state is more compact. 

The effective amplitude of gravitational waves observed 
from the most optimistic direction (which is parallel to the 
axis of the angular momentum) is proportional to $\sqrt{dE/df}$ 
in the manner 
\beqn
h_f && \equiv \sqrt{|\bar R_+|^2 + |\bar R_{\times}|^2}f \nonumber \\
&&=1.8 \times 10^{-21}
\biggl({dE/df \over 10^{51}~{\rm erg/Hz}}\biggr)^{1/2}
\biggl({100~{\rm Mpc} \over r}\biggr), \label{heff}
\eeqn
where $r$ denotes the distance from the source, and
$\bar R_{+,\times}$ are the Fourier spectrum of $R_{+,\times}$. 
Equation (\ref{heff}) implies that the 
effective amplitude of the peak is about 4--5 times larger
than that at 1 kHz. Furthermore, the amplitude of the 
peak in reality should be larger than that presented here, 
since we stopped simulations at $t \sim 10$ ms to save the computational time, 
and hence, the integration time 
$\sim 10$ ms is much shorter than the realistic value. Extrapolating 
the decrease rate of the angular momentum, the
hypermassive neutron star will dissipate sufficient angular momentum
by gravitational radiation until a black hole is formed.
As indicated in Secs. \ref{sec:gen} and \ref{sec:dedt}, 
the lifetime would be $\sim 30$--50 ms 
for models SLy1313a and SLy125135a and $\sim 50$--100 ms for model SLy1212b. 
Thus, we may expect that the emission will continue for such
time scale and the effective amplitude of the peak of $f \sim 3$--4 kHz
would be in reality amplified by a factor of 
$\sim 3^{1/2}$--$10^{1/2} \approx 2$--3 to be 
$\sim 3$--$5 \times 10^{-21}$ at a distance of 100 Mpc. 
Although the sensitivity of laser interferometric gravitational
wave detectors for $f > 1$ kHz is limited by the shot noise of the laser,
this value is larger than the planned noise level of the advanced
laser interferometer $\approx 10^{-21.5}(f/1~{\rm kHz})^{3/2}$ \cite{KIP}. 
It will be interesting to search for such quasiperiodic signal of
high frequency if the chirp signal of gravitational waves from inspiraling
binary neutron stars of distance $r \alt 100$ Mpc are detected
in the near future.

Detection of the quasiperiodic gravitational waves will demonstrate that 
a hypermassive neutron star of a lifetime much longer than 10 ms 
is formed after the merger. 
Since the total mass of the binary should be determined by 
the data analysis for the chirp signal emitted in the inspiral phase
\cite{CF}, the detection of the quasiperiodic gravitational waves
will provide the lower bound of 
the binary mass for the prompt formation of a black hole $M_{\rm thr}$. 
As found in this paper, the value of $M_{\rm thr}$ depends 
sensitively on the equations of state.
Furthermore, the values of $M_{\rm thr}$
($M_{\rm thr}\sim 2.7M_{\odot}$ and 
$\sim 2.5M_{\odot}$ for the SLy and FPS equations of state, respectively) 
are very close to the total mass of the binary neutron stars
observed so far \cite{Stairs}. 
Therefore, the merge of mass $\sim M_{\rm thr}$ is likely
to happen frequently, and thus, 
the detection of gravitational waves from 
hypermassive neutron stars will lead to constraining 
the equations of state for neutron stars. 
For example, if quasiperiodic gravitational waves 
are detected from a hypermassive neutron star formed after the 
merger of a binary neutron star of mass 
$M = 2.6M_{\odot}$, the FPS equation of state should be rejected. 
As this example shows, the merit of this method is that 
only one detection will significantly constrain
the equations of state. The further detail about this method is
described in \cite{S05}. 

\subsubsection{Calibration of radiation reaction}

Figure \ref{FIG18} shows evolution of the ADM mass and the angular momentum
computed in a finite domain by Eqs. (\ref{eqm00}) and (\ref{eqj00}) 
as well as the violation of the Hamiltonian constraint
$H$ defined in Eq. (\ref{vioham}) for models SLy1313a and SLy1414a.
The solid curves in the upper two panels denote $M$ and $J$ while
the dashed curves are 
$M_0-\Delta E$ and $J_0-\Delta J$ which are computed from the
emitted energy and angular momentum of gravitational waves. 
The ADM mass and angular momentum computed by two methods should be identical
because of the presence of the conservation laws. The figure indicates that 
the conservation holds within $\sim 2$\% error for the ADM mass and
angular momentum (except for the case that a black hole is present).
This implies that radiation reaction of gravitational waves is 
taken into account within $\sim 2\%$ in our numerical simulation.

The error in the angular momentum conservation is generated
mainly in the late inspiral phase with $t \alt 2$ ms 
in which $L$ is smaller than the wavelength of gravitational
waves and the radiation reaction cannot be evaluated accurately. 
To improve the accuracy for the conservation in this phase, 
it is required to take a sufficiently large value of $L$
that is larger than the wavelength. On the other hand, 
the magnitude of the error does not change much after the
formation of the hypermassive neutron stars for 
$t \agt 2$ ms as found in Fig. \ref{FIG18}(a). This implies that
the radiation reaction of gravitational waves to the angular momentum
for the formed hypermassive neutron stars is computed within 1\% error. 

The bottom panels show that the violation of the Hamiltonian
constraint is of order 0.01 in the absence of black holes.
Also noteworthy is that the violation does not grow but remain small 
in the absence of black holes. This strongly indicates that
simulations will be continued for an arbitrarily long duration 
for spacetimes of no black hole.
On the other hand, the computation crashes soon after the formation of 
a black hole for model SLy1414a. This is mainly due to the fact that
the resolution around the black hole is too poor.
If one is interested in the longterm evolution of the formed black hole, 
it is obviously necessary to improve the resolution around the black hole
to overcome this problem.  
In the current simulation, the radius of the apparent horizon
is covered only by $\sim 5$ grid points. The axisymmetric simulations
for black hole formation (e.g., \cite{shiba2d,S03})
have experimentally shown that more than 10 grid points
for the radius of the apparent horizon will be necessary 
to follow evolution of the formed black hole for $30M \sim 0.4$ ms.
To perform such a better-resolved and longterm simulation
with $L \agt 0.5 \lambda_0$, the grid size more than (1500, 1500, 750)
is required, implying that a powerful
supercomputer, in which the memory and the computational speed
are by a factor of $\agt 10$ as large as those of 
the present computational resources, is necessary. 
If the computational resources are not improved in 
the near future, adopting the adaptive mesh refinement technique 
will be inevitable for following the evolution of the black hole 
\cite{AMR}. 

\section{Summary}

We performed fully general relativistic simulations 
for the merger of binary neutron stars adopting realistic equations
of state. Since the stiffness is significantly different from that
in the $\Gamma=2$ equation of state adopted in the 
previous works (e.g., \cite{STU}), several new features have emerged. 
The following is the summary of the results obtained in this paper: 

\noindent
1: If the ADM mass of the system is larger than $\sim 2.7 M_{\odot}$
($\sim 2.5 M_{\odot}$), 
a black hole is promptly formed in the SLy (FPS) equation of state.
Otherwise, a hypermassive neutron star of ellipsoidal shape is formed.
This indicates that the threshold mass depends on the
equations of state and the values are very close to those
for observed binary neutron stars. 

\noindent
2: In the black hole formation, most of mass elements are
swallowed into the horizon, and hence, 
the disk mass around the black hole 
is much smaller than 1\% of the total baryon rest-mass as far as the
mass ratio $Q_M$ is larger than 0.9. Although the disk is hot
with the thermal energy $\sim 10$--20 MeV,
the total thermal energy which is available for the neutrino emission
is expected to be at most $\sim 10^{50}$ erg. Since the
pair annihilation of the neutrino and antineutrino to the
electron-positron pair would be $< 10^{-4}$ \cite{RJ},
it seems to be very difficult to generate cosmological gamma-ray bursts
in this system. 

\noindent
3: The nondimensional angular momentum parameter
($J/M^2$) of the formed Kerr black hole 
is in the range between 0.7 and 0.8. Then, for the system of mass
$\sim 2.8M_{\odot}$, the frequency of gravitational waves
associated with quasinormal mode ringing of $l=m=2$ modes 
would be $\sim 6.5$--7 kHz, which is too high for gravitational waves
to be detected by laser interferometric detectors. 

\noindent
4: The hypermassive neutron stars formed after the merger
have a large ellipticity with the axial ratio $\sim 0.5$.
They rotate with the period of $\sim 0.5$--1 ms, and thus, 
become strong emitters of quasiperiodic gravitational waves of
a rather high frequency $f \sim 3$--4 kHz. Although the frequency is
far out of the best sensitive frequency range 
of the laser interferometric gravitational wave detectors, 
the effective amplitude of gravitational waves is 
very high as several $\times 10^{-21}$ at 
a distance of $r \sim 100$ Mpc. Thus, if the merger happens for
$r < 100$ Mpc, such gravitational waves may be detectable by
advanced laser interferometers. The detection of these quasiperiodic
gravitational waves will be used for constraining the equations of
state for nuclear matter. 

\noindent
5: Because of the larger emission rate of gravitational waves, 
the angular momentum of the hypermassive neutron star is dissipated 
in a fairly short time scale $\alt 100$ ms for the mass
$M \sim 2.4$--$2.7 M_{\odot}$. Also, due to a high degree of
nonaxial symmetry, the angular momentum is transferred outward
by the hydrodynamic interaction. As a result of these effects, the 
hypermassive neutron stars collapse to a black hole within 100 ms.
This time scale is much shorter than
the viscous dissipation time scale and the transport time scale of
the angular momentum by magnetic fields \cite{BSS}. 
Therefore, the gravitational radiation or the outward angular momentum
transfer by the hydrodynamic 
interaction plays the most important role. 

\noindent
6: The thermal energy of the outer region of the hypermassive
neutron stars is high as $\sim 10$--20 MeV, and the total
emission rate of the neutrino energy is estimated as
$\sim 10^{53}~{\rm erg/s}$. The thermal energy is 
generated by the shocks due to the multiple collisions between the
spiral arms and the oscillating hypermassive neutron star.
Thus, the hypermassive neutron star will be a strong emitter of
neutrinos. However, the emission time scale is $\sim 1$--10 s which
is much shorter than the lifetime $< 100$ ms. This implies that
the neutrino cooling plays a minor role in the evolution of
the hypermassive neutron star. 

\noindent
7: The mass difference with the mass ratio $Q_M \sim 0.9$ does not
modify the dynamics of the merger and the outcome after the merger
significantly from that with $Q_M =1$.  This disagrees with the
previous result which was obtained in the simulations performed with the
$\Gamma=2$ equation of state \cite{STU}. The reason is that with the
realistic equations of state, the radius of neutron stars is small as
$\sim 11$--12 km depending weakly on the mass in contrast to that in
the $\Gamma=2$ equations of state.

\acknowledgments

Numerical computations were performed on the FACOM VPP5000 machines 
at the data processing center of NAOJ. 
This work was in part supported by Monbukagakusho 
Grant (Nos. 15037204, 15740142, and 16029202).

\end{document}